\documentstyle[preprint,prd,aps,amsfonts,epsf]{revtex} 

\newcommand{\mevc} {\ifmmode {\rm MeV}/c \else MeV$/c$\fi}
\newcommand{\mevcc} {\ifmmode {\rm MeV}/c^2 \else MeV$/c^2$\fi}
\newcommand{\gevc} {\ifmmode {\rm GeV}/c \else GeV$/c$\fi}
\newcommand{\gevcc} {\ifmmode {\rm GeV}/c^2 \else GeV$/c^2$\fi}
\newcommand{\ra}   {\rightarrow}

\newcommand{\jpsi} {\ifmmode J/\psi \else $J/\psi$\fi}
\newcommand{\vtd}  {\ifmmode |V_{td}| \else $|V_{td}|$\fi}
\newcommand{\vtb}  {\ifmmode |V_{tb}| \else $|V_{tb}|$\fi}
\newcommand{\vts}  {\ifmmode |V_{ts}| \else $|V_{ts}|$\fi}
\newcommand{\vcb}  {\ifmmode |V_{cb}| \else $|V_{cb}|$\fi}

\newcommand{\SC}   {sample composition}

\newcommand{\xs} {\ifmmode x_{\mbox{\sl s}}
                       \else $x_{\mbox{\sl s}}$\fi}
\newcommand{\xd} {\ifmmode x_d \else $x_d$\fi}
\newcommand{\lxy} {\ifmmode L_{\rm xy} \else $L_{\rm xy}$\fi}
\newcommand{\LxyB} {\ifmmode L_{\rm xy}^B \else $L_{\rm xy}^B$\fi}
\newcommand{\LxyD} {\ifmmode L_{\rm xy}^D \else $L_{\rm xy}^D$\fi}
\newcommand{\ctau} {\ifmmode c\tau \else $c\tau$\fi}
\newcommand{\Pt} {\ifmmode p_{\rm t} \else $p_{\rm t}$\fi}
\newcommand{\ed} {\ifmmode \varepsilon {\cal D}^2 
		    \else $\varepsilon {\cal D}^2$\fi}
\newcommand{\ptrel} {\ifmmode p_{\rm t}^{\rm rel} 
                       \else $p_{\rm t}^{\rm rel}$\fi}

\newcommand{\As}  {\ifmmode {\cal A} \else ${\cal A}$\fi}
\newcommand{\Dil}{\ifmmode {\cal D} \else ${\cal D}$\fi}
\newcommand{\Do} {\ifmmode {\cal D}_0 \else ${\cal D}_0$\fi}
\newcommand{\Dx} {\ifmmode {\cal D}_+ \else ${\cal D}_+$\fi}
\newcommand{\dmd}{\ifmmode \Delta m_d \else $\Delta m_d$\fi}
\newcommand{\Kf}  {\ifmmode {\cal K} \else ${\cal K}$\fi}
\newcommand{\Kfact}{${\cal K}$-factor}

\newcommand{\Bds} {\ifmmode B^{**} \else $B^{**}$\fi}
\newcommand{\Dds} {\ifmmode D^{**} \else $D^{**}$\fi}
\newcommand{\pids}{\ifmmode \pi_{**} \else $\pi_{**}$\fi}

\newcommand{\tbx}{\ifmmode \tau_{+} \else $\tau_{+}$\fi}
\newcommand{\tbo}{\ifmmode \tau_{0} \else $\tau_{0}$\fi}

\newcommand{\edkl}{\ifmmode \epsilon^D_{kh} \else $\epsilon^D_{kh}$\fi}
\newcommand{\fbkl}{\ifmmode \phi_{kh} \else $\phi_{kh}$\fi}

\newcommand{\edkh}{\ifmmode \epsilon^D_{kh} \else $\epsilon^D_{kh}$\fi}
\newcommand{\fbkh}{\ifmmode \phi_{kh} \else $\phi_{kh}$\fi}

\newcommand{\bgam}   {\ifmmode \beta\gamma \else $\beta\gamma$\fi}

\newcommand{\SCpars} {$f^{**}$, $R_f$, $\tau_{+}/\tau_{0}$ and $\tau_{0}$}

\newcommand{\postscript}[2] {\setlength{\epsfxsize}{#2\hsize} 
  \centerline{\epsfbox{#1}}}

\newcommand{\kp} {\ifmmode \ell^+ \overline{D}{^0}, 
  \overline{D}{^0} \to K\pi \else $\ell^+ \overline{D}{^0}$, 
  $\overline{D}{^0} \to K^+\pi^-$\fi}
\newcommand{\kpp} {\ifmmode \ell^+ D^-, D^- \to K\pi\pi \else 
  $\ell^+ D^-$, $D^- \to K^+\pi^-\pi^-$\fi}
\newcommand{\kps} {\ifmmode \ell^+ D^{*-}, \overline{D}{^0} \to K\pi \else 
  $\ell^+ D^{*-}$, $\overline{D}{^0} \to K^+\pi^-$\fi}
\newcommand{\ktps} {\ifmmode \ell^+ D^{*-}, \overline{D}{^0} \to K3\pi \else 
  $\ell^+ D^{*-}$, $\overline{D}{^0} \to K^+\pi^-\pi^+\pi^-$\fi}
\newcommand{\kpzs} {\ifmmode \ell^+ D^{*-}, \overline{D}{^0} \to K\pi\pi^0 
  \else $\ell^+ D^{*-}$, $\overline{D}{^0} \to K^+\pi^-\pi^0$\fi}

\begin{document}

\draft
\preprint{\begin{minipage}[t]{3in} 
                \flushright
                FERMILAB-PUB-98/188-E     
        \\      CDF/PUB/BOTTOM/CDFR/4572  
        \\      PRD Version 1.0           
        \\      \today 
        \\  $\;$
        \\  $\;$
                \end{minipage} }

\title{\boldmath Measurement of the $B^0_d$-$\overline{B}{^0_d}$ 
flavor oscillation frequency and study of same side flavor tagging 
of $B$ mesons  in $p\bar p$ collisions}

\tightenlines

\def\r#1{\ignorespaces $^{#1}$}
\font\eightit=cmti8

\author{ 
\hfilneg
\begin{sloppypar}
\noindent
F.~Abe,\r {17} H.~Akimoto,\r {39}
A.~Akopian,\r {31} M.~G.~Albrow,\r 7 A.~Amadon,\r 5 S.~R.~Amendolia,\r {27} 
D.~Amidei,\r {20} J.~Antos,\r {33} S.~Aota,\r {37}
G.~Apollinari,\r {31}
% ...
\end{sloppypar}
\vskip .026in
\begin{center}
(CDF Collaboration)
\end{center}
%
\vskip .526in
\begin{center}
\r 1  {\eightit Argonne National Laboratory, Argonne, Illinois 60439} \\
\r 2  {\eightit Istituto Nazionale di Fisica Nucleare, University of Bologna,
I-40127 Bologna, Italy} \\
\r 3  {\eightit Brandeis University, Waltham, Massachusetts 02254} \\
\r 4  {\eightit University of California at Los Angeles, Los 
Angeles, California  90024} \\  
\r 5  {\eightit University of Chicago, Chicago, Illinois 60637} \\
\r 6  {\eightit Duke University, Durham, North Carolina  27708} \\
\r 7  {\eightit Fermi National Accelerator Laboratory, Batavia, Illinois 
60510} \\
\r 8  {\eightit University of Florida, Gainesville, Florida  32611} \\
\r 9  {\eightit Laboratori Nazionali di Frascati, Istituto Nazionale di Fisica
               Nucleare, I-00044 Frascati, Italy} \\
\r {10} {\eightit University of Geneva, CH-1211 Geneva 4, Switzerland} \\
\r {11} {\eightit Harvard University, Cambridge, Massachusetts 02138} \\
\r {12} {\eightit Hiroshima University, Higashi-Hiroshima 724, Japan} \\
\r {13} {\eightit University of Illinois, Urbana, Illinois 61801} \\
\r {14} {\eightit Institute of Particle Physics, McGill University, Montreal 
H3A 2T8, and University of Toronto,\\ Toronto M5S 1A7, Canada} \\
\r {15} {\eightit The Johns Hopkins University, Baltimore, Maryland 21218} \\
\r {16} {\eightit Institut f\"{u}r Experimentelle Kernphysik, 
Universit\"{a}t Karlsruhe, 76128 Karlsruhe, Germany} \\
\r {17} {\eightit National Laboratory for High Energy Physics (KEK), Tsukuba, 
Ibaraki 305, Japan} \\
\r {18} {\eightit Ernest Orlando Lawrence Berkeley National Laboratory, 
Berkeley, California 94720} \\
\r {19} {\eightit Massachusetts Institute of Technology, Cambridge,
Massachusetts  02139} \\   
\r {20} {\eightit University of Michigan, Ann Arbor, Michigan 48109} \\
\r {21} {\eightit Michigan State University, East Lansing, Michigan  48824} \\
\r {22} {\eightit University of New Mexico, Albuquerque, New Mexico 87131} \\
\r {23} {\eightit The Ohio State University, Columbus, Ohio  43210} \\
\r {24} {\eightit Osaka City University, Osaka 588, Japan} \\
\r {25} {\eightit Universita di Padova, Istituto Nazionale di Fisica 
          Nucleare, Sezione di Padova, I-35131 Padova, Italy} \\
\r {26} {\eightit University of Pennsylvania, Philadelphia, 
        Pennsylvania 19104} \\   
\r {27} {\eightit Istituto Nazionale di Fisica Nucleare, University and Scuola
               Normale Superiore of Pisa, I-56100 Pisa, Italy} \\
\r {28} {\eightit University of Pittsburgh, Pittsburgh, Pennsylvania 15260} \\
\r {29} {\eightit Purdue University, West Lafayette, Indiana 47907} \\
\r {30} {\eightit University of Rochester, Rochester, New York 14627} \\
\r {31} {\eightit Rockefeller University, New York, New York 10021} \\
\r {32} {\eightit Rutgers University, Piscataway, New Jersey 08855} \\
\r {33} {\eightit Academia Sinica, Taipei, Taiwan 11530, Republic of China} \\
\r {34} {\eightit Texas A\&M University, College Station, Texas 77843} \\
\r {35} {\eightit Texas Tech University, Lubbock, Texas 79409} \\
\r {36} {\eightit Istituto Nazionale di Fisica Nucleare, University of Trieste/
Udine, Italy} \\
\r {37} {\eightit University of Tsukuba, Tsukuba, Ibaraki 315, Japan} \\
\r {38} {\eightit Tufts University, Medford, Massachusetts 02155} \\
\r {39} {\eightit Waseda University, Tokyo 169, Japan} \\
\r {40} {\eightit University of Wisconsin, Madison, Wisconsin 53706} \\
\r {41} {\eightit Yale University, New Haven, Connecticut 06520} \\
\end{center}
}

\date{\today}

\maketitle

\begin{abstract}
$B^0_d$-$\overline{B}{}^0_d$ oscillations are observed in 
``self-tagged'' samples of partially reconstructed $B$ mesons decaying 
into a lepton and a charmed meson collected in $p\bar p$ collisions
at $\sqrt{s} = 1.8$ TeV.
A flavor tagging technique is employed which relies 
upon the correlation between the flavor of $B$ mesons 
and the charge of nearby particles.
We measure the flavor oscillation frequency to be 
$\Delta m_d = 0.471 ^{+0.078}_{-0.068} \pm 0.034 {\; \rm ps}^{-1}$.
The tagging method is also demonstrated in exclusive samples of
$B^+_u \rightarrow J/\psi K^+$ and $B^0_d \rightarrow J/\psi K^{*0}(892)$,
where similar flavor-charge correlations are observed.
The tagging characteristics of the various samples are compared with 
each other, and with Monte Carlo simulations.
\end{abstract}

\narrowtext
\tightenlines

\section{Introduction} 
\label{sec:intro}

The study of $B$ mesons has been important
for understanding the relationships between the
weak interaction and the mass eigenstates of quarks, described in
the Standard Model by the Cabibbo-Kobayashi-Maskawa (CKM) matrix\cite{CKM}. 
Early studies were based on
branching fraction and lifetime measurements.
However, since the observations of $B^0$-$\overline{B}{}^0$
mixing, first in an unresolved mixture of $B^0_d$ and $B^0_s$
by UA1~\cite{UA1mix},
and then specifically for the $B^0_d$ by ARGUS~\cite{ARGUSmix}, 
a new window on the CKM matrix
was opened. 
$B^0$ mixing, analogous to $K^0$ mixing,
is possible via higher order weak interactions,
and is governed by the mass difference $\Delta m$ between the
two mass eigenstates. Unlike the $K^0$ system,
the $B^0$ mixing amplitude is dominated by the exchange 
of virtual top quarks,
and so provides a view of weak charged current transitions between 
a top quark and the quarks composing the $B^0_{d,s}$.

Mixing measurements are predicated upon 
identifying the  ``flavor'' of the $B^0$ meson
at its time of formation and again when it decays,
where by ``flavor'' we mean whether the meson contained 
a $b$ or $\bar{b}$ quark. 
Determination of the initial flavor is the primary difficulty, as
knowledge of the decay flavor is usually a byproduct of
the $B$ reconstruction, even if it is only partial. 

The effective size of flavor tagged $B$ samples
is a critical limitation of current measurements, 
especially for exclusive $B$ reconstructions.
This fact has motivated efforts to develop
a variety of tagging techniques to fully exploit existing data.
There has been considerable progress in recent years in utilizing
a variety of tagging methods and $B^0$ samples,
as illustrated by the
diversity of mixing measurements~\cite{ref:mixing}.
Even though a new generation of high statistics $B$ experiments
will soon come on-line~\cite{NewEx},
many tagging-based studies---such as $CP$ violation
in $B$ mesons---will still be statistics limited.
Thus, improvements in tagging capabilities
will be valuable in the next generation of $B$ experiments
as well as for the current ones.

We have reported in an earlier {\em Letter}~\cite{LepDPRL}
the development and application of a ``self-tagging'' method 
based on the proposal~\cite{Rosner} that the electric charge of
particles produced ``near'' the reconstructed $B$ meson can 
be used to determine its initial flavor.
Such correlations, first observed in 
$e^+e^- \rightarrow Z^0 \rightarrow b\bar{b}$ events by
OPAL~\cite{OPALCorr},  are expected to arise from particles
produced from decays of the orbitally excited $B^{**}$ mesons,
as well as from the fragmentation chain that formed the $B$. 
We refer to this approach as ``Same Side Tagging'' (SST),
in contrast to other tagging methods which rely upon the other
$b$-hadron in the event.

We applied SST to a large
sample of $B_{u,d} \rightarrow \ell D^{(*)} X$ decays:
the expected time dependent flavor oscillation was observed,
and its frequency $\Delta m_d$ was measured with a precision
similar to other single tagging results.
In addition to the intrinsic interest 
of obtaining a supplementary measurement of $\Delta m_d$, this result
also demonstrated that this type of tagging method is
effective even in the complex environment of a hadron collider.
A variant of this approach has also been studied by
ALEPH~\cite{ALEPHSST} in exclusively reconstructed $B$'s at the
$Z^0$ pole.

In this paper we describe in detail the SST method we have 
developed and its previously reported application to
$B_{u,d} \rightarrow \ell D^{(*)} X$ decays. 
Experimental complications surrounding the use of these decays
are described in detail, {\it i.e.}, both the cross-talk between 
$B^+_u$ and $B^0_d$, and the contamination 
from tagging on $B$ decay products.
The value of $\Delta m_d$, as well as the purity 
of the flavor-charge correlations, are reported.

This paper extends the application of SST to
two fully reconstructed $B$ decays which offer another
test of its effectiveness:
$B^+_u \rightarrow J/\psi K^+$ and
$B^0_d \rightarrow J/\psi K^{*0}(892)$.\footnote{
Reference to a specific particle state implies the charge 
conjugate state as well; exceptions are clear from the context.} 
Although our samples are too small to yield precise tagging results,
they are the largest currently in existence
and serve as a prototype for tagging 
$B^0_d \rightarrow J/\psi K^0_S$~\cite{OPALks,CDFpsiKs}, 
the centerpiece of
future $CP$ violation studies with $B$ mesons~\cite{NewEx}.
The tagging results from the $J/\psi K$ samples are compared 
to those from $\ell D^{(*)}$, and also to Monte Carlo simulations.
The simulation offers further insights into the behavior 
of this SST method.

This paper is structured as follows. We review the relevant 
aspects of our detector and data collection in Sec.~\ref{Sec:det}.
Section~\ref{Sec:mix} summarizes $B^0$-$\overline{B}{}^0$ mixing, 
and is followed by some remarks on tagging 
and a description of our specific SST method in Sec.~\ref{Sec:sst}.
Same Side Tagging is applied to the $\ell D^{(*)}$ sample in
Sec.~\ref{Sec:lepd}, which includes discussion of $B$ reconstruction,
sample composition, proper decay time measurement and corrections, 
the tagging asymmetries,
and finally extraction of $\Delta m_d$ and tagging dilutions.
This completes our main result.

Having established the technique in $\ell D^{(*)}$, we extend 
SST to the exclusive $J/\psi K$ modes in Sec.~\ref{Sec:psik}. 
We discuss the sample selection,
the fitting method, and the resultant tagging dilutions.
Special attention is given to handling tagging biases.
Finally in Sec.~\ref{Sec:compr} we present some checks of our
measurements and compare the
behavior of this tagger in these two different types of
$B$ decays. Aspects of the data are also compared to
Monte Carlo simulations, and the behavior of this SST method
is discussed. We close with a few remarks concerning
future applications of this type of SST method.

\section{The CDF detector and data collection}
\label{Sec:det}
\subsection{Apparatus}
\label{SubSec:det}
The data discussed here were collected using the CDF detector
in the Tevatron Run I period 
during 1992-1996, and comprise approximately 110 pb$^{-1}$ of integrated
luminosity of $p\bar{p}$ collisions at $\sqrt{s} = 1.8$ TeV.
Details of the CDF detector have been previously published~\cite{CDFDET,SVX},
and only the features relevant to this analysis are reviewed here:
the tracking system by which charged particles are identified and their
momenta precisely measured, the central calorimeters for 
electron identification, and the muon chambers for muon identification. 
Our coordinate system is such that the (spherical) polar angle $\theta$ is
measured from the outgoing proton direction ($+z$-axis) and the azimuthal
angle $\phi$ from the plane of the Tevatron.

The tracking system consists of three detectors immersed in a 1.4 T magnetic 
field generated by a superconducting solenoid 1.5 m in radius. The innermost
tracking device is a silicon microstrip vertex detector (SVX)~\cite{SVX},
which provides spatial measurements projected onto the plane transverse 
to the beam line. The SVX active
region is 51 cm long and composed of two 25 cm long cylindrical barrels. 
Each barrel has four layers of silicon strip detectors, 
ranging in radius $r$ from
3.0 to 7.9 cm from the beam line. The impact parameter resolution of the
SVX is $\sigma_d (p_T) = (13 + 40/p_T)$~$\mu$m, where $p_T$ is the 
transverse momentum of the track relative to the beam line
in GeV/c. The geometrical acceptance of the 
SVX is about 60\% for the data presented here due to the $\sim\!30$ cm
RMS spread of the $p\bar{p}$ interactions along the beam line.
Outside the SVX is a set of 
time projection chambers (VTX) which measure the position of the primary
interaction vertex along the $z$-axis, and is in turn surrounded by the
central tracking drift chamber (CTC). This 3 m long chamber radially spans
the range from 0.3 to 1.3 m, and covers the pseudorapidity interval
$|\eta | < 1.1$ ($\eta = - \ln[\tan(\theta/2)]$) 
relative to the nominal $p\bar{p}$ interaction point. 
The 84 radial wire layers of the CTC are organized into nine ``superlayers.''
Five ``axial'' superlayers consist of wires strung parallel to the beamline.
Interspersed between these five are four ``stereo'' superlayers in which
the wires are turned $3^\circ$; the two types of superlayers used together
yield three-dimensional charged track reconstruction.  Within each superlayer
the wires are further organized into ``cells'' which are rotated $45^\circ$
relative to the radial direction.  This rotation assists the resolution
of left-right ambiguities in track reconstruction.
The CTC and SVX combined provide a transverse momentum resolution of
$\sigma_{p_T}/p_T \approx \sqrt{(0.9 p_T)^2 + (6.6)^2}\times 10^{-3}$, with
$p_T$ in GeV/c.

Outside the magnet coil, and covering the pseudorapidity range of the
SVX-CTC system, are electromagnetic (CEM) and, behind them, hadronic (CHA)
calorimeters. They have a projective tower geometry with a segmentation
of $\Delta \phi \times \Delta \eta = 15^\circ \times 0.11$. 
The CEM is a lead-scintillator stack 18 radiation lengths thick. It
has a resolution of $13.5\%/\sqrt{E_T}$ plus a constant $2\%$ added
in quadrature, where $ E_T = E \sin(\theta)$, $E$ is the measured
energy of the cell in GeV, and $\theta$ is its polar angle. 
A layer of proportional
chambers (CES), embedded near shower maximum in the CEM, provides a more
precise measurement of electromagnetic shower profiles both in azimuth ($\phi$)
and along the beam ($z$) direction.
The CHA is an iron-scintillator calorimeter 4.5 interaction lengths thick,
and has a resolution of $50\%/\sqrt{E_T}$ plus a constant $3\%$ added
in quadrature.
 
The calorimeters also act as a hadron absorber for the muon chambers which
surround them. The central muon system (CMU), consisting of four layers of
drift chambers covering $|\eta| < 0.6$, can be reached by muons with 
$p_T$ in excess of $\sim\!1.4$ GeV/c. 
These are followed by 60 cm of additional
steel and another four layers of chambers referred to as the central muon
upgrade (CMP). The central muon extension (CMX) covers
approximately 71\% of the solid angle for $0.6 < |\eta| < 1.0$ with four
free-standing conical arches composed of drift chambers sandwiched  between
scintillator (for triggering).

The data samples of interest in this paper, inclusive electrons and muons, 
and dimuons in the mass region around the $J/\psi$, were collected using CDF's
three-level trigger system. The first two levels are hardware triggers,
and the third level is a software trigger based on offline reconstruction
code optimized for computational speed. Different elements of the trigger
have varying efficiency turn-on characteristics, generally dependent upon
track $p_T$'s or calorimeter $E_T$'s. The behavior of the trigger has
been extensively studied.  Since the analyses presented here are 
largely insensitive to trigger behavior, we refer the interested reader 
to Ref.~\cite{CDFTopEvidncePRD,CDFLifetimePRD}
for detailed discussion of the triggers and their performance.

\subsection{Inclusive lepton data set}
\label{SubSec:lepSample}

The inclusive lepton data set is composed of electron and muon triggers.
Electron identification is based on energy clusters in the CEM with an
associated CTC track. The principal single electron trigger required
a Level-2 trigger $E_T$ threshold of 8 GeV, and
an associated track with  $p_T > 7.5$ GeV/c.
The offline reconstruction requires tighter matching between the 
position of the CES cluster and the associated track
({\it i.e.},  $r| \Delta \phi | < 3.0$ cm and 
$ | \Delta z  | < 5.0$ cm).
The CEM cluster is also required to have a shower profile consistent
with an electron shower, 
{\it i.e.}, a longitudinal profile
with less than 4\% leakage in the hadron calorimeter,
and a lateral profile in the CEM and CES consistent with
electron test beam data.

Muon identification is based on matching CTC tracks with track segments in
the muon chambers.
The inclusive sample is based on a Level-2 trigger 
with a nominal $p_T$ threshold of 7.5 GeV/c.
Each muon chamber track is required to match its
associated CTC track.
Track segments in both CMU and CMP are required to reduce backgrounds.

The inclusive lepton triggers are the dominant contribution to our
sample. However, the offline selection does not explicitly 
require that these triggers be satisfied. 
All events with a lepton track of $p_T > 6.0$ GeV/c, 
and passing the above identification quality cuts, may enter this sample.
The contribution from other triggers is small, and
the bulk of events with lepton $p_T$ below the nominal 7.5 GeV/c
threshold arise when the lepton $p_T$ reconstructed offline is
 lower than that estimated by the trigger system.
Finally, only lepton candidates using SVX tracking information are
considered, so as to be able to do precision vertexing.

\subsection{$J/\psi$ data set}
\label{SubSec:psiSample}

The $J/\psi$ sample is based on a dimuon trigger.
The trigger and selection on each muon are similar to that 
for the inclusive muons described above, except for a lower 
nominal $p_T$ threshold of $\sim\!2$ GeV/c~\cite{CDFLifetimePRD}.  
The CMU-CMP requirement is also
relaxed: the muon candidates may be in any of the muon chambers
(CMU, CMP, or CMX), and in any combination. 
The Level-3 trigger requires the presence of two oppositely
charged muon candidates with combined invariant mass 
between 2.8 and 3.4 GeV/c$^2$.
In offline reconstruction we further impose tighter
track matching and require $p_T > 1.5$ GeV/c for
each muon.  We also require a minimum energy deposition
of 0.5 GeV for each muon in the hadron calorimeter,
as expected for a minimum ionizing particle. 
Again, the dimuon sample is not explicitly required 
to have passed the dimuon trigger.

At this stage, no SVX tracking requirement is imposed, and there
are about $400,000$ $J/\psi$'s reconstructed,
with a signal-to-noise of about 10:1. Only about half
of these are fully contained within the SVX.

\section{$B^0$-$\overline{B}{}^0$ mixing}
\label{Sec:mix}

\newcommand{\bnom}{N(t)_{B^0 \rightarrow B^0}}
\newcommand{\bmix}{N(t)_{B^0 \rightarrow \overline{B}{}^0}}

The phenomenon of $B^0$-$\overline{B}{}^0$ mixing, analogous to
$K^0$-$\overline{K}{}^0$ mixing, occurs via higher order weak interactions.
Starting with an initially pure sample of  $B^0$'s at proper time $t=0$,
the numbers of $B^0$ and $\overline{B}{^0}$ mesons decaying 
in the interval from $t$ to $t+dt$ 
are $dN(t)_{B^0 \rightarrow B^0}$ and $dN(t)_{B^0 \rightarrow \overline{B}{}^0}$
respectively; and they are given by
\begin{equation}
\frac{dN(t)_{B^0 \rightarrow B^0}}{dt} 
        = \frac{N(0)_{B^0}}
                {2\tau_0} e^{-t/\tau_0}(1+\cos\Delta m t)
        \label{eq:prob_nomix}
\end{equation}
\begin{equation}
\frac{dN(t)_{B^0 \rightarrow \overline{B}{}^0}}{dt}  
        = \frac{N(0)_{B^0}}{2\tau_0} e^{-t/\tau_0}(1-\cos\Delta m t),
        \label{eq:prob_mix}
\end{equation}
where $\tau_0$ is the average lifetime of the two neutral $B$ meson 
eigenstates, and $\Delta m$ is the mass difference between them.

To observe mixing one must experimentally
determine the flavor of the neutral $B$ meson at the times of
formation and decay, a process referred to
as ``flavor tagging.'' The flavor at decay is usually well
known from the observed decay products. The initial
flavor determination is more difficult
and is discussed in the next section.

In an experiment with no background and
perfect flavor tagging and lifetime reconstruction, 
the mixing frequency $\Delta m_d$ can be determined
from the asymmetry 
\begin{equation}
\As_0(t) \equiv \frac{{\frac{d}{dt}}\bnom
   -{\frac{d}{dt}}\bmix}{{\frac{d}{dt}}\bnom+{\frac{d}{dt}}\bmix} = 
    \cos \Delta m t.
\label{eq:mix_asym}
\end{equation}

If the flavor tag correctly identifies the $B^0$
flavor at production with only a probability ${\cal P}_0$,
then the amplitude of the measured
asymmetry $\As_0^{(meas)}(t)$ is reduced by a factor 
$\Dil_0 \equiv 2{\cal P}_0-1$,
called the ``dilution,'' {\it i.e.},
\begin{equation}
\As_0^{(meas)}(t) = (2{\cal P}_0-1)\cos\Delta m t = \Dil_0 \cos\Delta m t.
\label{eq:cosine}
\end{equation}
A parallel series of expressions may be written when tagging
$B^+_u$'s, but there is no time dependence, so
\begin{equation}
\As_+^{(meas)}(t) = (2{\cal P}_+-1) \equiv \Dil_+.
\label{eq:const}
\end{equation}
Tagging charged $B$'s can be used to infer the flavor of the
other $b$ hadron in the event, but in this paper it 
is principally of interest as a test of the tagging method.
The charged and neutral dilutions 
need not be equal, and $\Dil_+$ can not in general be used as
a measure for $\Dil_0$.

The uncertainty on a measurement of the asymmetry $\As$ from
a sample of $N$ (background-free) events is 
\begin{equation}
\sigma_\As^2=(1-\As^2\Dil^2) / 
        N\varepsilon \Dil^2\simeq 1/N\varepsilon \Dil^2,
  \label{eq:epsD2}
\end{equation}
where $\varepsilon$ is the efficiency to obtain a flavor tag for
the method being employed. The figure of merit,
\mbox{$\ed$}, is called the ``effective tagging efficiency''
of the method.

\section{Flavor tagging}
\label{Sec:sst}

\subsection{Tagging methods}
\label{SubSec:tagging}

There is now a considerable inventory of $B^0_d$ mixing measurements
available~\cite{ref:mixing}.
Most rely on determining the flavor of 
the second $b$-hadron in the event to infer the
initial flavor of the originally reconstructed $B$ meson.
Examples include lepton tagging~\cite{UA1mix} 
and jet-charge tagging~\cite{ALEPH1stJetQ}.
We refer to these as ``Opposite Side Tagging'' (OST) methods.
Reliance on the opposite-side $b$-hadron can have several 
disadvantages. 

At the Tevatron, once one $B$ meson is produced 
in the central rapidity region covered by CDF,
the second $b$-flavored hadron is present
only $\sim 40\%$ of the time in this region.
In the other $\sim 60\%$ of events the second $b$-hadron is
unavailable for tagging.
For lepton tagging, there is the additional
inefficiency arising from the semileptonic branching ratio of the $B$,
as well as the confusion from daughter charmed particles decaying to leptons.  
For jet-charge tagging, the purity of the flavor-tag decision is reduced
by the presence of charged tracks from the proton-antiproton remnants
and possible confusion with gluon (or light quark) jets.
Finally, tagging based on OST suffers from the inevitable
degradation arising from mixing of the second $b$-flavored hadron
when it is a $B^0_{}$.
In spite of these complications, OST methods have proven to be powerful
tagging methods in previous mixing measurements~\cite{lotsOfOST,CDFJETQ}.

A contrasting approach is ``Same Side Tagging'' (SST), which 
ignores the second $b$-flavored hadron and instead
considers flavor-charge correlations of charged particles produced along 
with the $B$ meson of interest.\footnote{Jet-charge tagging has been 
extended by  combining the opposite and same side jet-charges 
in $Z^0 \rightarrow b\bar{b}$~\cite{OPALSSTJetQ}. 
A same side jet-charge tag is clearly correlated with the
SST approach of this paper, but the philosophy is
different. The jet-charge method is based on a weighted average
of charged tracks reflecting the primary quark's
charge~{\cite{FieldFeynman}}, while the proposal 
of Ref.~\cite{Rosner} is based more on selecting a specific
charged particle to determine the flavor.}
Such correlations are expected \cite{Rosner} to arise from particles
produced in the fragmentation chain and from decays of 
\Bds\ mesons.

\begin{figure}
\centerline{
\epsfysize 7.5cm
\epsfbox{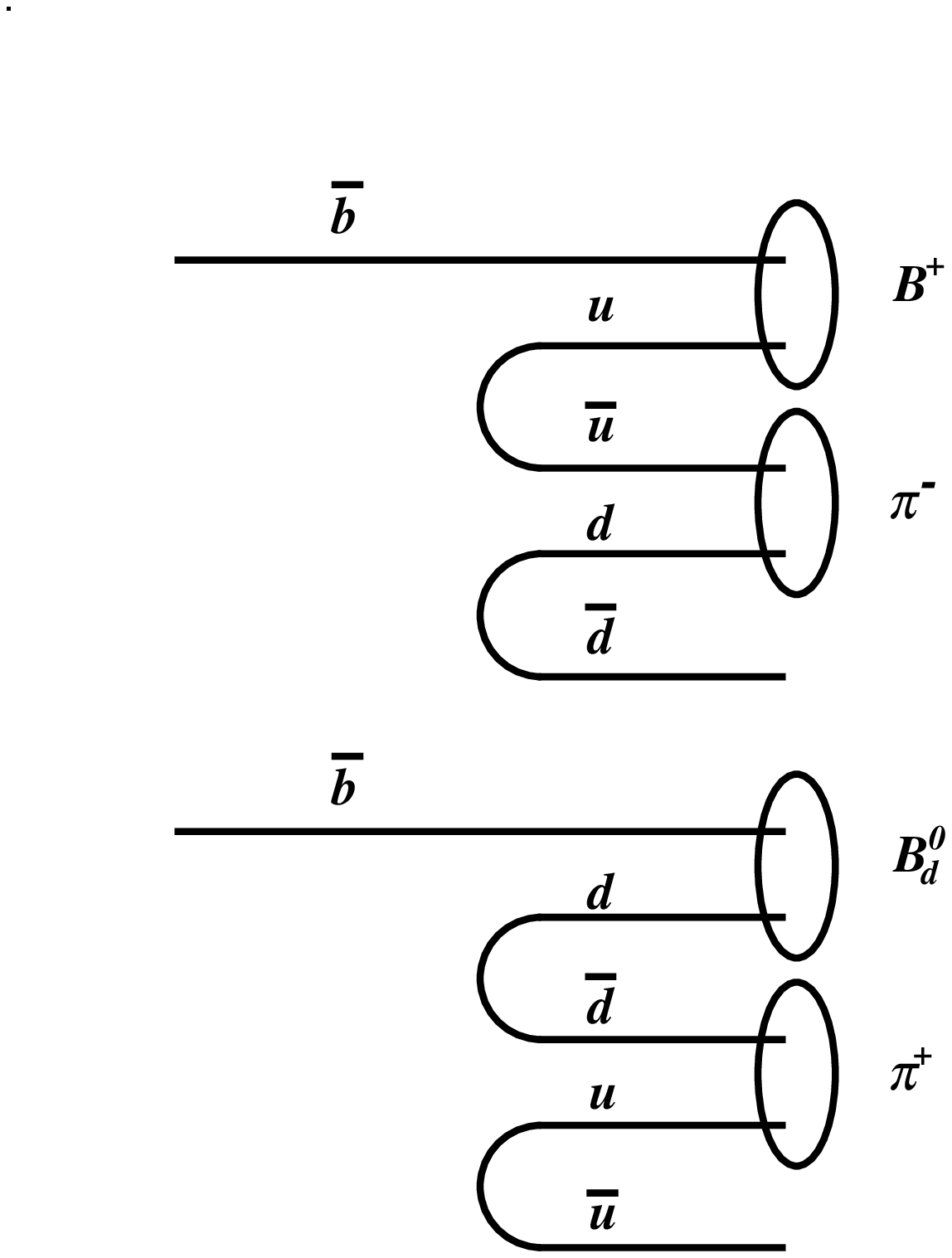}} 
\caption{A simplified picture of fragmentation paths for a $\bar b$ quark.}
\label{fig:sst_principle}
\label{fig:tagdiag}
\end{figure}

A simplified picture of the possible fragmentation
paths for a $\bar b$ quark is displayed in
Fig.~\ref{fig:tagdiag}.
If the $\bar b$ quark combines with a $u$~quark to form a
$B^+_u$, then the remaining $\bar u$ quark may combine with a
$d$~quark to form a $\pi^-$.
Alternatively, if the $\bar b$ quark fragments to form a $B^0_d$, the
correlated pion would be a $\pi^+$.
These correlations are the same as those produced from 
\Bds~decays, such as
$B^{**0}_d \ra B^{(*)+}_u \pi^-$ or
$B^{**+}_u \ra B^{(*)0}_d \pi^+$.
We do not attempt to
differentiate the sources of correlated pions.

In this simple picture of $B$-$\pi$ correlations, 
naive isospin considerations
imply that the tagging dilutions for $B^0_d$'s and  $B^+_u$'s should
be the same. However, this need  not be the case~\cite{RosnerDun},
and we make no such assumption.
Furthermore, we generically refer to the tagging particle
as a pion, although we do not attempt to 
experimentally identify it as one.

\subsection{Same Side Tagging algorithm}
\label{SubSec:sstag}
\label{sec:sst_algorithms}

\begin{figure}
  \vspace{-0.0cm}
\centerline{
\hspace{-0.0cm}
  \epsfysize 6.5cm
  \epsffile{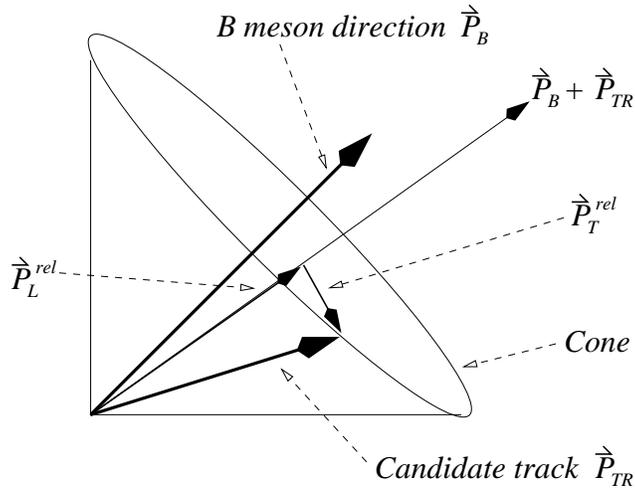}} 
\caption[]
{\small{
Schematic drawing of the momentum vectors determining $p_{T}^{rel}$
and $p_{L}^{rel}$
of an SST candidate.
}}
\label{fig:ptrel}
\end{figure}

General considerations of correlations between $B$ flavor
and particles produced 
in fragmentation offer only
qualitative guidance in constructing an SST algorithm.
String fragmentation models indicate that the velocity of fragmentation
particles are close to that of the $B$, and similarly
for pions from \Bds\ decays.
Motivated by this observation, a number of variables 
were studied for selecting a tagging track 
using data and Monte Carlo simulations, 
among them: 
({\it{i}})~the maximum $p_T$ track,
({\it{ii}})~the minimum $B$-track mass (using the pion mass),
({\it{iii}})~the minimum $\Delta R \equiv \sqrt{(\Delta \eta)^2 + (\Delta \phi)^2}$
               between the $B$ and track,
({\it{iv}})~the minimum of the track momentum component transverse to 
   the combined momentum of the $B$ ($\vec{P}_B$) plus track ($\vec{P}_{TR}$)
   momentum ($p_T^{rel}$, see Fig.~\ref{fig:ptrel}), and
({\it{v}}) the maximum of the track momentum component along the 
   $B$-track system  momentum ($p_L^{rel}$),
as well as several others.
We found that these five variables have similar performance, 
and moreover were highly correlated in selecting the same track as the tag.
Future studies with higher statistics samples may enable one
to optimize the choice,
but we were unable to identify one method as clearly superior.
We chose to use $p_T^{rel}$, as this variable was among the best for
correctly identifying the flavor ({\it i.e.}, had a large $\Dil$),
and it seemed less vulnerable to tagging 
on $B$ decay products missed in partial $B$ reconstructions 
(Sec.~\ref{sec:determine-xi}).\footnote{If
not for this issue, our studies tended to favor $p_L^{rel}$, 
essentially the same variable employed in Ref.~\cite{ALEPHSST}
for tagging exclusively reconstructed $B$ samples.}

For our specific SST algorithm, we consider all charged particles 
that  pass through all stereo layers of the CTC and
are within the $\eta$-$\phi$ cone of radius 
fraction$\Delta R  = 0.7$ centered
along the direction of the $B$ meson.  If the $B$ is
partially reconstructed, we approximate this direction with the
momentum sum from the partial reconstruction.
Tracks are required to be consistent with 
having originated from the fragmentation chain or the decay of
\Bds\ mesons, {\it i.e.},~coming from the primary vertex of the
$p\bar{p}$-interaction.  
This translates into the demand that tracks must have
at least 3 out of 4 SVX hits, $d_0/\sigma_{0}<3$ 
where $d_0$ is the distance of closest approach of the track trajectory to
the primary vertex when projected onto the plane
transverse to the beam line ($r$-$\phi$ plane) and $\sigma_{0}$ 
is the estimated error on $d_0$, and 
the 
closest approach in $z$ must be within 5 cm of the
primary vertex.

Due to chamber design, the CTC is known to have a lower 
reconstruction efficiency for negative tracks 
compared to positive ones at low $p_T$ (Sec.~\ref{SubSubSec:TagBias}).
To suppress this bias, all candidate tracks must have a $p_T$
above a threshold of $p_T(SST) = 400$~MeV/$c$.

At this point, more than one candidate tag may be available for
a given $B$. 
To select {\it the} tag, we choose
the candidate track with the smallest $p_T^{rel}$.

A $B$ is tagged if there is at least one track
that satisfies these selection requirements.
The fraction of $B$ candidates with a tag is the tagging efficiency,
and it is  $\sim\!60$-$70\%$  for this algorithm in our data.

\section{Flavor oscillations in the~lepton$+$charm 
sample}
\label{Sec:lepd}

We apply our SST method to a sample of $B_{u,d}$ decays 
to a lepton plus charmed meson.
We form the asymmetry, analogous to Eq.~(\ref{eq:mix_asym}),
between the decay flavor and the charge of the tag track,
and we fit this asymmetry using
a $\chi^2$ minimization to obtain $\Delta m_d$. 
As a by-product, the  tagging dilutions are also determined.
As we are henceforth concerned specifically with
$B^+_u$ and $B^0_d$, the subscripts are
suppressed for the remainder of this paper.

The incomplete reconstruction of the $B$'s introduces
several complications: 
({\it{i}}) missing decay products means that the precise $\beta\gamma$-factor
	to compute the proper decay time is not known;
({\it{ii}}) a missed charged decay product results in a $B^+$ being
	classified as a $B^0$ and vice versa; and
({\it{iii}}) a missed charged decay product may be chosen as the tag,
	biasing the asymmetry.
The latter two issues are the principal subtleties of this analysis,
and necessitate careful consideration of the
composition of the sample. Not all the branching ratio information required 
is well known, and when not, we rely internally on our data set.
Because the unknown sample composition parameters depend themselves
on other sample composition parameters we use an enlarged $\chi^2$
function to fit globally for $\Delta m_d$ and the unknown
composition parameters.

We first describe the sample selection and then discuss issues of
sample composition. The proper time measurement, and the corrections
for missing particles, is fairly standard, 
but $B^0 \leftrightarrow B^+$ cross-talk
introduces additional corrections.
We then discuss the measured and expected flavor-charge asymmetry
given the complications of the sample composition, including
the biases of tagging on $B$ decay products.
We finally discuss the $\chi^2$ fit, results, 
and the effects of systematic uncertainties
on $\Delta m_d$ and the tagging dilutions.

\subsection{$B$ candidate selection}
\label{sec:datasets}
\label{SubSec:lepd_recon}

We use partially reconstructed $B$'s  consisting 
of a lepton and a charmed meson.  
A particular $B$ reconstruction does not necessarily
arise from a unique sequence of bottom and charm decay modes
when there are unidentified decay products (Sec.~\ref{sec:SC}).
We therefore refer to the various $B$ reconstructions as
``decay signatures,'' and use the predominant decay sequence 
as a label. The samples of $B^0$'s consist of four decay signatures, 
one $B^0 \to \nu\ell^+ D^-$ signature and
three $B^0 \to \nu\ell^+ D^{*-}$:
\begin{eqnarray}
B^0 & \to & \nu \ell^+ D^-,\> \; D^- \to K^+ \pi^- \pi^-.  \label{dec:kpp}\\
B^0 & \to & \nu \ell^+ D^{*-},\> D^{*-} \to \overline{D}{^0} \pi_*^-,\>
 	\overline{D}{^0} \to K^+\pi^- \label{dec:kps} \\
B^0 & \to & \nu \ell^+ D^{*-},\> D^{*-} \to \overline{D}{^0} \pi_*^-,\>
 	\overline{D}{^0} \to K^+ \pi^- \pi^- \pi^+ \label{dec:k3ps} \\
B^0 & \to & \nu \ell^+ D^{*-},\> D^{*-} \to \overline{D}^0 \pi_*^-,\>
 	\overline{D}{^0} \to K^+\pi^-\pi^0, \label{dec:kpzs}
\end{eqnarray}
where we adopt the convention that a $\pi$ from a $D^*$ or $D^{**}$
decay is labeled by a ``$*$'' or ``$**$'' subscript.
For the $B^+$'s, we use only one decay signature:
\begin{equation}
B^+ \to \nu \ell^+ \overline{D}{^0},\> \overline{D}{^0} \to K^+\pi^-.
	\label{dec:kp}
\end{equation}
As noted above, the decay signatures do not represent a specific sequence
of decays; they in fact include several sequences, 
for instance, Eq.~(\ref{dec:kp}) includes the decay chain 
$B^+ \rightarrow \nu \ell^+ \overline{D}{^{*0}}$ followed 
by $\overline{D}{^{*0}} \rightarrow \overline{D}{^0} \pi^0_*$ and
$\overline{D}{^0} \rightarrow K^+ \pi^-$, where the $\pi^0_*$ is not identified.

The $B$ selection starts with the inclusive lepton 
($e$ and $\mu$) samples 
of Sec.~\ref{SubSec:lepSample}.
The tracks of the  $D^{(*)}$ daughters 
must lie within a cone of $\Delta R = 1.0$ around the lepton, 
pass through all nine CTC superlayers, have
enough hits for good track reconstruction, and
satisfy a $p_T$ requirement (see Table~\ref{tab:Selection-Cuts}).
All tracks (except one in the case of $D^0 \to K^+\pi^-\pi^-\pi^+$) 
must use SVX information, and they must also be consistent 
with originating in the vicinity of the same primary vertex.
The candidate tracks must form an invariant mass
in a loose window around the nominal $D$ mass, where all
permutations of mass assignments 
consistent with the charm hypothesis are attempted.

The candidate tracks are combined in a fit
constraining them to a $D$ decay vertex;
$\chi^2$ and mass window cuts are imposed. 
With the $D$ vertex established, we select the primary vertex 
from those\footnote{It is
 not uncommon to have multiple $p\bar{p}$ interactions 
 in a single bunch crossing at the higher Tevatron luminosities.}
reconstructed in the VTX
as the one nearest in $z$ to the $D$.
The transverse coordinates of the primary vertex
are obtained from a $z$ dependent average 
beam position, as  measured by the SVX 
over a large number of collisions recorded under identical
Tevatron operating conditions. 
We require the $D$ tracks to be displaced from this primary vertex 
($d_0/\sigma_0$ cut in Table~\ref{tab:Selection-Cuts}),
and the projected transverse distance $\lxy(D)$ between the $D$ vertex
and the primary vertex
to be greater than its uncertainty $\sigma_{\lxy}$ 
($\lxy(D)/\sigma_{\lxy}$ cut in Table~\ref{tab:Selection-Cuts}).
The projected distance $\lxy(D)$ is defined as
\begin{equation}
   \lxy(D) \equiv \frac{(\vec{x}_D-\vec{x}_{prim}) \cdot \vec{p}_T(D)}
                {\mid \vec{p}_T(D) \mid},
	\label{eq:Lxy-Def}
\end{equation}
where the two vertices are given by the position vectors
$\vec{x}_{prim}$ and $\vec{x}_D$, and the $D$ transverse momentum
is $\vec{p}_T(D)$.

\begin{figure}
\postscript{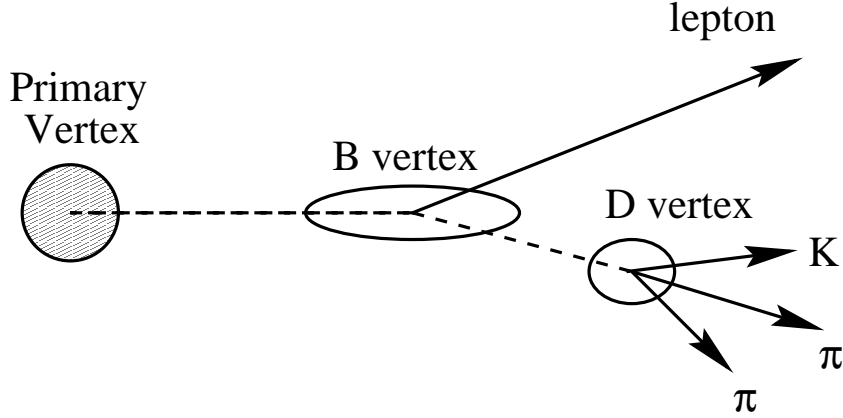}{0.7}
	\begin{center}
	\parbox{5in}{\caption{\small 
		A typical $B \to \nu\ell D$ event topology, where
		a $B$ meson is produced at the event primary vertex, 
                and decays 
                after traveling a short distance
                into a lepton, a $D$ meson, and a
		neutrino (undetected, and not shown).  The $D$
		later decays into a kaon and one or more pions.
		\label{fig:default-decay}
	}}
	\end{center}
\end{figure}

We next find the $B$ vertex. For the $B^0 \to \nu \ell^+ D^{*-}$ signature
the lepton and the  $\pi_*^-$ from the $D^{*-}$ decay both 
come from the $B$ decay point. We fit for the $B$ vertex by
intersecting the lepton and the $\pi_*^-$ tracks,
and require that the $D$ points back to it.  
For the $B^0\to \nu\ell^+ D^-$ or $B^+ \to \nu\ell^+\overline{D}{^0}$
signatures there is no additional track emerging from the $B$ vertex. 
The $D$ is projected back to
the lepton track and their intersection determines the $B$ vertex,
as sketched in Fig.~\ref{fig:default-decay}.
A loose cut is applied to the $D$ proper decay length relative to
the $B$ vertex ($ct_D$ in Table~\ref{tab:Selection-Cuts}).
The charges of the lepton and the charm candidates are required 
to be consistent with the decay of a single $B$, {\em i.e.},
a $\ell^\pm K^\pm$ correlation.

The $B^0 \to \nu\ell^+ D^{*-}$ decays followed by 
$D^{*-} \to  \overline{D}{^0} \pi_*^- $
also contribute to the $\ell^+ \overline{D}{^0}$ samples.  
The separation between $B^0$ and $B^+$ is improved by 
removing all $\ell^+ \overline{D}{^0}$  candidates
that also participate in the $\ell^+ D^{*-}$ reconstruction.  
We define a $D^{*-}$ candidate as a valid $\overline{D}{^0}$ candidate with 
a $\pi_*^-$ candidate that makes the mass difference 
$m(\overline{D}{^0} \pi_*^-)-m(\overline{D}{^0})$
consistent with the known  mass difference between
the $D^{*-}$ and $\overline{D}{^0}$~\cite{PDG}.
Since the $m(\overline{D}{^0} \pi_*^-)-m(\overline{D}{^0})$ distribution
for real $D^{*-}$'s is very narrow ($\sim\!1$ MeV), this removal 
is very efficient once the $\pi_*^-$ is 
reconstructed.

\label{subsec:mass-plots}

\begin{figure}
\postscript{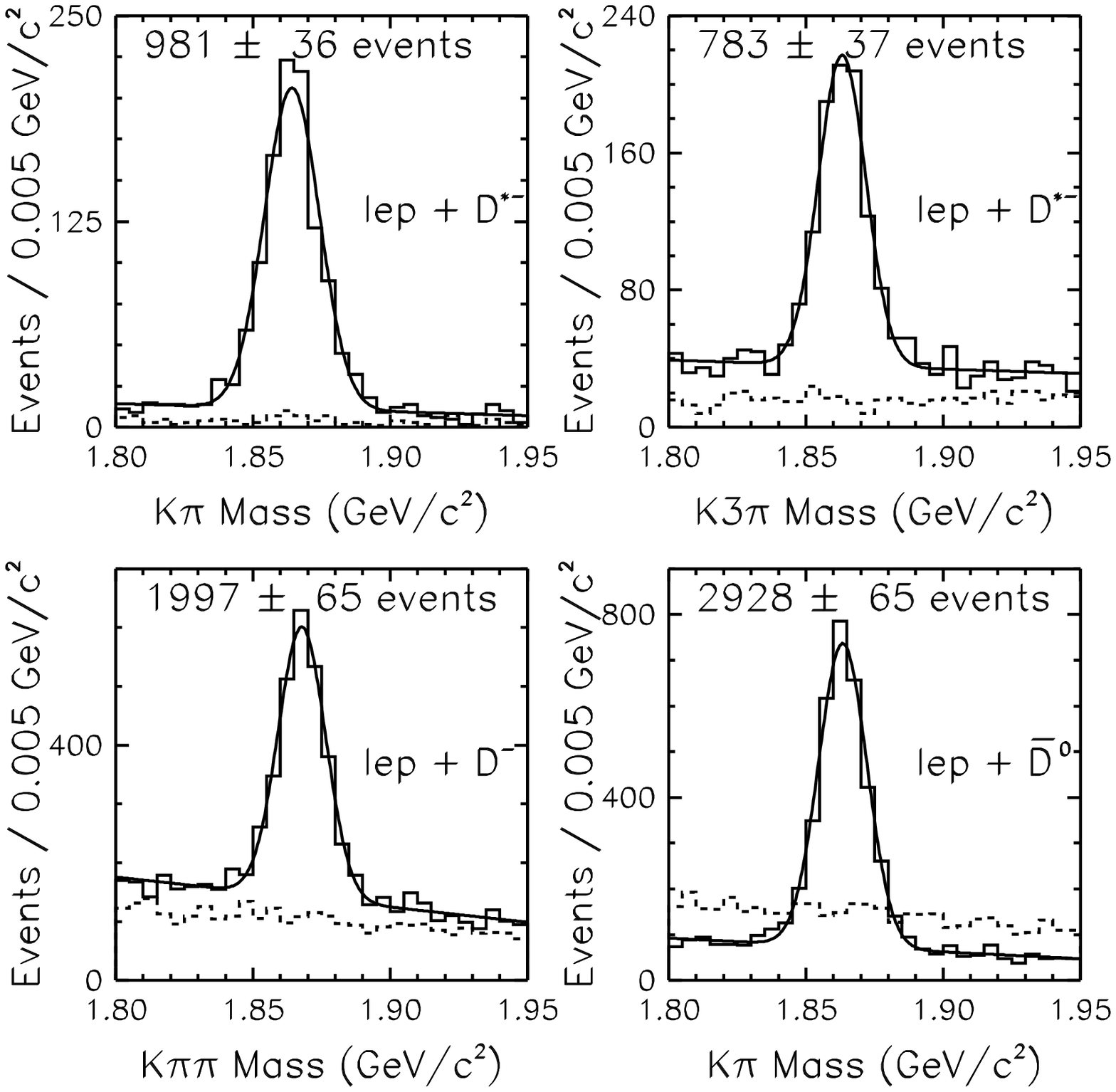}{0.7}
	\begin{center}
	\parbox{5in}{\caption{\small 
		The mass distributions of the fully reconstructed
		$D$ candidates (solid histogram)
		for: \kps\ (upper left);
		{\ktps} (upper right); \kpp\ (lower left) and 
		\kp (lower right).
		The dashed histograms are the distributions of the wrong-sign
		($\ell^\pm K^\mp$) candidates.
		The number of signal events from the fit (solid curve)
		is indicated in each plot.
		\label{fig:mass4}
	}}
	\end{center}
\end{figure}

The numbers of $B$ candidates are extracted from a fit of the charm
mass distributions. Figure~\ref{fig:mass4} shows the invariant mass 
distributions (solid histogram) for the four channels of exclusively 
reconstructed charm. The signal components of the $D$ 
mass distributions are modeled by Gaussians, 
and the combinatorial backgrounds by
linear functions (solid curves in Fig.~\ref{fig:mass4}).

The dashed histograms in Fig.~\ref{fig:mass4} represent the ``$D$'' 
mass distributions for $B$ candidates where the lepton and
the kaon have the wrong charge correlation
($\ell^\pm K^\mp$).  These ``wrong-sign'' events 
can be combinatorial background, as well as cases where there was a
real $D$  and a fake lepton.  The absence of a
peak in the wrong-sign ``$D$'' mass distribution demonstrates that 
the right-sign sample is a clean signal of $\ell D^{(*)}$ pairs coming 
from single $B$ mesons.

\begin{figure}
\postscript{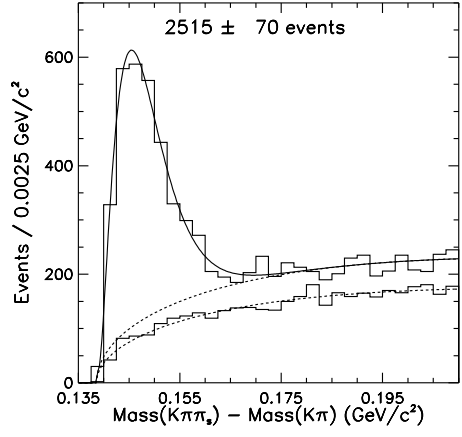}{0.4}
	\begin{center}
	\parbox{5in}{\caption{\small 
        The distribution of     
        $m(K\pi\pi_*)-m(K\pi)$
	for the signature $D^{*-} \to \overline{D}{^0} \pi_*^-$,
	$\overline{D}{^0} \to K^+\pi^-\pi^0$ 
	($\pi^0$ not reconstructed).
        The upper histogram is the distribution 
        of the right-sign $\ell D$ candidates, and
	the lower histogram is for the wrong-sign combinations.
        The solid curve is the fit of the right-sign distribution,
        with the fitted background component shown by the upper dashed curve.
        This background shape is obtained from a fit (lower dashed curve)
	of the wrong-sign data.
	\label{fig:kpzs-comb}
	}}
	\end{center}
\end{figure}

In the case of the decay \kpzs, the $\pi^0$ is lost, and the $K\pi$ 
invariant mass distribution has a broad excess below the $D$ mass.
However, in the $m(K \pi \pi_*)-m(K \pi)$ distribution a relatively
narrow peak appears at the value  $m(D^{*-})-m(\overline{D}{^0})$,
as seen in Fig.~\ref{fig:kpzs-comb}.
We parameterize the combinatorial background     
by  the shape of the wrong-sign ($\ell^\pm K^\mp$) distribution
(lower dashed curve). This shape, combined with 
the signal function, is then fit to the right-sign ($\ell^\pm K^\pm$) data,
and is shown by the solid curve in Fig.~\ref{fig:kpzs-comb}.

This completes our sample selection, which
has yielded almost 10,000 $B$ mesons.
However, before we can use them, several other issues must
be addressed.

\subsection{The composition of the $\ell D^{(*)}$ sample}
\label{sec:SC}

\subsubsection{$B^0 \leftrightarrow B^+$  cross-talk}
\label{sec:sources-of-cross-talk}

As noted earlier, the SST correlation depends on whether the
$B$ meson was charged or neutral.  However, only the ground state
charm mesons and one $D^{*}$ decay mode were reconstructed in the
previous section, and the existence of the intermediate
resonances $D^{*}$ and $D^{**}$ introduce contamination from $B^{+}$ decays
into $B^{0}$ decay signatures, and vice versa, when charged decay
daughters are unidentified or unreconstructed. We disentangle this cross-talk
by relating the charged and neutral $B$ fractions to the number 
of reconstructed charm mesons through relative branching ratios and 
reconstruction efficiencies.  This section details this connection.

There are two causes of the $B^0 \leftrightarrow B^+$ cross-talk in this
analysis:
\begin{itemize}
\item {\bf Missing the $\pi^{-}_{*}$ from the $D^{*-}$ decay.}  
	For example,
\begin{equation}
B^+ \to \nu \ell^+ \overline{D}{^0}
	\label{dec:original}
\end{equation}
can be mimicked by the decay sequence
$$
B^0 \to \nu \ell^+ D^{*-}
$$
\begin{equation}
D^{*-} \to \overline{D}{^0} \pi_*^-
	\label{dec:soft-pi}
\end{equation}
if the $\pi^-_*$ is not part of the reconstruction. 

\item {\bf $B$ decays to $D^{**}$-mesons.}
The decay chain
$$
B^0 \to \nu \ell^+ D^{**-}
$$
\begin{equation}
D^{**-} \to \overline{D}{^0} \pi_{**}^-
	\label{dec:pi**}
\end{equation}
will also mimic the $\ell^+\overline{D}{^0}$ signature of the $B^+$
when the $\pi^-_{**}$ is unidentified.
\end{itemize}

The first case
is of concern as it is not unusual for the momentum of
$\pi^-_*$ to fall below our $p_T$ cut.
The $\pi^-_*$ tends to be soft
because of the small energy release in the decay,
coupled with the modest boost of most of our $B$'s.
We identify the $\pi^-_*$
only with some efficiency $\epsilon(\pi_*)$.

In the other case,
we do not attempt to find the $\pi_{**}$ 
from the $D^{**}$.  There are four expected 
orbitally excited $D^{**}$
resonances (see Table~\ref{tab:d**-states}), 
some of which decay into $D\pi$, others to $D^*\pi$,
and one to both.  The total decay rate to these states is not 
well known, and the proportions of the four possible $D^{**}$ states 
are almost totally unknown.  There is evidence
that the $D_1(2420)$ and $D_2^*(2460)$ states are produced at some
level in $B$ decays~\cite{ref:exp-D**}.
There may also be non-resonant $D^{(*)}\pi$
production ($B \to \nu\ell D^{(*)}\pi$)~\cite{ref:exp-D**},
which has the same cross-talk effect.
It would be extraordinarily difficult to distinguish these decays
from the two $D^{**}$ resonances which are predicted to be broad by
Heavy Quark Effective Theory~\cite{ref:HQET}.
We therefore subsume the effects of four  $D^{**}$
resonances, as well
as the four-body semileptonic decay of the $B$ meson, 
into our treatment of {``$D^{**}$''}'s.

\begin{figure}
	\postscript{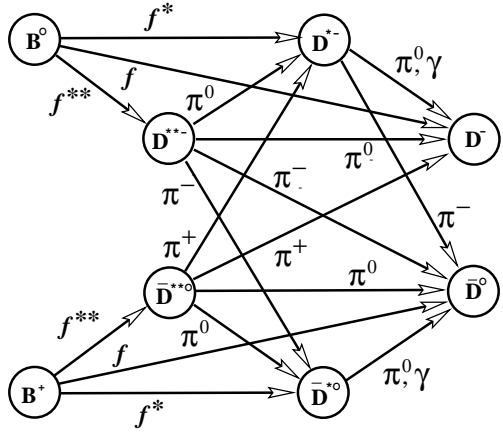}{0.4}
	\begin{center}
	\parbox{5in}{\caption{\small 
		The state diagram for all possible $B \to \ell D^{(*)}X$
		transitions.  The $f$, $f^*$, and $f^{**}$ 
		are semileptonic $B$ branching ratios to charm mesons.
	\label{fig:states}
	}}
	\end{center}
\end{figure}

The complete picture of the decay chains is more complicated, since
both $B^0$ and $B^+$ decay into {``$D^{**}$''}'s, and $D^{**-}$ and 
$\overline{D}{^{**0}}$ decay into both $D^{*-}$ and $\overline{D}{^{*0}}$, 
as well as $D^{-}$ and $\overline{D}{^0}$.
The full complexity of the decays is illustrated in the
state diagram shown in Fig.~\ref{fig:states}.  
To reiterate our terminology,
a specific sequence of decays in Fig.~\ref{fig:states} 
is called a {\em ``decay chain,''}
and the  reconstructed final state is a {\em ``decay signature.''}
Several decay chains may contribute to the same decay signature.
Decay chains in which the $B$ decays 
directly into a decay signature
({\it i.e.}, no particles except the neutrino are missed) are
called {\em ``direct decay chains.''}
Equations~(\ref{dec:original}), (\ref{dec:soft-pi}),
and~(\ref{dec:pi**}) are examples of decay chains;
Eq.~(\ref{dec:original}) is also a direct decay chain.
Each of the five decay signatures considered in this analysis consists of
several decay chains: 
three for every $\ell^+ D^{*-}$, nine for the $\ell^+ D^-$,
and twelve for the $\ell^+ D^0$.

\subsubsection{Determining the sample composition}
\label{sec:sc-equations}

Due to the $B^0 \leftrightarrow B^+$ cross-talk,
a simple computation of the time-dependent charge-flavor asymmetry 
of Eq.~(\ref{eq:mix_asym}) for a $\ell D^{(*)}$ decay signature will result 
in a weighted average of the $B^0$ and $B^+$ asymmetries 
[Eqs.~(\ref{eq:cosine}) and~(\ref{eq:const})]. The weighting is determined 
by the fractional contributions of $B^0$ and $B^+$ decays to that decay
signature; we call these fractions the ``sample composition.''
The fraction of $B^0$ and $B^+$ decays in a decay signature 
is essentially determined by the branching ratios and 
reconstruction efficiencies for each decay chain contributing to that signature.
Since only fractions are involved, it is convenient to use 
ratios of branching ratios and relative efficiencies.
These quantities, along with the  $B^0$ and $B^+$ lifetimes,
fully describe the sample composition as a function of proper time
and are referred to as the ``sample composition parameters.''  
We now discuss how we determine the fractions of $B^0$ and $B^+$ mesons
contributing to a signature, given
our choice of sample composition parameters.

\newcommand{\Br}[2]{{{\cal B}(B^{#2} \to {#1} \in k)}}
\newcommand{\eff}[2]{{\epsilon(B^{#2}\to {#1} \in k)}}
\newcommand{\BrNOk}[2]{{{\cal B}(B^{#2}  \to {#1}  )}}
\newcommand{\effNOk}[2]{{\epsilon(B^{#2}\to {#1}  )}}

We tabulate all possible decay chains that feed into each signature,
and classify them as originating from a $B^0$ or $B^+$.
For compactness, we label decay signatures by $k$, and decay chains by $h$.
The symbol ``$B^0\to h\in k$'' is interpreted as
``the decay chain $h$ originates from
a $B^0$ and contributes to the decay signature $k$.''
We write the fraction of neutral and charged mesons contributing to a
decay signature $k$ as
\begin{eqnarray}
{\cal F}_{k}^{0}(t) & \equiv & \frac{ \frac{d}{dt}N_{k}^{0}(t)}{\frac{d}{dt}\{
                                     N_{k}^{0}(t)+N_{k}^{+}(t)\} }
 	\label{eq:b0-frac-def2}\\
{\cal F}_{k}^{+}(t) & \equiv & \frac{ \frac{d}{dt}N_{k}^{+}(t)}{\frac{d}{dt}\{
                                     N_{k}^{0}(t)+N_{k}^{+}(t)\} },
 	\label{eq:b+-frac-def2}
\end{eqnarray}
where the $dN_{k}^{0,+}(t)$ are the numbers of events of signature $k$
originating from $B^{0}$ or $B^{+}$ which decayed in the proper
time interval from $t$ to $t+dt$.
These numbers are sums over all the
decay chains $h$ resulting in the signature $k$:
 \begin{eqnarray}
 \frac{dN_{k}^{0}(t)}{dt} & = & \frac{N_B e^{-t/\tbo} }{2\tbo} 
   \sum_{B^{0}\to h\in k}\BrNOk{h}{0}\effNOk{h}{0} 
 	\label{eq:b0-N-def1} \\
 \frac{dN_{k}^{+}(t)}{dt} & = & \frac{N_B e^{-t/\tbx}}{2\tbx}  
   \sum_{B^{+}\to h\in k}\BrNOk{h}{+}\effNOk{h}{+},
 	\label{eq:b+-N-def1}
 \end{eqnarray}
where we assume equal numbers of $B^0$'s and $B^+$'s are
produced ({\it i.e.}, $N_B = 2 N^+ = 2 N^0$); 
$\tau_{0,+}$ are the $B^{0,+}$ lifetimes, and
$\Br{h}{0,+}$ and $\eff{h}{0,+}$ 
are the branching ratios and
reconstruction efficiencies\footnote{We apply the term
``reconstruction'' only to those parts of the decay we identify;
neutrinos and decay products which are not part of the decay signature
are not included.} of a $B^{0,+}$
decaying through the chain $h$ and resulting in signature $k$.
The sums for the two mesons are different since they
are over a different set of decay chains for signature $k$.
Knowing all the branching ratios and efficiencies, we can calculate 
the sample fractions ${\cal F}_{k}^{0,+}$.

The efficiencies $\eff{h}{0,+}$ share common factors
across decay chains. Since only the ratios are needed in
Eqs.~(\ref{eq:b0-frac-def2}) and (\ref{eq:b+-frac-def2}),
we express the efficiencies
relative to the direct decay chain $d$, for signature $k$,
\begin{equation}
	\edkh \equiv \frac{\eff{h}{(h)}}{\eff{d}{(d)}}.
\end{equation}
The superscripts ``$(h)$'' and ``$(d)$'' represent the
charge of the $B$ which originated the $h$ and $d$ chains;
and the ``$D$'' superscript is a reminder that these relative efficiencies
are largely determined by the type of charm mesons in the decay chain.
For the direct decay chain $h = d$, and so $\epsilon^{D}_{kd}\equiv 1$.
We determine \edkl\ for each decay chain $h$ from a Monte Carlo simulation
as discussed in the next section. 

Similarly, only the branching ratios relative to the
the direct chain branching ratio are required here,
{\it i.e.}, $\Br{h}{(h)}/\Br{d}{(d)}$.
Rather than attempting to use each branching ratio explicitly,
not all of which are well known, we can re-express the required
ratios in terms of a few relative branching fractions by using a few
simplifying assumptions. We outline this process by considering
a specific example using the $B^0$ signature $k =$~``$\ell^+ D^{*-}$.''
The direct chain is $B^0 \to \nu\ell^+ D^{*-}$, and there are 
two ``$D^{**}$'' chains ($\pi_{**}$'s are unidentified):
$B^0 \to \nu\ell^+ D^{**-}$
followed by $D^{**-} \to D^{*-}\pi_{**}^0$, and
$B^+ \to \nu\ell^+ \overline{D}{^{**0}}$ with
$\overline{D}{^{**0}} \to D^{*-}\pi_{**}^+$.
If we index 
these three chains by $d$, $a$ and $b$ respectively, 
then the branching ratios relative to the direct chain are:
\begin{equation}
\frac{\Br{a}{0}}{\Br{d}{0}} = 
\frac{ {\cal B}(B^0 \to \nu\ell^+ D^{**-})
		{\cal B}(D^{**-}\to D^{*-}\pi_{**}^0 ) }
     { {\cal B}(B^0 \to \nu\ell^+ D^{*-})  }
	\label{eq:br-ratio-a}
\end{equation}
and
\begin{equation}
\frac{\Br{b}{+}}{\Br{d}{0}} = 
\frac{ {\cal B}(B^+ \to \nu\ell^+ \overline{D}{^{**0}})
		{\cal B}(\overline{D}{^{**0}}\to D^{*-}\pi_{**}^+ ) }
     { {\cal B}(B^0 \to \nu\ell^+ D^{*-})  }.
	\label{eq:br-ratio-b}
\end{equation}

The ratio of semileptonic $B$ decays 
can conveniently be re-expressed using ratios  relative to 
the inclusive branching
fraction to the lowest-lying $D$ state, 
including decays via intermediate $D^*$ and $D^{**}$ states,
${\cal B}(B\to\nu\ell D X)$:
\begin{eqnarray}
f_{0,+}      & \equiv & \frac{{\cal B}(B^{0,+}\to\nu\ell D) }
	               {{\cal B}(B^{0,+}\to\nu\ell D X) } \\
f^*_{0,+}    & \equiv & \frac{{\cal B}(B^{0,+}\to\nu\ell D^{*})}
	               {{\cal B}(B^{0,+}\to\nu\ell D X) } \\
f^{**}_{0,+} & \equiv &\frac{{\cal B}(B^{0,+}\to\nu\ell D^{**}) }
	   	      {{\cal B}(B^{0,+}\to\nu\ell D X)      }.
	\label{eq:def-fxx}
\end{eqnarray}
We assume that all the charged and neutral ratios are equal,
{\it e.g.}, $f \equiv f_{0} = f_{+}$.
Since the $D^{*}$ and $D^{**}$ decay strongly they
all ultimately result in a $D X$ signature, 
and thus $f + f^* + f^{**} \equiv 1$. Because the
$B\to \nu\ell D^{**}$ fractions are the least well known, 
we elect as our two independent parameters:
\begin{eqnarray}
  R_{f}  & \equiv & f^*/f = 2.5\pm 0.6, \\
  f^{**} & \equiv & 1 - f - f^* = 0.36\pm 0.12,
\end{eqnarray}
where the values are derived  from 
world averages~\cite{PDG} and 
the $f + f^*$ average from CLEO~\cite{f**-old-ref}.

We may now express Eq.~(\ref{eq:br-ratio-a}) as
\begin{equation}
\frac{\Br{a}{0}}{\Br{d}{0}} = \frac{f^{**}}{f^*}
		{\cal B}(D^{**-}\to D^{*-}\pi_{**}^0 ),
	\label{eq:BrRat_Ex3}
\end{equation}
solely in terms of $f$'s and $D$ branching ratios.
On the other hand, in Eq.~(\ref{eq:br-ratio-b}) we can use
\begin{equation}
\frac{ {\cal B}(B^+\to\nu\ell^+\overline{D}{^{**0}}) }
     { {\cal B}(B^0\to\nu\ell^+     D^{*-})  }
= 
\frac{ f^{**} }
     { f^{* } }
\frac{{\cal B}(B^+\to\nu\ell^+\overline{D}{^0} X)}
     {{\cal B}(B^0\to\nu\ell^+    {D}^- X)},
	\label{eq:BrRat_Ex4}
\end{equation}
where the ratio of the inclusive branching fractions to semileptonic 
decays of $B^+$ relative to $B^0$  must be taken into account.

The ratio
${\cal B}(B^+\to\nu\ell^+\overline{D}{^0} X)/
                           {{\cal B}(B^0\to\nu\ell^+{D}^- X)}$
can be approximated by the ratio of the $B^+$ and $B^0$ inclusive
semileptonic branching ratios ${\cal B}_{sl}(B^+)/{\cal B}_{sl}(B^0)$.
According to the spectator model, the semileptonic width
$\Gamma_{sl}$ is expected to be the same for the $B^0$ and $B^+$.  
The ratio of the semileptonic branching ratios 
(${\cal B}_{sl} = \Gamma_{sl}/\Gamma_{tot}$) for the $B^+$ and $B^0$
is then proportional to the ratio of their lifetimes, {\it i.e.},
\begin{equation}
	\frac{{\cal B}_{sl}(B^+)}{{\cal B}_{sl}(B^0)} 
		= \frac{\Gamma_{sl}(B^+)/\Gamma_{tot}(B^+)}
		       {\Gamma_{sl}(B^0)/\Gamma_{tot}(B^0)}
		= \frac{\Gamma_{tot}(B^0)}{\Gamma_{tot}(B^+)}
		= \frac{\tbx}{\tbo}.
	\label{eq:tbx/tbo}
\end{equation}
This allows us to also rewrite Eq.~(\ref{eq:br-ratio-b})
in terms of $f$'s and $D$ branching ratios as
\begin{equation}
\frac{\Br{b}{+}}{\Br{d}{0}} = \frac{\tbx}{\tbo} \frac{f^{**}}{f^*}
		{\cal B}(\overline{D}{^{**0}}\to D^{*-}\pi_{**}^+),
	\label{eq:BrRat_Ex6}
\end{equation}
with the $B$ lifetimes as the only additional parameter.
We use the $B^0$ lifetime and the ratio of $B$ lifetimes as our
two independent parameters, with the values
\begin{eqnarray}
	c\tbo             =  468 \pm 18 \>  \mu{\rm{m}} \\ 
	\frac{\tbx}{\tbo} =  1.02 \pm 0.05 \> 
\end{eqnarray}
obtained from world averages~\cite{PDG}.

The final branching ratios required are those for the charm decays.
For ${\cal B}(\overline{D}{^{**}}\to D^{*}\pi_{**})$, we
 need the fraction of $D^{**}$ states
that decay via $D^*\pi_{**}$ or $D\pi_{**}$.
Isospin symmetry gives relative
exclusive branching ratios for a particular $D^{**}$ species decaying
to a $D$ or $D^{*}$, such as
\begin{equation}
  \frac{{\cal B}(D^{**-}\to D^{(*)-}      \pi_{**}^{0})}{
        {\cal B}(D^{**-}\to \overline{D}{^{(*)0}}\pi_{**}^{-})} 
= \frac{{\cal B}(\overline{D}{^{**0}}\to \overline{D}{^{(*)0}}\pi_{**}^0)}{
	{\cal B}(\overline{D}{^{**0}}\to D^{(*)-}\pi_{**}^+)}
= \frac{1}{2}.
	\label{eq:D-isospin}
\end{equation}
As noted before,
we use the symbol ``$D^{**}$'' to represent the sum
over all four $D^{**}$ states (Table~\ref{tab:d**-states}) 
as well as two non-resonant channels.
The various ``$D^{**}$''  states, however, decay differently
to $D$ and $D^{*}$.
Reference to a $B \rightarrow D^{**} \rightarrow D^{(*)}$
decay chain implies summing over all possible ``$D^{**}$'' states.
We use $P_V$ to denote the inclusive probability that a $D^{**}$ 
decay yields a $D^{*}$ as opposed to a $D$, and it is given by
\begin{equation}
P_V \equiv 	\frac	{ {\cal B}(B\to D^{**} \to D^*\pi_{**})}
			{ {\cal B}(B\to D^{**} \to D^*\pi_{**}) 
			+ {\cal B}(B\to D^{**} \to D  \pi_{**})}.
	\label{eq:def-P_V}
\end{equation}
$P_{V}$ also depends upon the relative fractions 
of the various $B\to D^{**}$ decays
since $P_{V}$ is an effective average over all the $D^{**}$ states.

Equation~(\ref{eq:BrRat_Ex3}) then becomes
\begin{equation}
\frac{\Br{a}{0}}{\Br{d}{0}} = \frac{f^{**}}{f^*}
		\left(\frac{1}{3}\right) P_V,
	\label{eq:BrRat_Ex8}
\end{equation}
where the ``1/3'' is the isospin factor [similar to Eq.~(\ref{eq:D-isospin})].
A parallel expression may be written  for Eq.~(\ref{eq:BrRat_Ex6}).
$P_V$ is poorly known and is often assumed to lie
between 0.34 and 0.78~\cite{Pv_theory}. However, 
it can be (weakly) constrained 
by our data, and we therefore let it vary as a free fit parameter
in our $\Delta m_{d}$ fit (Section~\ref{sec:dilution_correction}).

For the $\ell^+ D^-$ and $\ell^+ \overline{D}{^0}$ decay signatures, 
we also need  $D^{*}$ branching fractions.
The $D^{*0}$ always
decays to a $D^{0}$ with a $\pi^{0}_*$ or photon, and the signature
is always $\ell^+ \overline{D}{^{0}}$.
On the other hand the  $D^{*-}$ has two decay channels 
which feed into different signatures. These ratios are
well known~\cite{PDG}:
\begin{eqnarray}
  {\cal B}(D^{*-}\to\overline{D}{^{0}}\pi^{-}_*) & = & (68.3\pm 1.3)\% \\
  {\cal B}(D^{*-}\to D^{-}\pi^{0}_*) & = & (31.7\pm 0.8)\%.
\end{eqnarray}

From this $\ell^+ D^{*-}$ example we have seen the basic ingredients 
for determining the sample composition.
In order to use a general notation, we define the relative ratio:
\begin{equation}
	\fbkh \equiv \frac{ {\cal B}(B^{(h)}\to h\in k) }
			  { {\cal B}(B^{(d)}\to\nu\ell D X) }
		     \frac{ \tau_d }{ \tau_h  },
	\label{eq:fbkl-def}
\end{equation}
where by $B^{(d)}$ we denote the $B$ charge state
for the direct decay chain, and by $\tau_d$ its lifetime.
$\tau_h$ is the lifetime of the $B^{(h)}$ 
from which the chain $h$ originates.
In Eq.~(\ref{eq:fbkl-def}) the $\tau_d/\tau_h$ ratio is
included in order to cancel out the lifetime ratio
that may appear in the branching ratios ${\cal B}$ by Eq.~(\ref{eq:tbx/tbo}) 
({\it e.g.}, in Eq.~(\ref{eq:BrRat_Ex6}))
so that the \fbkh's depend only on branching ratios averaged over
both $B$ meson species.  
The \fbkh's are compiled in Table~\ref{tab:all-decay} 
along with the $\pi_*$ reconstruction efficiency factor, which is
discussed in the next section.

We then determine the sample composition fractions ${\cal F}_{k}^{0,+}(t)$
for signature $k$ from the numbers of $B^{0,+}$ mesons $dN_{k}^{0,+}(t)$
[Eqs.~(\ref{eq:b0-N-def1}) and~(\ref{eq:b+-N-def1})] as,
\begin{eqnarray}
\frac{d N_{k}^{0}(t)}{dt} & = & {\cal M}_{kd} e^{-t/\tbo}
  \sum_{B^0\to h\in k}\fbkh \edkh  \\
\frac{d N_{k}^{+}(t)}{dt} & = & {\cal M}_{kd} e^{-t/\tbx}
  \sum_{B^+\to h\in k}\fbkh \edkh,
\end{eqnarray}
with the normalization factor 
\begin{equation}
  {\cal M}_{kd} \equiv \frac{N_{B}{\cal B}(B^{(d)}\to\nu\ell DX)
	\epsilon(B^{(d)}\to d \in k)}{2 \tau_{d}}.
\end{equation}
When calculating the ratios for ${\cal F}_{k}^{0,+}(t)$
this factor cancels out.
It is with the  ${\cal F}_{k}^{0,+}(t)$ fractions that we fully quantify 
the sample composition.

\subsubsection{Reconstruction efficiencies}
\label{sec:deriving-epsilon}

We use a Monte Carlo simulation to calculate
the relative reconstruction efficiencies \edkh\ 
for each decay chain $h$ contributing to signature $k$
relative to the direct decay chain for  $k$. 
Many systematic effects cancel out in these ratios
of lepton+charm reconstruction efficiencies.
In fact, these ratios depend almost exclusively on the decay
kinematics, which are reliably simulated.

We use our single $B$ Monte Carlo generator (App.~\ref{sec:bgenerator-mc})
to produce and decay $B$ mesons, and we pass them through the standard
CDF detector simulation. We then apply the selection cuts and 
calculate the relative efficiencies. The \edkh\ vary from about
$0.2$ to $1.5$, with most of the variation arising from the effects of the
fixed lepton $p_T$ threshold on the reconstruction of the various
$D$ states~\cite{Petar}.

One last efficiency is needed.
The $D^{*-} \to \overline{D}{^0}\pi^-_*$
reconstruction efficiency includes a contribution
for the efficiency of finding the $\pi^-_*$, 
which cancels out in the ratios  \edkh.
Loss of the  $\pi^-_*$ makes 
$D^{*-} \to \overline{D}{^0}\pi^-_*$ look like a $\overline{D}{^0}$.
Since
$\overline{D}{^0}\pi^-_*$ candidates are removed from
the $\overline{D}{^0}$ sample,
we need to know the absolute efficiency 
$\epsilon(\pi_{*})$ to quantify the separation of 
the $D^{*-}$ and $\overline{D}{^0}$ signatures.

We use data rather than Monte Carlo to determine $\epsilon(\pi_{*})$,   
since the absolute detector response for such low $p_T$ particles
is difficult to model. We use a related quantity,
which can both be calculated from $\epsilon(\pi_{*})$ and 
other sample composition parameters, as well as be measured in data.  
This quantity is the fraction of $\ell^+ D^{*-}$ candidates 
reconstructed out of the entire  $\overline{D}{^0} \to K^+\pi^-$ sample 
({\it i.e.}, before the $D^{*}$ removal),
\begin{equation}
R^{*}\equiv\frac{N(\ell^+ D^{*-})}{N(\ell^+ \overline{D}{^0}(\pi_{*}^\pm))},
    \label{eq:count-R*}
\end{equation}
where $\ell^+ \overline{D}{^0}(\pi_*^\pm)$ signifies $\ell^+ \overline{D}{^0}$ 
candidates before $D^{*-}$ removal.
The measurement in data, $R^{*(meas)}$, is accomplished by fitting 
the $\ell^+ D^{*-}$ and $\ell^+ \overline{D}{^{0}}$ 
(without $D^{*-}$ removal) mass distributions
simultaneously, and returns $R^{*(meas)}=0.249\pm 0.008$.

The calculation of $R^{*}$ consists of summing over all the decay chains
which give the desired signatures.  Each term is weighted by reconstruction
efficiencies.  The denominator sums over all decay chains which have a 
$\overline{D}{^0}$ in their final states, including $D^{*-}$ decays:
\begin{eqnarray}
N(\ell^+ \overline{D}{^0}(\pi_*^\pm)) 
 & = & \int_{0}^{\infty}[\frac{d}{dt}N_{\ell^+\overline{D}{^0} X}^0(t) 
	               + \frac{d}{dt}N_{\ell^+\overline{D}{^0} X}^+(t)]dt \nonumber\\
 & = & 
   \left\{ \frac{N_B \epsilon(\nu\ell^+\overline{D}{^0}) 
	   {\cal B}(B^+\to\nu\ell^+ \overline{D}{^0} X)}
               {2\tbx} 
   \right\} \nonumber \\
  & \times & 
   \left[
	       \tbx \sum_{B^+\to h\in \ell^+\overline{D}{^0}}\fbkh \edkh \right. \nonumber \\
  & + & \left. \tbo \sum_{B^0\to h\in \ell^+\overline{D}{^0}}\fbkh \edkh
   \right],
    \label{eq:R-denom}
\end{eqnarray}
where the lifetime factors result from integrating the exponential
factors over time. 

The numerator of Eq.~(\ref{eq:count-R*}), on the other hand, 
only sums over those decay chains which
give a $D^{*-}$, and is then multiplied by $\epsilon(\pi_*)$ to make it
the number of $D^{*-}$'s which are fully reconstructed, {\it i.e.}, 
\begin{eqnarray}
N(\ell^+ D^{*-}) 
 & = & \int_{0}^{\infty}[\frac{d}{dt}N_{\ell^+ D^{*-} X}^0(t) 
	               + \frac{d}{dt}N_{\ell^+ D^{*-} X}^+(t)]dt \nonumber \\
 & = & 
   \left\{ \frac{N_B \epsilon(\nu\ell^+\overline{D}{^0}) 
	   {\cal B}(B^+\to\nu\ell^+ \overline{D}{^0} X)}
               {2\tbx} 
   \right\} \nonumber \\
  & \times & 
   \left[
          \tbx ( f^{**}\frac{2}{3}P_{V}\epsilon_{a}^{D} ) + 
          \tbo ( f^*\epsilon_{b}^D 
               + f^{**}\frac{1}{3}P_{V}\epsilon_{c}^{D} ) 
   \right] \nonumber \\
  & \times & {\cal B}(D^{*-}\to D^{0}\pi^{-}_*)\epsilon(\pi_*).
    \label{eq:R-numerat}
\end{eqnarray}
We have explicitly substituted the sample composition parameters
\fbkh's  from Table~\ref{tab:all-decay} 
in the square bracket term since it is relatively
simple in this case. 
The subscripts on the relative efficiencies refer to the following chains:
($a$)~$B^+ \rightarrow \ell^+ \overline{D}{^{**0}}$,
     $\overline{D}{^{**0}} \rightarrow D^{*-} \pi^+_{**}$; 
($b$)~$B^0 \rightarrow \ell^+ {D}{^{*-}}$; and                 
($c$)~$B^0 \rightarrow \ell^+ \overline{D}{^{**-}}$,        
     $ D^{**-} \rightarrow  D^{*-} \pi^0_{**}$.
All $ D^{*-}$'s decay to $\overline{D}{^{0}} \pi^-_*$.
We see that the ratio of these two expressions, the prediction for 
$R^{*}$, is directly proportional to $\epsilon(\pi_*)$, and only depends
upon previously defined sample composition parameters.  When the
sample composition dependent prediction for $R^{*}$ is constrained
to the value $R^{*(meas)}$ in the $\Delta m_{d}$ fit, we find that
$\epsilon(\pi_*)\approx 0.85$ (Sec.~\ref{sec:fit-result}).

\subsubsection{Summary of the sample composition}

The fractions of the $B^0$ and $B^+$ decays in
each of the five decay signatures are described by a set 
of \SC\ parameters. Among them, 
$R_f$, $f^{**}$, and $\tbx/\tbo$ are obtained from other experiments, 
and the \edkh\ are calculated from Monte Carlo simulation.
The parameter $\epsilon(\pi_*)$ is expressed in terms of 
the other \SC\ parameters 
(via ${R^*}$) and 
$R^{*(meas)}$ (obtained from the data).
The final parameter,  $P_V$, will be a free parameter
to be determined in the $\Delta m_d$ fit.

Measuring $B^0$-$\overline{B}{^0}$ oscillations also
requires the determination of the proper time of the $B$ decay.
This will be discussed next, but 
sample composition effects must be included there as well.

\subsection{Proper time of the $B$ decay}
\label{sec:bgcorr_resolutions}

\label{sec:meas-ct}

The true proper time $\hat{t}$ of a $B$ decay may be determined 
by using its projected transverse decay length 
relative to the primary vertex 
${\lxy}(B)$ (following Eq.~(\ref{eq:Lxy-Def})), by
\begin{equation}
 c\hat{t} = {\lxy}(B)\frac{m({B})}{p_{T}(B)}, 
\label{eq:ct-true}
\end{equation}
where $m({B})$ is the mass of the $B$ and $p_{T}(B)$ is its
transverse momentum.  Since the $B$ is only partially reconstructed here,
we use Monte Carlo-derived average corrections  ${\cal K}_{kh}$
relating the reconstructed parts of the transverse momentum 
$p_{T}({\ell D})$ to that of the complete  $p_{T}({B})$, {\it i.e.},
\begin{equation}
 {\cal K}_{kh}\equiv\left<\frac{p_{T}({\ell D})}{p_{T}({B})}\right>_{MC} 
\end{equation}
for decay chain $h$ contributing to signature $k$.

\begin{figure}
\postscript{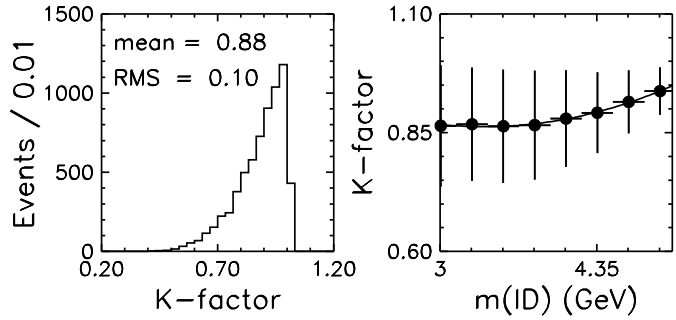}{0.6}
	\begin{center}\parbox{5in}{\caption{\small 
		The distribution of 
 		$\Kf \equiv p_T({\ell D})/p_T({B})$ (left),
                and
		$\Kf$ vs. $m(\ell D)$ (right) 
		with a fit of a quadratic function,
                for the direct decay 
		$\ell^+ D^{*-}$, $\overline{D}{^0} \rightarrow K^+ \pi^-$.
		The ``error'' bars in the 
		$\Kf$ vs. $m(\ell D)$ plot represent the RMS
		spread of 
		the $\Kf$-distribution in each bin, and not the error
                on the bin mean of $\Kf$.
		\label{fig:kps.phase1}
	}}\end{center}
\end{figure}

The ${\cal K}$-factors are determined from the same
simulation (App.~\ref{sec:bgenerator-mc}) as
the efficiencies \edkh.
An example of a  $p_T({\ell D})/p_{T}({B})$ distribution 
is shown in Fig.~\ref{fig:kps.phase1}.
The distribution is relatively well
concentrated because the lepton trigger threshold
favors decays where the neutrino takes only a small portion of
$p_{T}({B})$, thereby making  the $\ell D$ system
a fair representation of the $B$. 
The direct decay chains have means of $\sim\!85\%$, and 
RMS's of $\sim\!10\%$.
Also shown in Fig.~\ref{fig:kps.phase1} is the
mean of $p_T({\ell D})/p_{T}({B})$ as a function of
the $\ell D$ mass $m(\ell D)$; 
less of a correction is needed
the closer $m({\ell D})$ is to the $B$ mass.
We improve our resolution by using a $m({\ell D})$ dependent
correction on an event-by-event basis.

The correction factor varies with decay chain, so
the complete scale factor, ${\overline{\Kf}}^{0,+}_{k}$,   
for signature $k$
is a sample composition-weighted average of the ${\cal K}_{kh}$'s,
\begin{equation}
  \overline{\Kf}_k^0 = 
       \frac{  \sum_{B^0\to h} \fbkh\edkh \Kf_{kh} }
            {  \sum_{B^0\to h} \fbkh\edkh }
	\label{eq:k-ave}
\end{equation}
for $B^0$'s, and an analogous expression for  ${\overline{\Kf}}^{+}_{k}$.
In order to simplify averaging over the sample composition and
cancel systematic uncertainties, 
we replace $\Kf_{kh}$ in Eq.~(\ref{eq:k-ave})
by ${\Kf}_{kd} \times ({\Kf}_{kh}/{\Kf}_{kd})$,
where $d$ is the direct
chain contributing to $k$.
We factor ${\Kf}_{kd}$ outside the summation leaving
the ratio ${\Kf}_{kh}/{\Kf}_{kd}$ behind.
The set of factors we then use are the  ${\Kf}_{kd}$
with the  $m(\ell D)$ dependent corrections,
and the ${\Kf}_{kh}/{\Kf}_{kd}$ averaged over $m(\ell D)$
(where the $m(\ell D)$ dependence largely cancels out).

The factors ${\overline{\Kf}}^{0}_{k}$ and ${\overline{\Kf}}^{+}_{k}$
are different by virtue of the summation over different decay chains
for $B^0$'s and $B^+$'s.
The dependence of the sample composition ${\cal F}_{k}^{}$ 
on the lifetimes is accounted for by using the corrected times,
\begin{equation}
   ct_{k}^{0,+}\equiv 
   {\lxy}({B}) \, \frac{m({B})}{p_{T}({\ell D})} \, \overline{\Kf}_k^{0,+}
  \approx  c\hat{t}^{}_{}
	\label{eq:ct-kd2}
\end{equation}
as an estimate of the true proper time $\hat{t}^{}_{}$
in the sample composition fractions, {\it e.g.},
for Eq.~(\ref{eq:b0-frac-def2}) we write
\begin{eqnarray}
 {\cal F}_{k}^{0}(ct_{k}^{0},ct_{k}^{+})     & \equiv & 
        \frac{\frac{d}{dt}N_{k}^{0}(ct_{k}^{0})}{
              \frac{d}{dt}\{ N_{k}^{0}(ct_{k}^{0})+N_{k}^{+}(ct_{k}^{+})\} }.
 	\label{eq:b0-frac-def3}
\end{eqnarray}

The use of  $p_T({\ell D})$ rather than the true $p_T({B})$ 
smears the $ct$ distribution in addition to the {\it average} shift 
considered above. The difference between the reconstructed 
proper decay distance $ct_{k}^{0,+}$ and the true distance $c\hat{t}$ is
(suppressing most super- and subscripts)
\begin{eqnarray}
        \Delta ct & \equiv ct_{k}^{0,+}  - c\hat{t} 
                    = \Delta ( \lxy (B) /\beta_T \gamma ). 
                \label{eq:ct1}
                \label{eq:ct2a}
\end{eqnarray}
Approximating $1/\beta_T \gamma$ with its mean value
$\langle 1/\beta_T \gamma \rangle$ gives
\begin{equation}
    \Delta ct = \left\langle \frac{1}{\beta_T \gamma}\right\rangle \Delta\lxy (B)
                  + ct \frac{\Delta (1/\beta_T \gamma)}
                 {\langle 1/\beta_T \gamma\rangle},
                \label{eq:ct-resol-parts}
                \label{eq:ct2}
\end{equation}
which illustrates the effect of the reconstruction resolution 
via the $\Delta\lxy (B)$ term, 
and the additional smearing due to our average corrections
by $\Delta (1/\beta_T \gamma)$.

\begin{figure}
\postscript{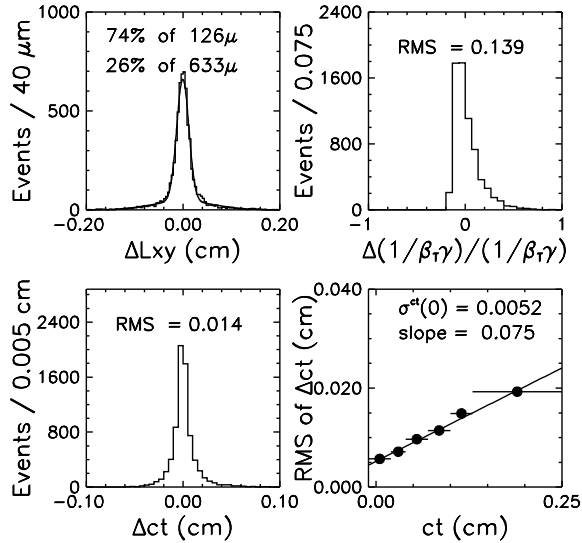}{0.5}
	\begin{center}\parbox{5in}{\caption{\small 
		The simulation for the direct chain $\ell D^{*-}$, with
		$\overline{D}{^0} \rightarrow K^+ \pi^-$:
		$\Delta \lxy(B)$ distribution, fitted with two Gaussians
                with relative fractions and RMS values listed 
                (top left);
		$\Delta (1/\beta_T\gamma)/(1/\beta_T\gamma)$ distribution 
		(top right); $\Delta ct$ distribution   (bottom left);
		and the
		RMS of the $\Delta ct$ distribution as a function of $ct$
                along with the linear fit for $\sigma^{ct}_{kd}(ct_{kd})$
		(bottom right).
		\label{fig:kps.phase2}
	}}\end{center}
\end{figure}

An example of the simulated $\Delta\lxy (B)$ 
distribution is shown in Fig.~\ref{fig:kps.phase2}. It has a 
Gaussian-like shape and an average resolution of a few hundred microns.
Also shown is the fractional $\beta_T \gamma$ distribution, which  
is sharply peaked (RMS $\sim\!14$\%), and is essentially a mirror 
image of  $p_T^{\ell D}/p_{T}^{B}$ (Fig.~\ref{fig:kps.phase1}).
The combined effect of both factors is shown 
by the $ \Delta ct$ distribution in  Fig.~\ref{fig:kps.phase2}, it
has an RMS of $140 \mu$m.

Given the linear dependence of $\Delta ct$ on the proper time 
in Eq.~(\ref{eq:ct2}), we parameterize the $ct$ resolution as
\begin{equation}
	\sigma^{ct}(ct) = \sigma^{ct}(0) + b \times ct.
		\label{eq:sigma(ct)}
\label{eq:lin-par-ct}
\end{equation}
We use the RMS spread of the $\Delta ct$ distribution 
for bin $ct$ as the resolution $\sigma^{ct}(ct)$, and
fit the RMS values of the various $ct$ bins
for the slope and offset of Eq.~(\ref{eq:lin-par-ct}).
The linear model works well as seen for the
sample chain shown in Fig.~\ref{fig:kps.phase2}.
This process is repeated for all five 
direct decay chains, and the
results are listed in Table~\ref{tab:sigma0+bct}.
Each chain has a somewhat different
slope, but the intercepts  
are similar to the intrinsic detector resolution 
of $\sim\!40$-$50\,\mu$m obtained when vertexing 
pairs of high  $p_T$ tracks at low $ct$~\cite{CDFLifetimePRD}.

The different decay chains that compose a decay signature are
topologically similar. 
Simulation shows that the $ct$ dependence of the $ct$ resolution
among the decay chains within a signature are very similar.
We make the (small) sample composition correction to
the $ct$ resolution for signature $k$ by approximating it as
\begin{equation}
  \sigma_k^{ct}(ct_{kd}) =\frac{\overline{\langle \sigma_k \rangle}_{ct} }
                          {\langle {\sigma}_{kd} \rangle_{ct} } 
                         \,\sigma_{kd}^{ct}(ct_{kd}),
\label{eq:sigma(ct)2}
\end{equation}
where $\sigma_{kd}^{ct}(ct_{kd})$ is the parameterization
of Eq.~(\ref{eq:sigma(ct)}) for the direct chain $d$,
and the bar indicates an average over contributing chains while
the angle brackets denote an average over $ct$.
Thus $\langle {\sigma}_{kd}\rangle_{ct}$ is the $ct$-averaged 
$ct$ resolution for direct chain $d$, and
$\overline{\langle \sigma_k \rangle}_{ct}$
is the sample composition-weighted average,
over all decay chains contributing to $k$,
of the  $ct$-averaged resolution.
The parameter  $\overline{\langle \sigma_k \rangle}_{ct}$
not only reflects
the different $ct$ resolutions of the various decay
chains, but also the fact that the earlier use of the average correction
factor $\overline{\Kf}_k^{0,+}$, rather than 
${\Kf}_{kh}^{}$, introduces additional smearing~\cite{Petar}.

\subsection{Tagging \boldmath $\ell D^{(*)}$ and the \SC}

\label{sec:tagging-lepD}

We apply the SST algorithm to the $\ell D^{(*)}$ sample
and find that $\sim\!70$\% of the events are tagged.
We classify events for each decay signature  as
having the ``unmixed'' lepton-tag charge combination
({\it e.g.}, $\ell^+ \pi^+$ for $B^0$'s and
$\ell^+ \pi^-$ for $B^+$'s), or the ``mixed'' one with the inverted
$\pi$ charge. Each set is further subdivided into 6 $ct_{kd}$ 
bins,\footnote{We cannot use  $ct^{0,+}_k$
to bin the data because the sample composition is not
completely known until after the binned data have been fit
(Sec~\ref{sec:dilution_correction}).}
where $ct_{kd}$ is the proper time 
obtained using the direct decay chain $d$ correction factor 
for signature $k$ (like Eq.~(\ref{eq:ct-kd2}), but only using 
${\cal K}_{kd}$).

The charm mass distribution for each of these $ct_{kd}$ subsamples 
is fit to a Gaussian signal plus linear background. 
The mean and width of the Gaussian, and the background slope, 
are all constrained to the same value
for all the subsamples of a given signature.
The fitted numbers of unmixed ($N_k^{U}(ct_{kd})$) 
and mixed ($N_k^{M}(ct_{kd})$) events 
for signature $k$ in the discrete $ct_{kd}$ bin
are then used to compute the measured asymmetry,
\begin{equation}
{\As}^{(meas)}_{k}(ct_{kd})\equiv\frac{N_k^{U}(ct_{kd})-N_k^{M}(ct_{kd})}
                                      {N_k^{U}(ct_{kd})+N_k^{M}(ct_{kd})}.
	\label{eq:A(RS-RS)}
\end{equation}
Numerically, the value of the $ct_{kd}$ bin center is chosen 
as the average over the candidates' $ct_{kd}$'s  in the bin, 
thus accounting for the nonuniform
$ct$ distribution from the exponential decay.

\label{sec:observed-as-and-sc}

Denoting the true asymmetries for $B^0$ and $B^+$ as $\As^0$ and $\As^+$,
one has for a pure, perfectly identified $B^0$ sample 
the ``predicted'' asymmetry $\overline{\As}_{k[0]}(ct) = \As^0(ct)$, 
where ``$k[0]$'' indicates that $k$ is a $B^0$ signature. 
When $k$ also includes $B^+$ decay chains, one has
\begin{equation}
\overline{\As}_{k[0]}(ct^{}_{kd}) = {\cal F}^0_k(ct^0_k,ct^+_k) \As^0(ct^0_k)
	        + {\cal F}^+_k(ct^+_k,ct^0_k) \{ -\As^+(ct^+_k) \} ,
	\label{eq:A(k,ct)-no-pids}
\end{equation}
for the prediction.
The true asymmetries are combined in a sample 
composition-weighted average, 
with  the fractional contributions ${\cal F}^{0,+}_k$ 
from Eqs.~(\ref{eq:b0-frac-def2}) and~(\ref{eq:b+-frac-def2}).
Furthermore, the observed proper time must now be corrected for 
the sample composition by using the $ct^{0,+}_k$ 
from Eq.~(\ref{eq:ct-kd2}).
The $\As^+$ term appears with a negative sign since the
charge of the flavor-correlated tag is reversed when
a $B^+$ is mistaken for a $B^0$.
A similar expression, albeit with signs reversed, holds for
a $B^+$ signature.

A further correction for $\overline{\As}_k$ is necessary because
there is the possibility of selecting the $\pids^\pm$ from
a $D^{**}$ decay [see Eqn.~(\ref{dec:pi**})] as a tag by mistake.
No attempt was made in the sample selection to identify $\pids^\pm$'s.
The lepton and a $\pids^\pm$ tag almost always\footnote{The
$B^+$ chain that cascades through 
$\overline{D}{^{**0}} \rightarrow D^{*-} \rightarrow \overline{D}{^{0}}$,
tags on the $\pi^+_{**}$, and loses the $\pi^-_{*}$ 
in the reconstruction, is an exception. However, this has 
a small contribution, {\it i.e.}, ${\cal F}^{+,**} \sim\!0.5$\%
in Eq.~(\ref{eq:A(k)}).} 
have 
the right charge correlation for an unmixed $B$, given
the apparent charge of the reconstructed $B$.
The $\pids^\pm$  contribution is quantified by the relative number 
of $D^{**}$'s present ($f^{**}$)
and the probability $\xi$ of selecting the $\pids^\pm$ as a tag in a tagged event
in which a $\pids^\pm$ was produced.
With this definition of $\xi$ the effect of the tagging algorithm is
separated from the $D^{**}$ branching ratios.

We split the $B^0$ and $B^+$ decays 
into those with and without a $\pids^\pm$, and
define ${\cal F}^{0,+,**}_k$ as the fraction
of decay signature $k$ in which a $\pids^\pm$ was produced.
${\cal F}^{0,**}_k$ is calculated in the same way as ${\cal F}^{0}_k$
in Eq.~(\ref{eq:b0-frac-def2}),
except that here the numerator is a sum only of the decay chains 
involving a $\pids^\pm$ from $B^0$.  ${\cal F}^{+,**}_k$ is
calculated analogously.
Only a fraction $\xi$ of $\pids^\pm$'s are selected as tags,
and we split the $B^0$ components 
into
${\cal F}^0_k - \xi{\cal F}^{0,**}_k$ and $\xi{\cal F}^{0,**}_k$,
and similarly for $B^+$'s. 
We then generalize Eq.~(\ref{eq:A(k,ct)-no-pids})
to include $\pids^\pm$ tags in the prediction for the measured
asymmetry:
\begin{eqnarray}
\overline{\cal A}_{k[0]} 
 & = &   \{ {\cal F}^0_k - \xi{\cal F}^{0,**}_k \} {\cal A}^0
       + \xi{\cal F}^{0,**}_k  \{ -1 \}  \nonumber\\
 & +   & \{ {\cal F}^+_k - \xi{\cal F}^{+,**}_k \} \{ -{\cal A}^+ \}
       + \xi{\cal F}^{+,**}_k  \{ +1 \},
\label{eq:A(k)}
\label{eq:A(k,ct)}
\end{eqnarray}
where the $-1$ ($+1$) asymmetry factors in the second (fourth) 
term reflect the perfect correlation of $\pids^\pm$ tags.

All relevant effects for a mixing measurement using SST
are contained in Eq.~(\ref{eq:A(k)}); it describes
the observed asymmetry $\overline{\cal A}_{k[0]}$  
given the true asymmetries ${\cal A}^{0,+}$, 
the $\pids^\pm$  tagging  probability $\xi$, and the sample composition
${\cal F}$'s.

\subsection{Determination of the $\pids^\pm$ fraction $\xi$}
\label{sec:determine-xi}

The $\pids^\pm$ tagging  probability $\xi$
depends on the tagging algorithm, the kinematics and geometry of 
the $B$ and $D^{**}$ decays, as well as the characteristics 
of the fragmentation and underlying event tracks.  
We use a full event simulation (App.~\ref{sec:default-PYTHIA-mc})
to model the decay kinematics and
geometry---which it does reliably---to obtain the $ct$ dependence
of $\xi$, denoted by  $\xi_{MC}(ct)$.
The decay kinematics and geometry determine the  $\xi$ shape, whereas
the relative competition between the $\pids^\pm$ and the other 
tracks to become the tag 
affects the overall $\pids^\pm$
tagging probability.
This observation enables us to use 
our data to determine the global normalization of $\xi$,
instead of relying on the simulation's
modeling of nearby tracks.
We therefore define
\begin{equation}
\xi(ct) = \xi_{norm} \cdot \xi_{MC}(ct),
	\label{eq:def-xi_norm}
\end{equation}
where  $\xi_{norm}$ is the normalization needed to 
scale the simulation to agree with the data.

\begin{figure}
\postscript{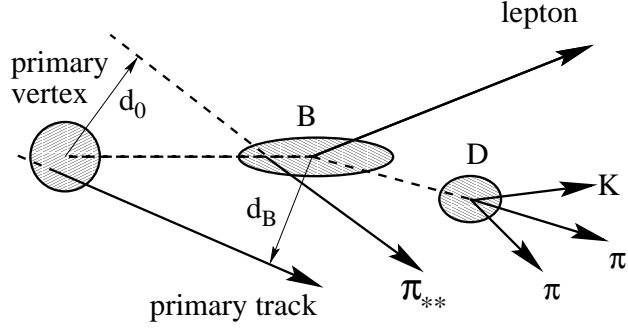}{0.5}
	\begin{center}
	\parbox{5in}{\caption{\small 
		A schematic representation 
		of a $B \rightarrow \nu \ell D^{**}$ decay.
		SST candidate tracks originate from the primary vertex,
		while the $\pids^\pm$ track originates from the $B$ 
		decay vertex.  
		The impact parameter of a track with respect to the
		primary vertex is $d_0$ while the impact parameter with
		respect to the $B$ vertex is given by $d_B$.  When the
		$B$ vertex and the primary vertex are well separated,
		the $\pids^\pm$ track usually has a large $d_0$ and
		a small $d_B$.  The converse is true for primary tracks.
		\label{fig:pixx-wrt-pv}
	}}
	\end{center}
\end{figure}

\begin{figure}
\postscript{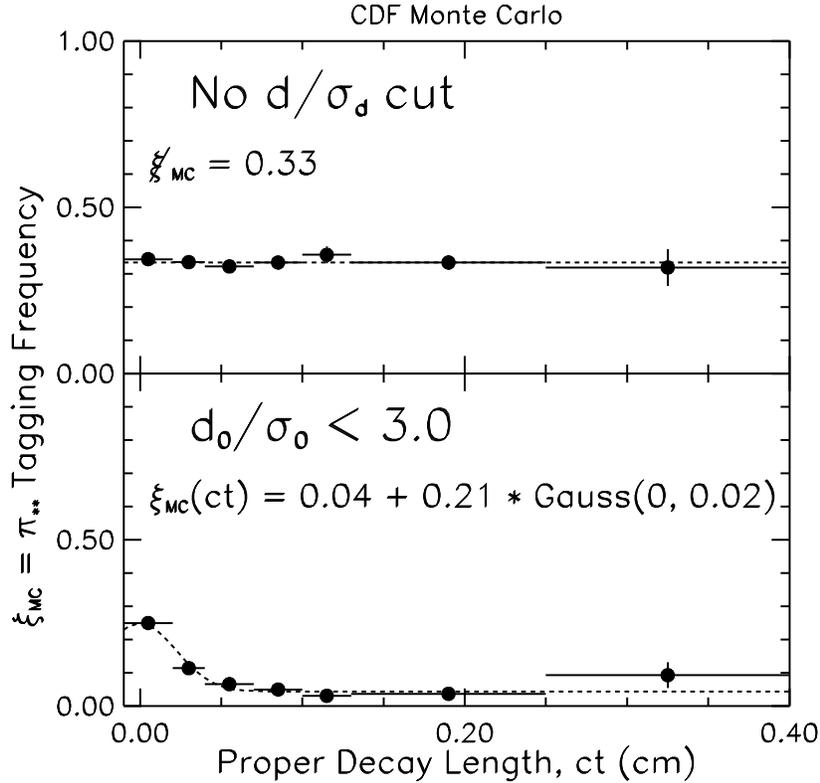}{0.7}
	\begin{center}
	\parbox{5in}{\caption{\small
		Monte Carlo calculation of
		$\xi_{MC}$ as a function of corrected proper time, 
                $ct^0_{kd}$,
                for decay signature $\ell^+ D^-$:
		no $d_0/\sigma_{0}$ cut (top), and 
		$d_0/\sigma_{0} < 3.0$ (bottom).
	\label{fig:xi_vs_ctau.eps}
	}}
	\end{center}
\end{figure}

The topology of a $B \rightarrow \nu \ell D^{**}$ decay chain is 
shown in Fig.~\ref{fig:pixx-wrt-pv}.
The $ct$ dependence of  $\xi$ is the result of the
impact parameter significance cut ($d_0/\sigma_{0} < 3.0$)
in the SST selection (Sec.~\ref{sec:sst_algorithms}).
By removing this cut, we remove the
$ct$ dependence from $\xi$.
Figure~\ref{fig:xi_vs_ctau.eps} shows $\xi_{MC}(ct)$
with the $d_0/\sigma_{0}$  cut removed (top), and
with the cut applied as normal (bottom).
Without the cut the distribution is flat, as expected,
and corresponds to a 33\% probability to tag on a $\pids^\pm$  
given that one is present.
Applying the $d_0/\sigma_{0}$ cut rejects most of the
$\pids^\pm$ tags, especially once $ct$ is beyond a few hundred microns.
The $\xi$ shape is modeled by a Gaussian, centered at zero, with
a constant term.

\begin{figure}
 \postscript{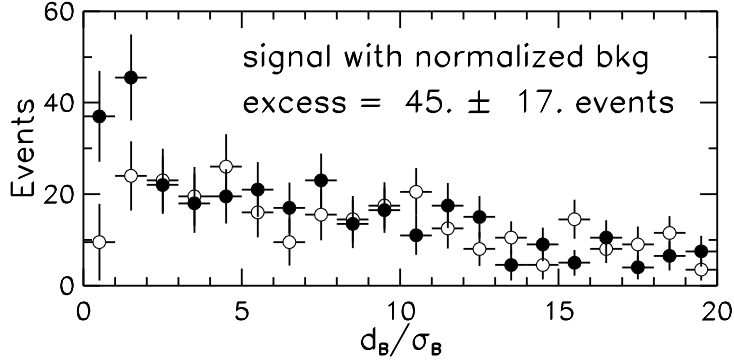}{0.6}
	\begin{center}
	\parbox{5in}{\caption{\small
		Impact parameter distribution with respect 
		to the $B$ vertex ($d_B/\sigma_{B}$) 
		for the \kpp\ decay signature from the data
		(no $d_0/\sigma_{0}$ cut, $D$ mass sideband subtracted).
                Right-sign tags are shown by the solid points.
		Wrong-sign tags (open circles) are renormalized
                to model the right-sign continuum at large impact
		parameter significance, as described in the text. 
		The right-sign excess near zero is due 
                to the $\pids^\pm$ tags.         
	\label{fig:it4r7.c4xx.kpp-0}
	}}
	\end{center}
\end{figure}

To determine the normalization, $\xi_{norm}$, 
we remove the $d_0/\sigma_{0}$ cut from the data 
(analogous to Fig.~\ref{fig:xi_vs_ctau.eps}, top), 
thereby eliminating the $ct$ dependence as well as 
enriching the sample in $\pids^\pm$  tags.
We divide the tagged events into
right-sign and wrong-sign tags, and make the distribution
of the impact parameter significances
{\it with respect to the $B$ vertex}  ($d_B/\sigma_{B}$
of Fig.~\ref{fig:pixx-wrt-pv}).
An example of such a distribution 
is shown in Fig.~\ref{fig:it4r7.c4xx.kpp-0}.
The excess of right-sign events near $d_B/\sigma_{B}=0$ is due to the
$\pids^\pm$ tags. Their number, $N(\pi^{tag}_{**})$,  
is determined by fitting the distribution 
to a Gaussian (centered at zero
and with a unit RMS) for the  $\pids^\pm$'s,
plus a $\pids^\pm$ background shape obtained from the wrong-sign distribution.
The wrong-sign distribution renormalized to the fit
result is overlaid onto the right-sign distribution in
Fig.~\ref{fig:it4r7.c4xx.kpp-0}. It is seen to agree 
very well with the right-sign distribution at large $d_B/\sigma_{B}$,
and displays a clear excess near zero.

We also count, again without the $d_0/\sigma_{0}$ cut,
the total number of tags $N(tags)$
and compute the fraction of $B$ decays where $\pids^\pm$'s
are tags,
\begin{equation}
	R^{**}_k \equiv
		\frac{ N(\mbox{$\pi^{tag}_{**}$ }) }
		     { N({tags}) }.
	\label{eq:count-ipsb}
\end{equation}
for signature $k$. The measured ratios are given 
in Table~\ref{tab:measured-R**}.
We extend our notation so that \mbox{$\not\!{\xi}_{MC}$} is 
$\xi_{MC}(ct)$ when the impact significance cut, $d_0/\sigma_0 < 3.0$, 
is removed. \mbox{$\not\!{\xi}_{MC}$} is then independent of $ct$.
The ${R}^{**}$'s are simply
\begin{equation}
  {R}^{**}_k = \xi_{norm}{\not\!{\xi}}_{MC}\int_0^\infty
               {\{{\cal F}^{0,**}_k(t) + {\cal F}^{+,**}_k(t) \}\, dt},
    \label{eq:R**_from_alpha**}
\end{equation}
for decay signature $k$.
Thus,  $\xi_{norm}$ is simply related to ${R}^{**}_k$,
the other sample composition parameters, and  \mbox{$\not\!{\xi}_{MC}$}.
Rather than attempting to compute an average $\xi_{norm}$,
we will constrain the five $R^{**}_{k}$ predictions
to the  measured ratios in the $\Delta m_{d}$ fit (Sec.~\ref{sec:eps_ipsb}).

\subsection{Fitting the asymmetries}
\label{sec:dilution_correction}

\subsubsection{The $\chi^2$ function}
\label{sec:time-dependence}

The observed tagging asymmetries can be predicted in terms
of the sample composition parameters and the true asymmetries.
The true asymmetry for the $B^+$ is constant in $ct$, 
while for $B^0$ it follows a cosine dependence,
and accounting for the $ct$ resolution one has
\begin{eqnarray}
{\cal A}^+(ct) & = & {\cal D}_+ 
	\label{eq:flat_line} \\
{\cal A}^0(ct)&=& \frac{ 
    G(ct;ct',\sigma^{ct})\otimes\{e^{-t'/\tbo}  {\cal D}_0\> \cos(\dmd \, ct')\} }
   {G(ct;ct',\sigma^{ct})\otimes e^{-t'/\tbo}  },
\label{eq:oscillation_werf}
\end{eqnarray} 
where $\otimes$ denotes convolution of the physical time dependence 
(cosine and/or exponential functions) with
the $ct$ resolution function $G(ct;ct',\sigma^{ct})$ over $ct'$.
The latter function is a normalized Gaussian of mean $ct'$ and 
RMS $\sigma^{ct}$.
However, the measured $ct$, and associated resolution, depends upon
the sample composition. Therefore, the proper times
used for the predicted asymmetries are 
the $ct^{0,+}_k$ [Eq.~(\ref{eq:ct-kd2})]
obtained using  the sample composition-averaged \Kfact.
For the resolution  $\sigma^{ct}$ we use 
the composition-weighted resolution $\sigma^{ct}_{k}(ct^{0,+}_k)$
from Eq.~(\ref{eq:sigma(ct)2}).

\label{sec:eps_ipsb}

We form a $\chi^2$ function to simultaneously 
fit \dmd, \Dx, and \Do\  over all $ct$-bins of all
decay signatures by comparing  the predictions 
$\overline{\As}_k(ct_{kd})$
calculated via Eq.~(\ref{eq:A(k,ct)}) against
the measured asymmetries $\As^{(meas)}_k(ct_{kd})$,
where $ct_{kd}$ is used for $\As^{(meas)}_k$ since
we were restricted  to the direct chain  $ct_{kd}$ when
binning the data (Sec.~\ref{sec:tagging-lepD}).
The $\overline{\As}_k(ct_{kd})$  asymmetry 
depends not only upon the
parameters $\dmd$, $\Do$, and $\Dx$, which are of direct interest, 
but also on $\tbo$, $\tbx/\tbo$, $f^{**}$, $P_V$, $\epsilon(\pi_*)$, 
and $\xi_{norm}$ through the ${\cal F}^{0,+}$'s.
The last two parameters are also expressed
as functions of $R^*$ and $R^{**}_k$, as well as the other composition 
parameters.
The comparison of $\As^{(meas)}_k$ and $\overline{\As}_k$
corresponds to the first summation in our $\chi^2$ function:
\begin{eqnarray}
\chi^2  & = & \sum_{k,ct_{kd}} \left( \frac
		{ \As^{(meas)}_k(ct_{kd}) - \overline{\As}_k(ct_{kd}) }
		{ \sigma^\As_k(ct_{kd}) } 
		\right)^2  \nonumber\\
	& + & \sum_j \left( \frac
		{ F^{(meas)}_j - 
		  \overline{F}_j(f^{**},P_V,R_f,\epsilon(\pi_s)\ldots) }
		{ \sigma^F_j }
		\right)^2,
    \label{eq:extended-chi2}
    \label{eq:grand-chi2}
\end{eqnarray}
where $k$ is an index that runs over the five decay signatures, and
$ct_{kd}$ symbolizes the summation over the proper time bins.

The second summation is over the set of fit input parameters:
$F^{(meas)}_j$ is the measured input value for
parameter~$j$, $\sigma^F_j$~is its error, and
the ``predicted'' value is
$\overline{F}_j(f^{**},P_V,R_f,\epsilon(\pi_*),\ldots)$.
This prediction is a function of the sample composition parameters, 
and in most cases it is a trivial substitution, such as
$\overline{F}_j(f^{**},P_V,R_f,\epsilon(\pi_*),\ldots) = {\tau}_{0}$
for the $B^0$ lifetime. However, $\epsilon(\pi_*)$, $\xi_{norm}$, 
and $P_V$ are not directly measured but are constrained in the fit 
by their appearance in the predictions for other measured quantities,
namely $R^*$ and $R^{**}_k$. Allowing $\epsilon(\pi_*)$, $\xi_{norm}$, 
and $P_V$ to float in the fit constrained by the measured sample
composition parameters was one of the motivations
for extending the $\chi^2$ function with the second summation.

The reconstruction efficiency for the $D^{*}$ pion $\epsilon(\pi_*)$
can be obtained (Sec.~\ref{sec:deriving-epsilon}) from  
$R^{*\>(meas)}$, measured in the data to be $0.249 \pm 0.008$.
Since the prediction $\overline{R^*}$ is a function
of the \SC\ parameters, $\epsilon(\pi_*)$ depends on
them also [Eqs.~(\ref{eq:count-R*})-(\ref{eq:R-numerat})]
and must be recomputed whenever the
\SC\ parameters change.  This recomputation naturally occurs in the $\chi^2$
minimization by allowing the composition parameters that determine
$\overline{R^*}$ to float, coupled with the constraint 
of the ``$F^{}_j$'' term 
\begin{equation}
      	\left( 
   \frac{ R^{*\>(meas)} - \overline{R^*}(\epsilon(\pi_*),f^{**},\ldots) }
        { \sigma^* } 
      	  \right)^2
	\label{eq:chi2-R*}
\end{equation}
in the $\chi^2$.

A similar approach holds for $\xi_{norm}$ and $P_V$ using the $R^{**}$'s.
In this case there are five $R^{**}_k$'s, one for each decay
signature, and a $F^{}_j$ term for each. 
The prediction for $R^{**}_k$ is proportional to $\xi_{norm}$ 
by Eq.~(\ref{eq:R**_from_alpha**}).
Because  $\xi_{norm}$ is common
across decay signatures, it is essentially determined 
by the average of all five $R^{**\>(meas)}_k$'s.
$P_V$, the relative $D^{**}$ decay rate to $D^{*}$ vs. $D$
[Eq.~(\ref{eq:def-P_V})],
is treated as completely unknown. However, it is also related
to the $R^{**}_k$'s.
If $P_V = 0$, there would be no  $D^{**} \to D^*$ decays, and
consequently no $\pi_{**}^-$'s in the $\ell^+ D^{*-}$ signatures, 
resulting in the corresponding $R^{**}_k = 0$.
The values of the $R^{**}_k$'s relative to each other
determine $P_V$. 
While the errors on $R^{**\>(meas)}_k$ are large 
(Table~\ref{tab:measured-R**}), 
and therefore $P_V$ is not tightly constrained,
this method is preferable to just using a theoretical estimate for $P_V$. 
Its incorporation into the $\chi^2$ function automatically
enables it---like the other parameters---to vary within 
the allowed experimental constraints
and propagates the associated uncertainty to the fit 
parameters.

\subsubsection{Result of the fit}
\label{sec:param-behavior}
\label{sec:fit-result}
\label{Sec:lepd_fit} 

The $\chi^2$ function is minimized
over the six $ct$ bins for all five decay signatures simultaneously,
letting the unknown parameters float freely, 
and the known inputs to vary within their errors.
The fit results in the following values 
\begin{eqnarray*}
\Delta m_d      & = &   0.471 ^{+0.083}_{-0.075} \mbox{ ps}^{-1}\\
{\cal D}_+      & = &   0.267 ^{+0.039}_{-0.034} \\
{\cal D}_0      & = &   0.181 ^{+0.036}_{-0.031} \\
\epsilon(\pi_*) & = &   0.845 ^{+0.073}_{-0.058}\\
\xi_{norm}      & = &   0.747 ^{+0.470}_{-0.292}\\
P_V             & = &   0.331 ^{+0.276}_{-0.298},
\end{eqnarray*}
with the nominal fit errors quoted. The $\chi^2$ of the fit
is 26.5 for 30 degrees of freedom.

The $\pi_*$ efficiency is quite high, thereby limiting one source
of the $B^0 \leftrightarrow B^+$  cross-talk.
The $D^{**}$ contribution to cross-talk is quantified by $P_V$,
which is on the low side of what is sometimes assumed~\cite{Pv_theory}.
Our value could be biased by our sample selection, but
in any case the errors are large.
The final sample composition yields: 
$\sim\!82\%$ of the $\ell^+\overline{D}{^0} X$ 
signature comes from $B^+$ decays, 
while $\sim\!80\%$ of the $\ell^+D^- X$  
and $\sim\!95\%$ of the $\ell^+ D^{*-} X$ originate from $B^0$.

\begin{figure}
  \postscript{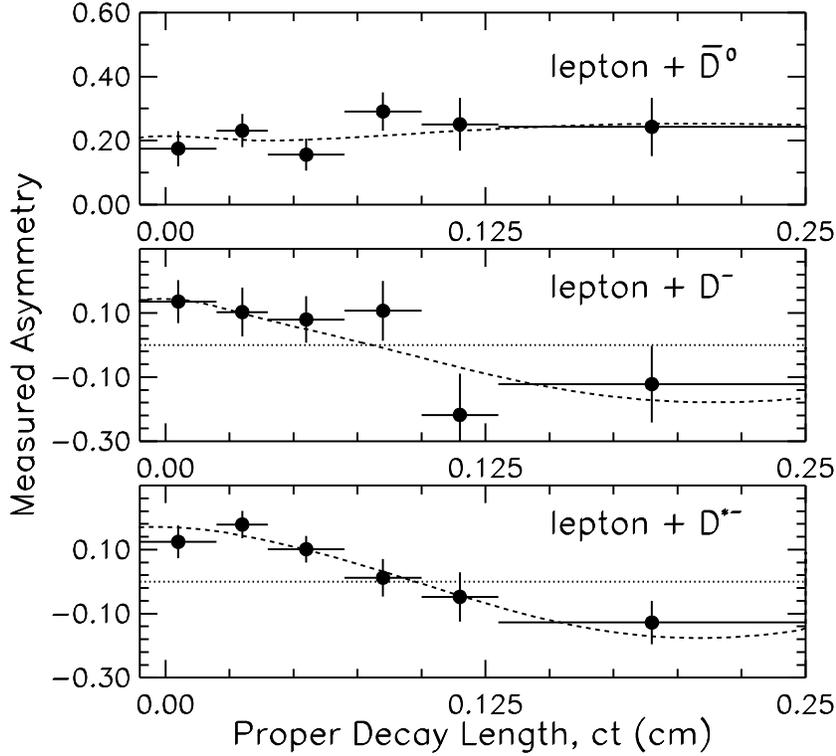}{0.7}
	\begin{center}
	\parbox{5in}{\caption{\small
	Measured asymmetries as a function of the corrected 
	proper decay length $ct_{kd}$
	for the decay signatures $\ell^+ \overline{D}{^0}$
	(dominated by $B^+$), 
	$\ell^+ D^-$, and the sum of all three $\ell^+ D^{*-}$
	(dominated by $B^0$). The three $\ell^+ D^{*-}$ signatures
	are combined only for display purposes.
	The dashed line is the result of the fit.
	\label{fig:result}
	}}
	\end{center}
\end{figure}

\begin{figure}
 \postscript{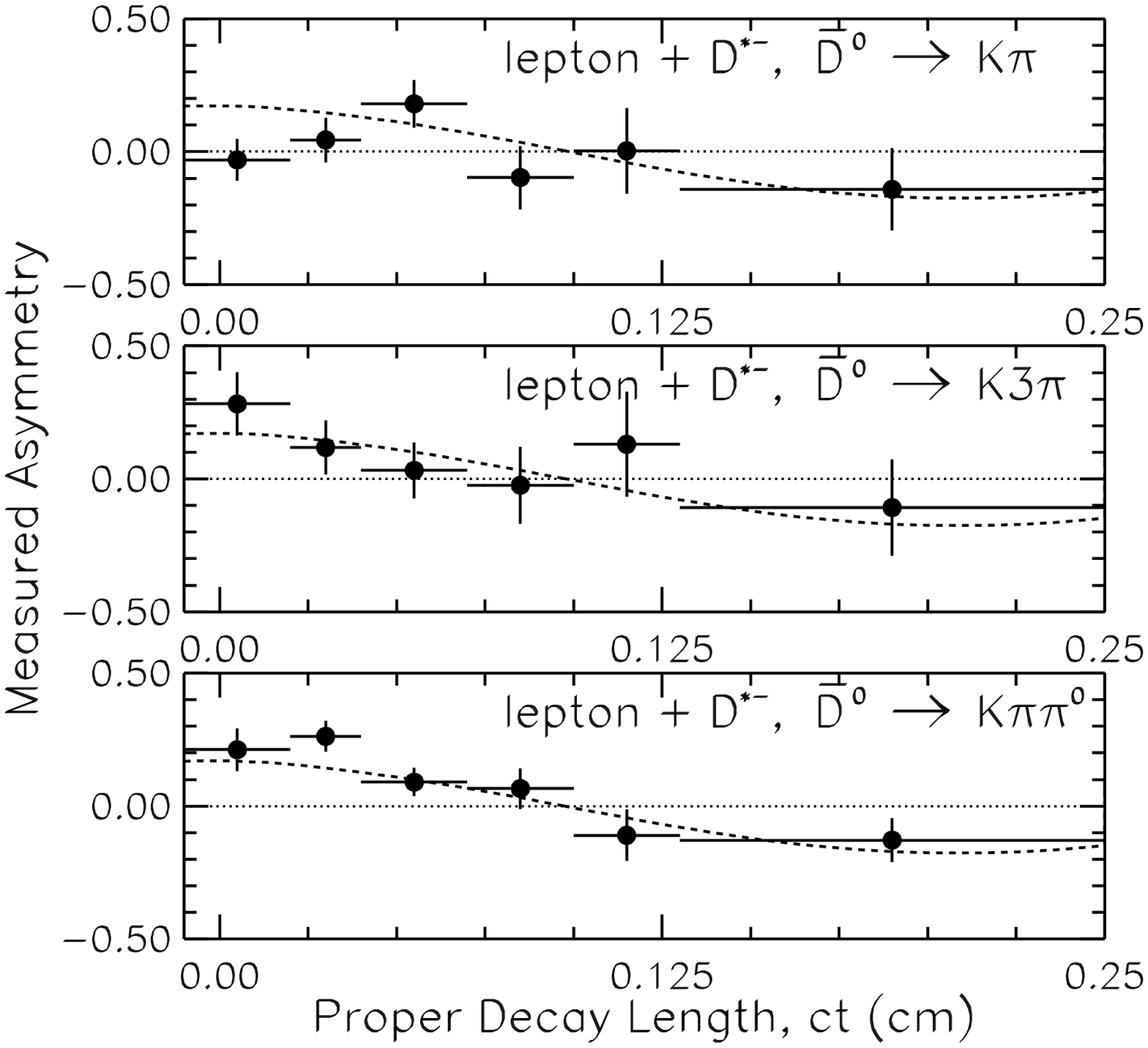}{0.6}
	\begin{center}
	\parbox{5in}{\caption{\small
	The breakdown of the measured asymmetries of $\ell^+ D^{*-}$ into 
	the three $\ell^+ D^{*-}$ decay signatures
	``\kps'', ``\ktps'' and ``\kpp''.  
	The result of the fit is overlaid.
	\label{fig:fit5.pretty-2}
	}}
	\end{center}
\end{figure}

Figure~\ref{fig:result} shows the result of the fit overlaid
on top of the measured asymmetries, where all three $\ell^+ D^{*-}$
signatures are combined.  Figure~\ref{fig:fit5.pretty-2} gives
the three $\ell^+ D^{*-}$ signatures separately.
The cross-talk is relatively modest, and the signatures
dominated by $B^0$ generally show a fairly clear oscillatory
behavior in the raw observed asymmetries.
For the $B^+$-dominated signature,  $\ell^+ \overline{D}{^0}$,
the raw asymmetry is compatible with being a constant 
(Fig.~\ref{fig:result}), but the residual effect of the
$B^0$ cross talk is visible in the fitted curve in the form
of a slight oscillation. The effect of the $B^+$ contamination
in the $B^0$ signatures amounts to a constant shift in the
asymmetry and is therefore not readily apparent.

The fit parameters constrained to {\it a priori} measured values
are shown in Table~\ref{tab:in-out} along with the value and error
output by the fit.
Except for $f^{**}$, these parameters are largely decoupled 
from the other fit parameters, and are virtually unchanged
from their input values. The data are more sensitive 
to the value of $f^{**}$ because it governs the
amount of cross-talk between $B^0$ and $B^+$.

The correlation coefficients of the fit parameters with 
\dmd, \Dx, and \Do\  are shown in Table~\ref{tab:correl-coef}.
We see that the lifetimes are largely decoupled from
other parameters. 
On the other hand, $f^{**}$, $\xi_{norm}$,
and $P_V$ are strongly coupled to \dmd, \Dx, and \Do, underscoring
the importance of the $\pids^\pm$ corrections to the analysis.
The correlation between \dmd\ and \Dx\ is stronger than 
between  \dmd\ and \Do. The reason for this stronger corrrelation
is that the effect of \Dx\ enters via
the $B^+$ contamination of the $B^0$ signatures
and is manifested by a downward translation of the $B^0$ oscillation
in Fig.~\ref{fig:result}. As the oscillation is translated down,
the intercept with zero asymmetry moves to shorter times, thereby
decreasing \dmd. On the other hand, if \Do\ is varied, the oscillation
amplitude varies symmetrically about the vertical axis and
\dmd\ is weakly affected.

\subsection{Statistical and systematic uncertainties}

We fit for \dmd, \Dx, and \Do\ using a $\chi^2$ function 
which also incorporates the sample composition parameters.
The errors it returns are
a combination of statistical and systematic effects,
yet the errors only partially account for the systematic uncertainties.
The sources of the uncertainty 
can be divided into statistical and three systematic categories:
\begin{itemize}
\item {\bf Statistical:} the error that is directly correlated
        with the $\ell D^{(*)}$ sample size.
\item {\bf Correlated Systematics:} parameters of the fit 
	(\SCpars), coupled
	to  \dmd, \Dx, and \Do\  through the \SC\ [Eq.~(\ref{eq:A(k,ct)})].
	These parameters are not correlated among themselves; only their
	effects on \dmd\, \Dx, and \Do\ are.

\item {\bf Uncorrelated Systematics:} systematic uncertainties 
	caused by imperfect simulation models of the physics processes or
         the detector.

\item {\bf Systematics from Physics Backgrounds:} uncertainties due 
        to other physics
	processes that contribute to $B \to \nu\ell D^{(*)}$ data samples
	that have been hitherto neglected.  
\end{itemize}
We determine the uncertainties from each of these four categories
in turn to estimate the statistical and systematic uncertainty
for our final result.

\label{sec:correl-syst}

We separate the statistical and correlated systematic errors of
the original fit by repeating the fit with
the \SC\ parameters fixed to the results of the 
original fit, and only six variables (\dmd, \Dx, \Do, 
$\epsilon(\pi_s)$, $\xi_{norm}$, and $P_V$) floating.
The errors from the six-variable fit are just 
statistical ($\sigma_{stat}$), while the errors from the
full fit are the combined statistical and correlated 
systematic errors ($\sigma_{stat+C.S.}$).
In a Gaussian approximation, we estimate the
correlated systematic error, $\sigma_{C.S.}$, by
\begin{equation}
     \sigma_{C.S.} \equiv \sqrt{ \sigma_{stat+C.S.}^2 - \sigma_{stat}^2 }\, ;
  \label{eq:sigma-SC}
\end{equation}
and find, for example,
\begin{eqnarray*}
{\dmd}  & = &  0.471^{+0.078}_{-0.068} \, ^{+0.030}_{-0.031} \;{\rm ps}^{-1}, 
\end{eqnarray*}
where the first error is purely statistical and the second is
the correlated systematic. This correlated error is
listed in Table~\ref{tab:systematics} under ``Sample Composition,''
and is by far the dominant systematic uncertainty.

\label{sec:uncorrel-syst}

The ``uncorrelated'' uncertainties include the contributions 
from the uncertainty in the Monte Carlo modeling, 
which are also listed in Table~\ref{tab:systematics}.
An uncertainty in the $b$-quark production 
spectrum (App.~\ref{sec:bgenerator-mc}) carries over into the
determination of the \Kfact\ distributions.
The systematic uncertainty was estimated by weighting the
generated $p_T(B)$ distribution by a power law factor
whose range was obtained from a $b$ cross section analysis 
using an inclusive electron sample.
The shifts in fit parameters using this weighted $p_T$ spectrum
are taken as the associated uncertainty.

The isolation requirement of the inclusive electron trigger
({\it i.e.}, no matching cluster in the hadronic calorimeter), 
could, if poorly simulated, bias the decay kinematics 
of the selected $B$'s, and result in an erroneous ${\cal K}$-factor.
Since this requirement is not present in the inclusive muon trigger,
we use the difference obtained in the fit when using the
electron sample composition parameters versus those of 
the muon for this uncertainty.

Various calculations ({\it e.g.}, of efficiencies, \Kfact{s}\ldots)
are sensitive to the differences in the $B$ decay dynamics 
to a $D$,  $D^*$, or $D^{**}$.
The systematic uncertainty due to the decay model is obtained by repeating
the analysis where the decays are governed by phase space
instead.

The $L_{xy}$ resolution used in this analysis is from the
CDF detector simulation. 
The systematic uncertainty is obtained by
varying the intrinsic resolution 
by $\pm 20$\%, and the resultant shifts are taken as the uncertainty.

The last uncorrelated systematic error is due to the shape of $\xi_{MC}(ct)$,
the time dependence of the probability to tag on
a $\pi_{**}$ from $D^{**}$ decay. An alternative shape for $\xi_{MC}(ct)$
is obtained by using the $\ell D^{*-}$ signature instead of
$\ell D^-$, and using a variant of the CDF detector simulation.
The $\xi_{MC}(ct)$ is again well described by 
a Gaussian plus a constant, but the new RMS
of the Gaussian is $400 \,\mu$m, which is twice the nominal value. The shifts
in the fit resulting from this wider $\xi_{MC}(ct)$ 
are taken as the systematic uncertainty.

\label{sec:Phys-syst}

Our results may also be affected by
physics backgrounds not included in the sample composition
which result in $\nu\ell D^{(*)} X$ with the correct correlation 
of $\ell$ and $D^{(*)}$:
\begin{itemize}
  \item $B \to D_s^{(*)}D^{(*)}X$, followed by $D_s^{(*)} \to \nu\ell X$;
  \item $B_s \to \nu\ell D_s^{**}$, followed by $D_s^{**} \to D^{(*)}K$;
  \item gluon splitting $g \to c\bar{c}$, followed by $c \to \ell X$ and
		$\bar{c} \to D^{(*)}Y$.
\end{itemize}
The fractional contributions of the first two processes to our
sample are estimated~\cite{Petar} by Monte Carlo simulation 
(App.~\ref{sec:bgenerator-mc}). The fractions,
listed in Table~\ref{tab:combined-frac}, are small.

Because of the uncertainty in accurately 
predicting the rate and other characteristics of gluon splitting,
we use data rather than simulation
to set an upper bound on this contribution.
For this background the apparent $B$ vertex is
reconstructed from two different charm decays,
so the reconstructed $D$ will
have a broad $ct_D$ distribution, including cases where the
$D$ apparently decayed {\it before} the ``$B$.''
There is some (statistically
marginal) evidence of right sign ($\ell^\pm K^\pm$) 
signal in the $ct_D < -1$ mm region in the data.
We use the size of the far negative $ct_D$ tail to
constrain the potential size of the gluon 
contribution~\cite{Petar}. Because of the large statistical
uncertainty, we take as the upper bound on the gluon contribution
the central value of our fitted fraction plus twice the
statistical error on the fraction (Table~\ref{tab:combined-frac}).
Doubling the statistical error is {\it ad hoc}, but we wished
to be conservative in accounting for this poorly constrained
process.

The effect of each physics process on the asymmetry is accounted 
for by adding two new terms to the predicted asymmetry 
of Eq.~(\ref{eq:A(k)}), one for tagging on fragmentation tracks 
${\cal A}^{frag}$,
and another for tagging on decay products ${\cal A}^{decay}$.
We can determine each of these asymmetries, or their upper bounds,
and combined with the composition fractions repeat the fits
to the observed asymmetry~\cite{Petar}. The shift in fit output under
each of these processes is taken as their contribution
to the systematic uncertainty.

Examination of Table~\ref{tab:systematics} shows that
by far the largest contribution to the systematic uncertainty
comes from the input \SC\ parameters.  
The combined systematic uncertainty is obtained by
adding the individual contributions in quadrature.
The combined  systematics are still smaller than the statistical uncertainties, 
especially in the case of \dmd.  
As a mixing measurement, the application of
Same Side Tagging on the $\ell D^{(*)}$ sample is still limited by
statistics.

\subsection{Summary of the $\ell D^{(*)}$ analysis}

We have applied our SST method to a partially reconstructed
$B \rightarrow \ell D^{(*)}$ sample and accounted for
the effects of $B^0 \leftrightarrow B^+$ cross talk in the sample composition.
The flavor oscillation is readily apparent, and the oscillation frequency
and dilutions are found to be 
\begin{eqnarray}
{\dmd}  & = &  0.471^{+0.078}_{-0.068} \, \pm 0.034 \;{\rm ps}^{-1}\\ 
\label{eq:final_lepD_dmd}
{\Dx}   & = &  0.267 \pm {0.032} \; ^{+0.024}_{-0.015}  
                                            \label{eq:final_lepD_Dx}\\ 
{\Do}   & = &  0.181 \pm {0.028} \; ^{+0.023}_{-0.015}  
\label{eq:final_lepD_D0}
\end{eqnarray}
where the first error is statistical and the second is the
combined systematic. 
Our \dmd\ value compares well with a recent world average of
$ 0.484 \pm 0.026 \; {\rm ps}^{-1}$~\cite{PDG97}. 
We will discuss the dilutions further in Sec.~\ref{Sec:compr}.

\section{Testing SST in $B \rightarrow J/\psi K$ Samples}
\label{Sec:psik}
 
Having demonstrated that our SST algorithm is capable of 
revealing the $B^0$-$\overline{B}{^0}$ flavor oscillation 
in a large lepton+charm sample, we extend its use to the exclusive modes,
$B \rightarrow  J/\psi K^+$ and $J/\psi K^{*0}(892)$, where 
one may further test this method. 
The SST dilutions should, ignoring some experimental biases,
be independent of decay mode, and the $J/\psi K$ samples
should yield results comparable to the $\ell D^{(*)}$ analysis.
These $J/\psi K$ samples are too small to provide precise 
tagging measurements,  but they provide an experimental opportunity 
to study flavor tagging in this type of exclusive mode. 
This study is especially interesting because it serves as a model 
for tagging $B^0 \rightarrow J/\psi K^0_S$, which we consider
in Ref.~\cite{CDFpsiKs}.

\subsection{Reconstruction and tagging of \boldmath $B \rightarrow J/\psi K$}
\label{SubSec:psik}

Our $J/\psi K$ samples have appeared
in a number of previous publications, in whole or part, 
on measurements of $B$
masses~\cite{CDFmass}, lifetimes~\cite{CDFLifetimePRD,CDFlife}, 
branching ratios~\cite{CDFbr}, and 
production cross sections~\cite{CDFcross}.
The reconstruction criteria are
somewhat different here; we wished to
maximize the effective statistics and were less concerned 
about accurately modeling efficiencies or triggers.

$B$ reconstruction begins by forming charged particle 
combinations with $J/\psi$ candidates (Sec.~\ref{SubSec:psiSample}). 
Since we require a well measured $B$
vertex, at least two particles of the decay
must be reconstructed in the SVX with loose quality
requirements (principally that the track used hits on at 
least 3 out of 4 SVX layers). For the $J/\psi K^+$
a single particle, assumed to be a kaon, 
with $p_T > 1.75$ GeV/c is combined with the $J/\psi$.
The $J/\psi K^{*0}$ reconstruction uses
pairs of oppositely charged particles, each with $p_T > 0.5$ GeV/c.
The pair is accepted as a $K^{*0}$ candidate if a 
vertex-constrained $\chi^2$ fit---considering both permutations 
of $K^\pm$ and $\pi^\mp$ mass assignments---yields
a mass within 80 MeV/c$^2$ of the $K^{*0}$ mass
(896 MeV/c$^2$), and has $p_T(K^{*0}) > 3.0$ GeV/c.
The fit includes $dE/dx$ energy loss corrections appropriate
for the mass assignments.
If both permutations satisfy these requirements, the assignment
closest to the $K^{*0}$ mass is selected. 
The high $p_T(K^{*0})$ cut is necessary to reduce the
larger combinatoric background for $J/\psi K^{*0}$.

The particle(s) making the $K^+$ ($K^{*0}$) are combined
with the $\mu^+\mu^-$ pair in a multiparticle $\chi^2$ 
fit for the $B$ with
the $\mu^+\mu^-$ mass constrained to the world average $J/\psi$ mass,
all daughter particles originating
from a common vertex, and the entire system constrained to point 
to the $p\bar{p}$ interaction vertex. 
A run-averaged interaction vertex is used as was done for
the $\ell D^{(*)}$ sample (Sec.~\ref{sec:datasets}).
The RMS spread of the transverse beam profile is
taken to be 40 $\mu$m. The resulting $B$ candidate must
have $p_T(B) > 4.5$ GeV/c.

Rather than cutting on the $\chi^2$ from the multiparticle fit,
we cut on only the portion 
coming from the transverse ($r$-$\phi$) track parameters 
({\it i.e.}, curvature, azimuthal angle, 
and impact parameter~\cite{CDFmass}). 
The $B$ pointing resolution to the primary vertex in the $r$-$z$ plane 
is very coarse in CDF, providing little separation between
signal and non-pointing backgrounds.
Including the $z$-$\cot \theta$ contributions to the $\chi^2$
tends to smear the separation that is available 
from the precise transverse measurements.
We require the transverse tracking terms 
of the $\chi^2$ sum to be less than 20, and similarly
that the transverse components 
of the vertex  $\chi^2$ sum to be less than 4.
Although formally {\it ad hoc},  we found these cuts to be
a little better at discriminating signal from background
than the full  $\chi^2$. 
However, the final tagging analysis is
insensitive to the type and value of the $\chi^2$ cuts used
in the $B$ reconstruction.

Finally, if there are multiple $B$ candidates in the same event, 
the one with the smaller transverse track parameter $\chi^2$ 
is taken.

These $B$ samples are used 
in a likelihood fit (Sec.~\ref{SubSec:fitter})  employing
a normalized mass variable $M_N$, and so we discuss the selection results
in terms of this variable. $M_N$ is defined as
$(M_{FIT} - M_0)/\sigma_{FIT}$, where $M_{FIT}$ is the
mass of the $B$ candidate from the fit described above, 
$M_0$ is the central value of the $B$ mass peak (5.277 GeV/c$^2$),\footnote{The 
mean is systematically low by 2 MeV/c$^2$ compared with the world 
average mass because we do not make all the detailed corrections 
of Ref.~\cite{CDFmass}.} and
$\sigma_{FIT}$ is the mass error from the fit.
Over the range of $\mid M_N \mid < 20$ we have a total, signal plus 
background, of  $12564$ events in the $J/\psi K^+$ sample and
$2339$ for the $J/\psi K^{*0}$. 

\begin{figure}
\centerline{
\epsfysize 8.0cm 
\epsfbox{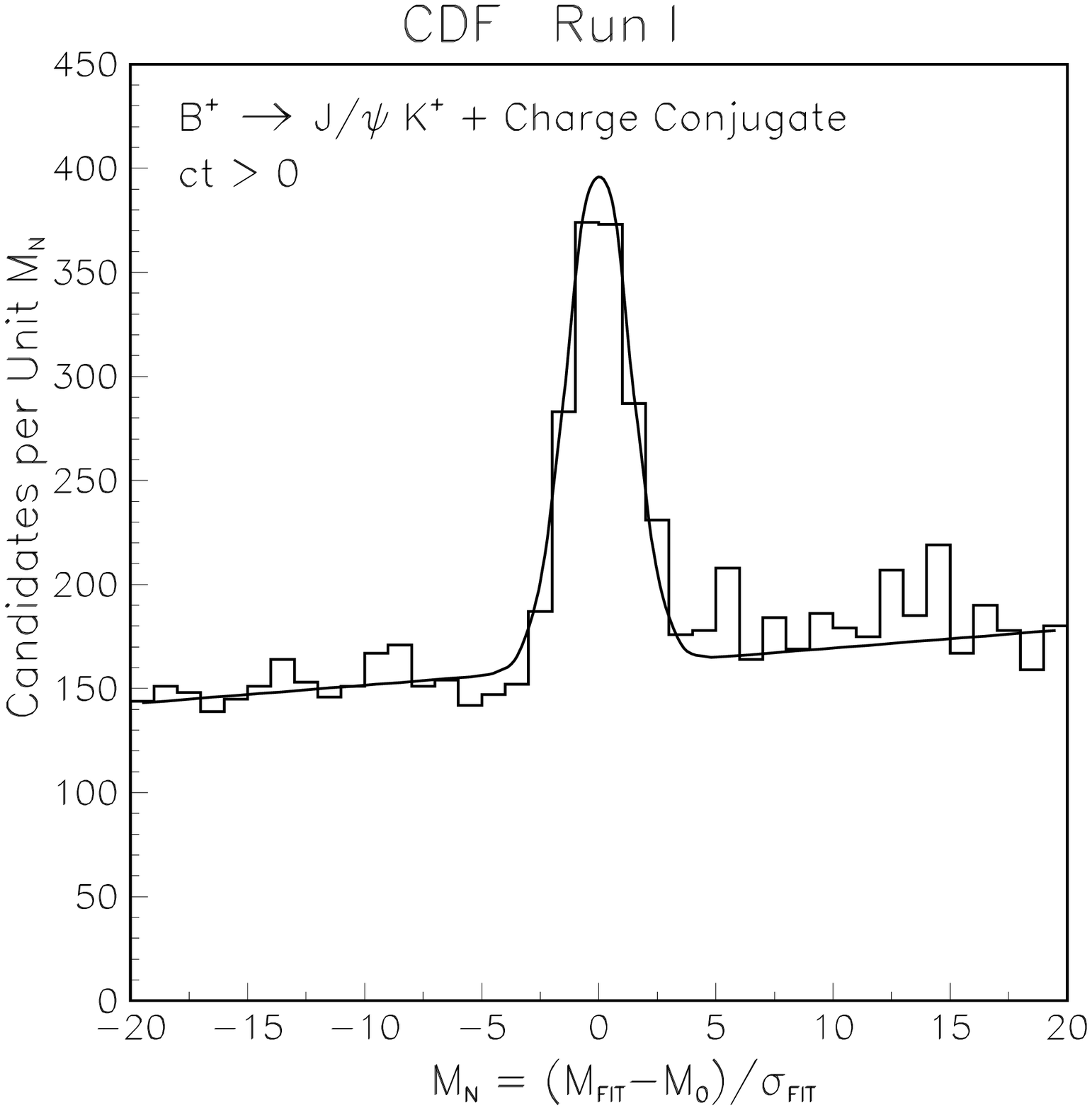}
\epsfysize 8.0cm 
\epsfbox{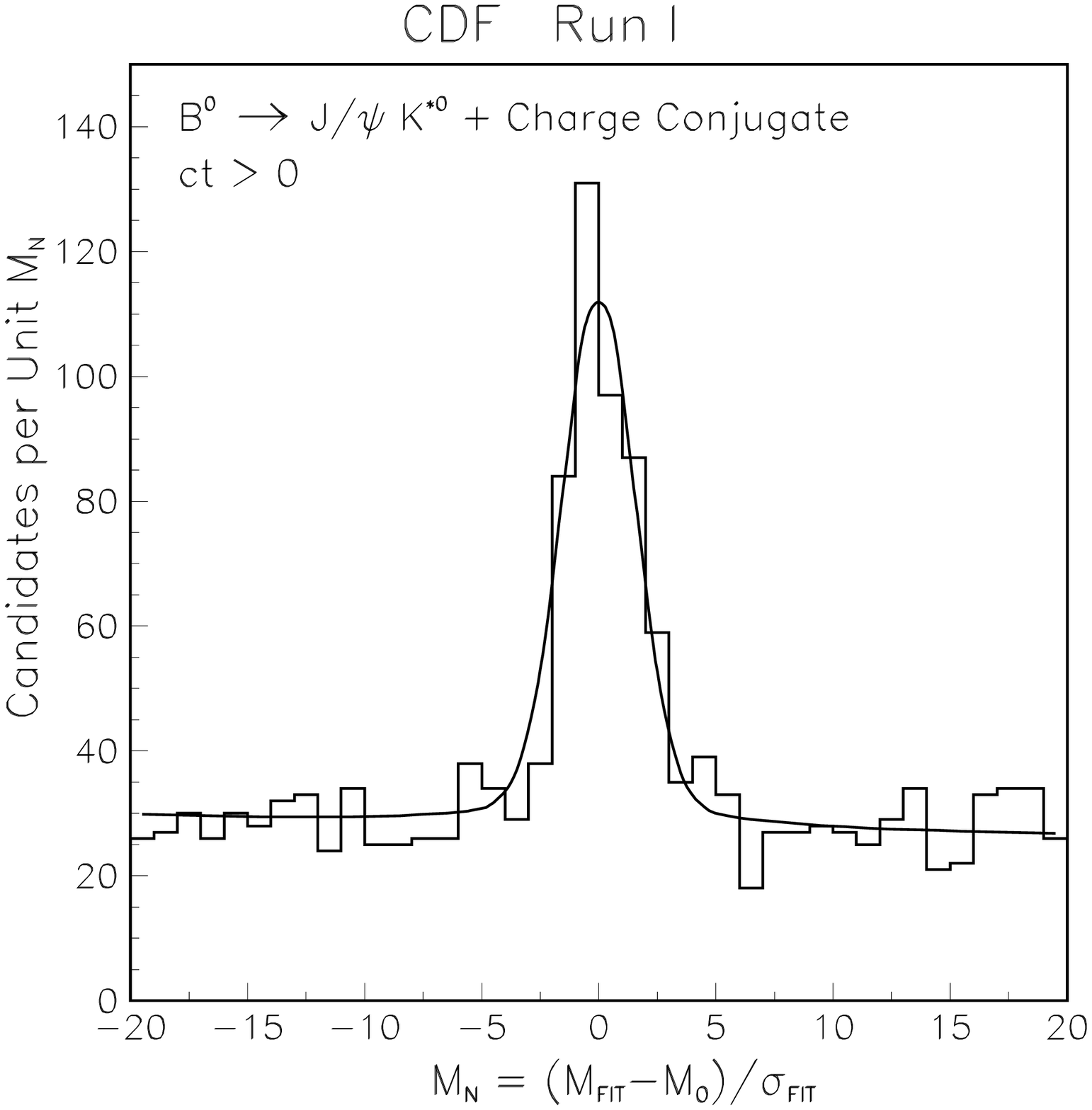}
}
\caption{Normalized mass distribution for $J/\psi K^+$ and
$J/\psi K^{*0}$  candidates with reconstructed $ct > 0$. The smooth curve is
a fit from the full likelihood function
of a Gaussian signal plus linear background parameterization.}
\label{fig:PsiKMass}
\end{figure}

Figure~\ref{fig:PsiKMass}
shows the normalized mass distributions for candidates 
with reconstructed $ct > 0$ [Eq.~(\ref{eq:ct-true})]. Also shown is the result of
the likelihood fit performed in Sec.~\ref{SubSec:fitPsiK},
where the mass is modeled by a Gaussian signal plus linear
background. The fit
yields $846 \pm 35$ $B^+$'s and $365 \pm 22$ $B^0$'s (for all $ct$).

\begin{figure}
\centerline{
\epsfysize 8.6cm 
\epsfbox{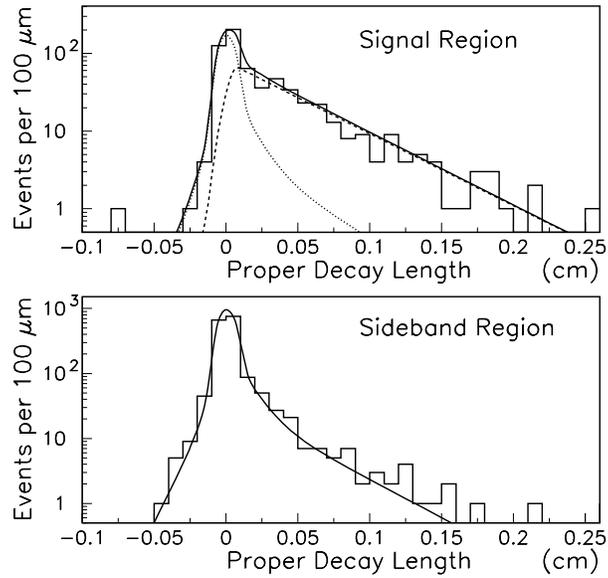}
}
\caption{The $J/\psi K^{*0}$ lifetime distributions 
for the signal (top) and sideband (bottom) regions.
Superimposed on the data are the
likelihood fit results (solid line). In the signal region, 
the $B$ component is shown by the dashed line
and the prompt plus long-lived background by the dotted line.
}
\label{fig:PsiKstrctau}
\end{figure}

Events with $ct < 0$ are dominated by  background 
(see Fig.~\ref{fig:PsiKstrctau} in Sec.~\ref{SubSec:fitPsiK}),
and the mass distributions show no clear signal.
However, these events help constrain the
background behavior and are kept as part of the analysis.

These two samples are then tagged with the SST
criteria of Sec.~\ref{SubSec:sstag}. We find about 63\% of the
$J/\psi K^+$ and $J/\psi K^{*0}$ signal events
are tagged.

\subsection{Likelihood function for the $J/\psi K$ samples}
\label{SubSec:fitter}

The tagging correlations of the  $B \rightarrow J/\psi K$ samples 
have the same physical time dependence 
as the $\ell D^{(*)}$ samples
[Eqs.~(\ref{eq:cosine}) and~(\ref{eq:const})]
but without the complications of sample composition and  
average $\beta\gamma$-corrections.
Maximal use of the smaller $J/\psi K$ samples motivated a more
sophisticated approach than used to fit the $\ell D^{(*)}$ data.

An unbinned likelihood function is devised to simultaneously
fit over various measured event properties
and obtain the SST dilutions for the $J/\psi K$ samples.
The likelihood function incorporates
the candidate's proper decay time and invariant mass 
to facilitate separation of signal and background.
It is also generalized to consider tagging biases. These
are relatively unimportant in mixing measurements 
which  only use the relative flavor-charge
asymmetry but are critical for $CP$ violation measurements
where the effect appears as an absolute charge asymmetry
of the tag.
Although our focus is on the charge-flavor correlations of SST,
this generalized approach serves as a prototype for $CP$ violation
studies~\cite{CDFpsiKs}.         

The likelihood function to be maximized is given by
\begin{eqnarray}
   {\cal L} & = &    \prod_{i=1}^{N}
     [ f_{B}{\cal L}_{B} + 
(1-f_{B}) ( f_{L}{\cal L}_{L} +
     (1 - f_{L}){\cal L}_{P} ) ]
\label{eq:FullLikelihood}
\end{eqnarray}
where the product is over all $N$ events in the mass window
$|M_{N}|<20$.
The subscripts $B$, $L$, and $P$ respectively indicate terms associated
with the $B$-meson signal, long-lived backgrounds, and prompt backgrounds.
The fraction of events that are $B$ signal is $f_{B}$.
The backgrounds are subdivided into two classes:  
``long-lived,''  which are  those  consistent with a non-zero lifetime, and
``prompt,'' which are those  consistent  with zero lifetime.
The fraction of long-lived backgrounds, which are
predominantely real $B$'s that have been misreconstructed,
is given by $f_{L}$.

The ${\cal L}_{\phi}$ ($\phi=B$, $P$, and $L$) are functions 
describing the relative probability
for obtaining the following measured values in an event:
the normalized mass ($M_N$),
the proper decay time 
   and its uncertainty ($t$ and $\sigma_{t}$),
the reconstructed
   decay flavor $r$ ($r$ is $+1$ for $B^{+}$ and $B^{0}$, and
   $-1$ for $B^{-}$ and $\overline{B}{^0}$),
and the tag track sign
   $s$  ($s$ is $+1$ for a positive track,
   $-1$ for a negative track, and 0 if there was no tag).

The density function for the $J/\psi K^{+}$ signal describing
the mass and $t$ dependence, 
the relative numbers of $B^+$ and $B^-$,
the flavor-charge tagging correlation,
and the tagging efficiency is
\begin{eqnarray}
  {\cal L}_{B} & = & G(M_{N};0,X)\;G(t;t',Y\sigma_{t})\;\otimes\;
  E(t';\tau_{B}) 
  \times \nonumber \\
  & & 
                \left(\frac{1+rR_{B}}{2}\right)
                \left(\frac{1-r\kappa_{B}(s)\Dil_{B}}{2}\right)
  {\cal E}_{B}(s). 
\label{eqn:L_B}
\end{eqnarray}
$G(x;\mu,\sigma)$ is a normalized Gaussian distribution in 
$x\in(-\infty,+\infty)$ with
mean $\mu$ and RMS $\sigma$, and $E(x;\tau)$ is a normalized exponential
distribution in $x\in [0,+\infty)$ with mean $\tau$.  
The first factor in ${\cal L}_{B}$ is the shape of the 
mass distribution ($M_{N}$), 
where $X$ is a scale factor for the mass error.  
The second factor is the Gaussian resolution of the
reconstructed $t$, including a resolution scale factor $Y$.
The $\otimes$ denotes convolution over $t'$, in this case 
with an exponential distribution $E$ 
of lifetime $\tau_{B} = \tau_{+}$.
The resolution scales, $X$ and $Y$, are
extra degrees of freedom  to monitor
our description of the errors.

The density function next contains two asymmetry factors. The first
is the probability of reconstructing the observed meson flavor  $r$, and
depends upon 
\begin{equation}
 R_B = \frac{N(B^+)-N(B^-)}{N(B^+)+N(B^-)}.
\label{eq:ReconAsym}
\end{equation}
This first factor decouples other flavor-related asymmetries
from a ``reconstruction asymmetry'' $R_B$
by accounting for the different numbers ($N$) of $B^{+}$'s and $B^{-}$'s 
that may be reconstructed
due to  a detector bias, or simply a statistical
fluctuation in the relative yield.

The second asymmetry factor represents the probability of obtaining 
a tag of sign $s$ given the reconstructed flavor $r$.
The strength of the flavor-charge ($r$-$s$) 
correlation is the usual dilution, $\Dil_{B} \equiv \Dil_{+}$.
The effective tag for a track of sign $s$ is $\kappa_{B}(s)$.
Since the $B$-$\pi$ correlation is between
$B^{\pm}$ ($r=\pm 1$) and $\pi^{\mp}$ ($s=\mp 1$),
the  $r \kappa_{B}(s)$  term appears with a negative sign.
Finally, the efficiency to obtain such a tag is ${\cal E}_{B}(s)$.

This formulation with $\Dil_{B}$, $\kappa_{B}(s)$, and ${\cal E}_{B}(s)$
is able to account for the general situation where the tagging method 
suffers from intrinsic tagging asymmetries, as might be caused by
detector biases. Tagging asymmetries may result in different dilutions 
and efficiencies for the two $B$ flavors. We define $\Dil_{B}$
to be the flavor-averaged dilution, and we are able incorporate all 
tagging asymmetry effects in $\kappa_{B}(s)$ and ${\cal E}_{B}(s)$.
In the absence of tagging biases, $\kappa_{B}(s)$ is simply the 
charge of the tagging track ($\kappa_{B}(s) = s$), and
${\cal E}_{B}(+1) = {\cal E}_{B}(-1) = \frac{}{}\epsilon_{B}$, 
where $\epsilon_{B}$ is the flavor-averaged tagging efficiency.

The ``charge asymmetry corrected'' tag $\kappa_{B}(s)$ and
efficiency ${\cal E}_{B}(s)$ are determined using
two new parameters: $\alpha_B$, which is the
charge asymmetry in selecting a tag track, {\it i.e.},
a bias in selecting positive {vs.} negative tags; and
$\delta_B$, which is the $b$-flavor asymmetry in finding a tag track, 
{\it i.e.}, having different efficiencies to tag $b$ {vs.} $\bar{b}$ mesons.
For convenience we sometimes
normalize the latter by the dilution, $\gamma_B \equiv \delta_B/D_B$.
The derivation of $\kappa_{B}(s)$ and ${\cal E}_{B}(s)$
may be found in App.~\ref{app:tagasym},
along with a complete characterization of a generic tagging method.
The actual determination of the tagging biases in our detector is 
discussed in Sec.~\ref{SubSubSec:TagBias}.

The density function for the $J/\psi K^{*0}$ signal takes the same basic
form as for $J/\psi K^{+}$, but it now incorporates 
the time-dependent flavor oscillation
$\As^0(t')=\cos(\Delta m_{d} t')$.
Since no particle identification is used,
some $J/\psi K^{*0}$ events enter the sample with the
correct $K$-$\pi$ mass assignments ``swapped'' (Sec.~\ref{SubSec:inputs}),
for which the apparent flavor is inverted.
The density function is divided into 
unswapped and swapped parts,
with the reconstructed flavor $r$ for the swapped
events appearing with a reversed sign.
The complete expression is
\begin{eqnarray}
  {\cal L}_{B} & = & (1-f_{S}) G(M_{N};0,X)\;G(t;t',Y\sigma_{t})\;\otimes\;
                     \left[E(t';\tau_{B})\right. \times \nonumber \\
  & & \left. \left(\frac{1+rR_{B}}{2}\right)
                \left(\frac{1+r\kappa_{B}(s)\Dil_{B}{\As_0}(t')}{2}\right) 
                 {\cal E}_{B}(s)\right] \nonumber \nonumber \\
  & + & f_{S} G(M_{N};\mu_{S},X_{S})\;G(t;t',Y\sigma_{t})\;\otimes\;
             \left[E(t';\tau_{B})\right. \times \nonumber \\
  & & \left. \left(\frac{1-rR_{B}}{2}\right)
                \left(\frac{1-r\kappa_{B}(s)\Dil_{B}{\As_0}(t')}{2}\right)
                 {\cal E}_{B}(s)\right],
\end{eqnarray}
where $f_{S}$ is the fraction of swapped events, 
and $\mu_S$ and $X_{S}$
are the mean and RMS of the normalized mass 
distribution for the swapped events.
The rest of the parameters parallel those of the $J/\psi K^+$,
but with $\Dil_{B} = \Dil_{0}$.

The density function for the long-lived background for both decay modes
is similar to the signal except for a linear mass distribution, 
the presence of three exponential
lifetime distributions,
and the lack of mixing:
\begin{eqnarray}
  {\cal L}_{L} &=&\left(\frac{1+\zeta_{L}M_{N}}{2W}\right)\;G(t;t',Y\sigma_{t})
  \;\otimes \nonumber \\
  & & \{f_{N}E(-t';\tau_{L2})+(1-f_{N})[f_{\tau 2}E(t';\tau_{L2})+\nonumber \\
  & & (1-f_{\tau 2})E(t';\tau_{L1})]\} \times \nonumber \\
  & & \left(\frac{1+rR_{L}}{2}\right)
  \left(\frac{1+r\kappa_{L}(s)\Dil_{L}}{2}\right){\cal E}_{L}(s).
 \label{eqn:L_LL}
\end{eqnarray}
The linear mass distribution is parameterized by a slope $\zeta_{L}$ 
and the width of the mass window $|M_{N}| < W = 20 $.
The long-lived background consists of positive- and negative-$t$
components, with a fraction $f_{N}$ in the negative exponential
(with lifetime $\tau_{L2}$).
The positive-$t$ background  is described by two exponentials,
one with a large lifetime $\tau_{L1}$,
plus a short one of $\tau_{L2}$. The latter lifetime is fixed
to be the same as for the negative-$t$ tail. The fraction of positive-$t$ 
events ($1-f_{N}$) which compose the short lifetime
exponential is $f_{\tau 2}$.

The background may also possess a reconstruction asymmetry $R_L$, 
or a charge correlation between the tag and what is assumed to be
the $K^+$ or $K^{*0}$, {\it i.e.}, a dilution $\Dil_L$.
The background asymmetry description parallels that of the signal 
with  $R_\phi$, $D_\phi$, $\kappa_\phi$, and ${\cal E}_\phi$ 
defined independently for each event class 
($\phi = B$, $L$, and $P$).

The prompt background density function is
\begin{eqnarray}
 {\cal L}_{P} & = &
    \left(\frac{1+\zeta_{P}M_{N}}{2W}\right)\;G(t;0,Y\sigma_{t}) \times \nonumber \\
  & & \left(\frac{1+rR_{P}}{2}\right)
    \left(\frac{1+r\kappa_{P}(s)\Dil_{P}}{2}\right){\cal E}_{P}(s),
 \label{eqn:L_Prmpt}
\end{eqnarray}
with the same variable definitions as before except that they
apply to the prompt background. 
There is only a dependence on the proper time
through the $t$ resolution, and thus no convolution is needed.

When summed together and multiplied over all the selected events in a
particular dataset, the density functions form a properly normalized
likelihood function.

\subsection{Input likelihood parameters}
\label{SubSec:inputs}

A number of the likelihood parameters are more accurately
obtained from sources other than our  $J/\psi K$ data.
In this section we will discuss which parameters are fixed in the fit, and
their sources, values, and uncertainties.

\subsubsection{$B$ meson parameters}
\label{SubSubSec:Bdecay}

The likelihood function relies on the temporal properties of
the $B$ decay, and these are best obtained from world averages.
Since we wish to
measure the tagging dilution, and not the oscillation frequency,
we include  $\Delta m_d$ in this list.
We use the following averages from the Particle Data Group~\cite{PDG},
\begin{eqnarray*}
\Delta m_{d}  & = & 0.474\pm 0.031\;{\rm ps}^{-1}\\
\tau_{{+}} & = & 1.62\pm 0.06\;{\rm ps} \\
\tau_{{0}} & = & 1.56\pm 0.06\;{\rm ps}.
\end{eqnarray*}

\subsubsection{Incorrect $K-\pi$ assignment}
\label{SubSubSec:KPiSwap}

The $J/\psi K^{*0}$ events include 
real $B^{0}\rightarrow J/\psi K^{*0}$ decays,
but with the incorrect $K$-$\pi$ mass assignment.  
A Monte Carlo sample of $B^{0}$ decays 
(App.~\ref{sec:bgenerator-mc}) was generated and then processed 
as data.  
The reconstruction tries both assignments, 
and if both versions pass the selection criteria, the one with
the $K\pi$ mass nearest the $K^{*0}$'s is chosen.
The events with the $K$ and $\pi$ swapped 
have an $M_N$ distribution which is roughly 
Gaussian with mean $\mu_{S}=-0.5$ and RMS $X_{S}=4.9$.  
The area of the swapped Gaussian is $9.8$\% of that for the unswapped 
distribution. The kinematic dependence of the swapped events 
on $p_T(B)$ has also been studied~\cite{Ken}. 
The fraction of swapped events is constant within a few
percent over our range of $p_T(B)$, but the mean and RMS of $M_N$
show some systematic variation.

The swapped component is difficult to fit in the $J/\psi K^{*0}$ data
because 
it is difficult to distinguish these events from the combined
shapes of the narrow central Gaussian and linear background.
The likelihood fit therefore
fixes the input parameters $\mu_{S}$, $X_{S}$, and $f_{S}$
to the values from the simulation.
We allow for a 100\% variation in the fraction,
$f_{S} = 0.1 \pm 0.1$, and 
assign uncertainties to the other swapping parameters
which covers the range observed when spanning the $p_T(B)$ interval
of the data,
{\it i.e.},   $\mu_{S} = -0.5 \pm 0.5$ and
$X_{S} = 5.0 \pm 2.0$.

\subsubsection{Tagging biases}
\label{SubSubSec:TagBias}

A tagging method may display
two sorts of inherent asymmetries  (App.~\ref{app:tagasym}):
selecting one charge more often than the other as a tag
($\alpha$), or having a greater efficiency to tag 
on one $b$-flavor over the other ($\delta$).

The charge asymmetry of the tags
for a flavor symmetric sample is
\begin{eqnarray}
\alpha = \frac{N_{+} - N_{-}}{N_{+} + N_{-}},
\label{eq:alpha}
\end{eqnarray}
where $N_{+}$ ($N_{-}$) is the number of positive 
(negative) tags.
Since we reconstruct the decay flavor, we can correct for
any flavor asymmetry in our samples and determine $\alpha$
from the data. This is done for the
backgrounds by letting $\alpha_L$ and $\alpha_P$ float in the
likelihood fit. 

However, 
$\alpha$ appears in the likelihood function partly as a factor 
$\Dil/(1\pm  \alpha)$ [Eqs.~(\ref{eqn:L_B}) and~(\ref{eqn:Corr_tag})].
We can essentially eliminate the influence of the $\Dil_B$-$\alpha_B$ coupling
in the fitted $\Dil_B$ by fixing $\alpha_{B}$
to an independently measured value;  we
obtain a better constraint on $\alpha_{B}$ as well.
We do this
by using a large inclusive sample of non-prompt $J/\psi$'s,
{\it i.e.}, a flavor-symmetric $b$ sample. 
This is the sample of  $J/\psi$'s described 
in Sec.~\ref{SubSec:psiSample} with the following additional requirements:
both muons are in the SVX, and
the $J/\psi$ projected flight distance 
$\lxy(J/\psi)$ [Eq.~(\ref{eq:Lxy-Def})] exceeds $200\;\mu m$.
This last cut results in a sample which is more than $90\%$ pure 
$b$ hadron decays.
We also require $p_T(J/\psi) > 4\;{\rm GeV}/c$
so that the  $p_T$'s are similar
to that of the $J/\psi K$ samples.

We have looked for tagging asymmetry dependencies in a variety
of variables, including the $p_T$ and $\lxy$
of the $J/\psi$, and found  only
two variables of interest.
First, there is an $\alpha$-dependence on the $p_{T}$ 
of the tagging track, with more positive tracks reconstructed 
than negative ones at low  $p_{T}$.
This is due to the charge asymmetry inherent in the design of the CTC
(wire planes are oriented along the direction of positive tracks).
Proton spallation from the beampipe might contribute an additional
$p_{T}$-dependent asymmetry,
but this effect has been largely eliminated by the
impact parameter significance cut on the SST candidates.  
The second variable is 
the number of good $p\bar{p}$  interaction vertices 
$n_{V}$ found by the VTX. 
The number of vertices is an indicator 
of the total hit occupancy 
in the CTC, which influences the tracking efficiency.

Characteristics of the $J/\psi K$ samples, or
the criteria of the tagging method, 
could modify the biases in the tracking asymmetry.
We therefore compare the charge asymmetry in four types of
$J/\psi$ ($\lxy > 200 \, \mu$m) 
subsamples (the $J/\psi$ is used for the $b$ direction): 
({\it i})~the SST tags,
({\it ii})~SST candidates ({\it i.e.}, no  $p_T^{rel}$ selection),
({\it iii})~SST candidates passing a
            $b$-vertex veto, and
({\it iv}) all tracks satisfying the SST cuts except for being
            in a ``side cone'' {\it away} from the $J/\psi$-axis.
The first case is the most direct extension of our analysis, but it is
contaminated by tagging on $B$ daughters.
The second case has greater statistics as 
there are multiple track entries per $J/\psi$,
but more importantly there is no bias associated with the $p_T^{rel}$ cut.
The third sample suppresses tagging
on $B$ daughters---such tags are impossible
with the exclusive $J/\psi K$ signal---by requiring
that the SST candidate impact parameter significance 
relative to the $J/\psi$ vertex, $d_\psi/\sigma_{\psi}$,
is greater than 2.  
In the final case we select tracks with the basic SST cuts, but 
which are divorced from the $b$ by using all 
tracks in a $1.0 < \Delta R < 2.1$ ``side cone''
relative to the $J/\psi$.

\begin{figure}
\centerline{
\epsfysize 9.2cm 
\epsffile{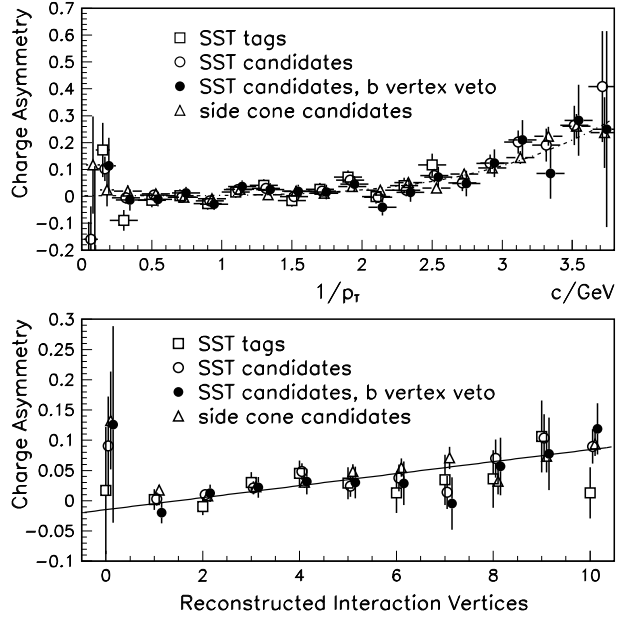}}
\caption{Charge asymmetry ($\alpha$)
dependence on the track's $1/p_{T}$ (top) and 
number of primary vertices $n_{V}$ in the event (bottom) for: 
Same Side Tags (squares), 
SST candidate tracks (open circles),
SST candidate tracks with $b$-vertex veto (solid circles), and
tracks in a cone away from the $J/\psi$ direction (triangles).
The solid curves are the results of 
least-squares fits of the asymmetry parameterization (see text)
to the SST candidates with the $b$-vertex veto.
}
\label{a:ptnvc0}
\end{figure}

Since we wish to study a $p_T$ dependence, and
only sample ({\it i}) selects a unique track per $J/\psi$, 
we relax the 400 MeV/c SST cut on samples ({\it ii})-({\it iv}).
The upper plot in Fig.~\ref{a:ptnvc0} shows the charge asymmetry
as a function of the track's inverse $p_T$ for these four samples.
The asymmetry is fairly small at $1/p_T = 2.5$ (GeV/c)$^{-1}$ 
(the nominal SST $p_T$  cut of 400 MeV/c)
but then rises significantly.
The asymmetry as a function of number of vertices $n_V$ 
is shown in the lower plot in Fig.~\ref{a:ptnvc0}.
Neither variable shows any significant differences 
across the four samples.

With all four samples so similar, we consider the asymmetry
to be independent of the tagging, and
choose 
sample ({\it iii}) to determine a parameterization 
for the tag asymmetry $\alpha_B$.
The $p_T$ dependence of the asymmetry is well described 
by  $p_T^{-4}$ and a linear function for the number of vertices.
In terms of both variables we write
\begin{eqnarray}
  \alpha_B(p_{T},n_{V}) & = & \{ a_{1}(n_{V}-n_{0})+b_{1} \} (p_{T}^{-4} -
   p_{T0}^{-4})  \nonumber \\ 
   & + & \{ a_{2}(n_{V}-n_{0}) + b_{2}\} ,
\label{eq:alpha_param} 
\end{eqnarray}
where we have included offsets $p_{T0}^{-4}$ and $n_{0}$, fixed to
20.0 (GeV/c)$^{-4}$ and 3.0 respectively,
to remove the correlation between the $a$'s and $b$'s
when fitting for them.

We determine these parameters by making subsamples
for each integer value of $n_{V}$, and fitting them for the 
coefficient of the $p_T^{-4}-p_{T0}^{-4}$ term and a constant
offset. The series of these coefficients and offsets  are
then fit for the linear $n_{V}$ dependence.
We obtain the values
\begin{eqnarray*}
  a_{1} & = & (3.9\pm 1.8)\times 10^{-4} \; {\rm (GeV/c)^4}\\
  b_{1} & = & (1.3\pm 0.4)\times 10^{-3} \; {\rm (GeV/c)^4}\\
  a_{2} & = & (1.4\pm 0.4)\times 10^{-2}\\
  b_{2} & = & (2.6\pm 0.8)\times 10^{-2},
\end{eqnarray*}
which give the curves in Fig.~\ref{a:ptnvc0}.
If we consider the tags in the  $J/\psi$ data [case ({\it i})],
we find the average tag asymmetry is
($1.6 \pm 0.7$)\%.
In the limit as $1/p_T \rightarrow 0$ the asymmetry 
parameterization gives
($0.14 \pm 0.86$)\%.
We use this parameterization to describe the tagging asymmetry
for the $B$ signal $\alpha_B$
in the likelihood fit (Sec.~\ref{SubSec:fitPsiK}).

As well as an intrinsic bias towards positive 
or negative tags $\alpha$, the SST could also have a bias $\delta$
to tag $b$ and $\bar{b}$ mesons with different efficiencies
(see App.~\ref{app:tagasym}), {\it i.e.},
\begin{eqnarray}
\delta = \frac{\epsilon(\bar{b}) - \epsilon({b})}
              {\epsilon(\bar{b}) + \epsilon({b})},
\label{eq:delta}
\end{eqnarray}
where $\epsilon$ is the efficiency  to tag on a given flavor.
It is more convenient to express its ratio relative to the dilution
in the likelihood,
so we often use $\gamma \equiv \delta/\Dil$.
We can constrain $\delta$ from the data, and do so for the backgrounds
by letting $\delta_{P,L}$ float in the likelihood.

In the likelihood function $\gamma_B$ appears partly 
as a factor $(1\pm  \gamma_B)\Dil_B$.
As with $\alpha_B$, an independent determination of $\gamma_B$
is preferable in order to decouple it from the dilution.
The inclusive $J/\psi$'s cannot be used since we have no
knowledge of the $b$ flavor. The higher statistics  $\ell D^{(*)}$
data indicate that $\gamma_B$ is less than 15\%.
However, as discussed in Appendix~\ref{app:gamma_alpha},
we can improve upon this constraint by about a factor of three
by considering the behavior of $\gamma_B/\alpha_B$.
We find $\gamma_B/\alpha_B$ spans the range from 0.0 to about 2.5,
and we use $\gamma_{B}/\alpha_{B} = 1.0^{+1.5} _{-1.0}$ in
the likelihood fit.

This completes the list of input parameters fixed in the likelihood fit,
and we now proceed to fitting the data.

\subsection{Fitting the $J/\psi K^+$ and $J/\psi K^{*0}$  samples}
\label{SubSec:fitPsiK}
\label{SubSec:fitPsiKst}

We use the likelihood function to fit the $J/\psi K$  samples 
with the parameters discussed in the last section fixed, 
and allow the others to float freely, {\it i.e.},
\begin{itemize}
 \item $f_B$, the fraction of events which are signal,
 \item $f_L$, the fraction of background which is long-lived,
 \item $f_N$, the fraction of the long-lived background in the
       negative lifetime tail,
 \item $f_{\tau2}$, the fraction of positive long-lived background 
       with lifetime $\tau_{L2}$,
 \item $\zeta_P$ and $\zeta_L$, the slopes of the prompt and 
       long-lived backgrounds in normalized mass,
 \item $X$, the error-scale factor for the normalized mass,
 \item $Y$, the error-scale factor for the decay time,
 \item $\tau_{L1}$,  the large lifetime of the positive long-lived background,
 \item $\tau_{L2}$,  the small lifetime used for positive and negative 
         long-lived backgrounds,
 \item $\epsilon_B$, $\epsilon_P$, and $\epsilon_L$, the efficiencies
       for tagging signal, and prompt and long-lived backgrounds,
 \item $R_B$, $R_P$, and $R_L$, the reconstruction asymmetries of the signal, 
        and prompt and long-lived backgrounds, 
 \item $\alpha_P$ and $\alpha_L$, the tagging asymmetries of prompt and 
       long-lived backgrounds,
 \item $\delta_P$ and $\delta_L$, the tagging efficiency asymmetries 
       of prompt and long-lived backgrounds, and
 \item $\Dil_B$, $\Dil_P$,  and $\Dil_L$, the dilutions for the signal, 
       and  the prompt and long-lived backgrounds.
\end{itemize}
Each decay mode is separately fit by minimizing
the negative log-likelihood function.

The fit results are compared to the $M_N$ distribution in
Fig.~\ref{fig:PsiKMass}.
The fitted proper decay lengths are shown 
in Fig.~\ref{fig:PsiKstrctau} for the $J/\psi K^{*0}$, where
we have defined---for display purposes only---the signal region as $|M_N| < 3$,
and $3 < |M_N| < 20$ as sidebands.
The data are well described by the fits.

\begin{figure}
\centerline{
\epsfysize 7.7cm 
\epsfbox{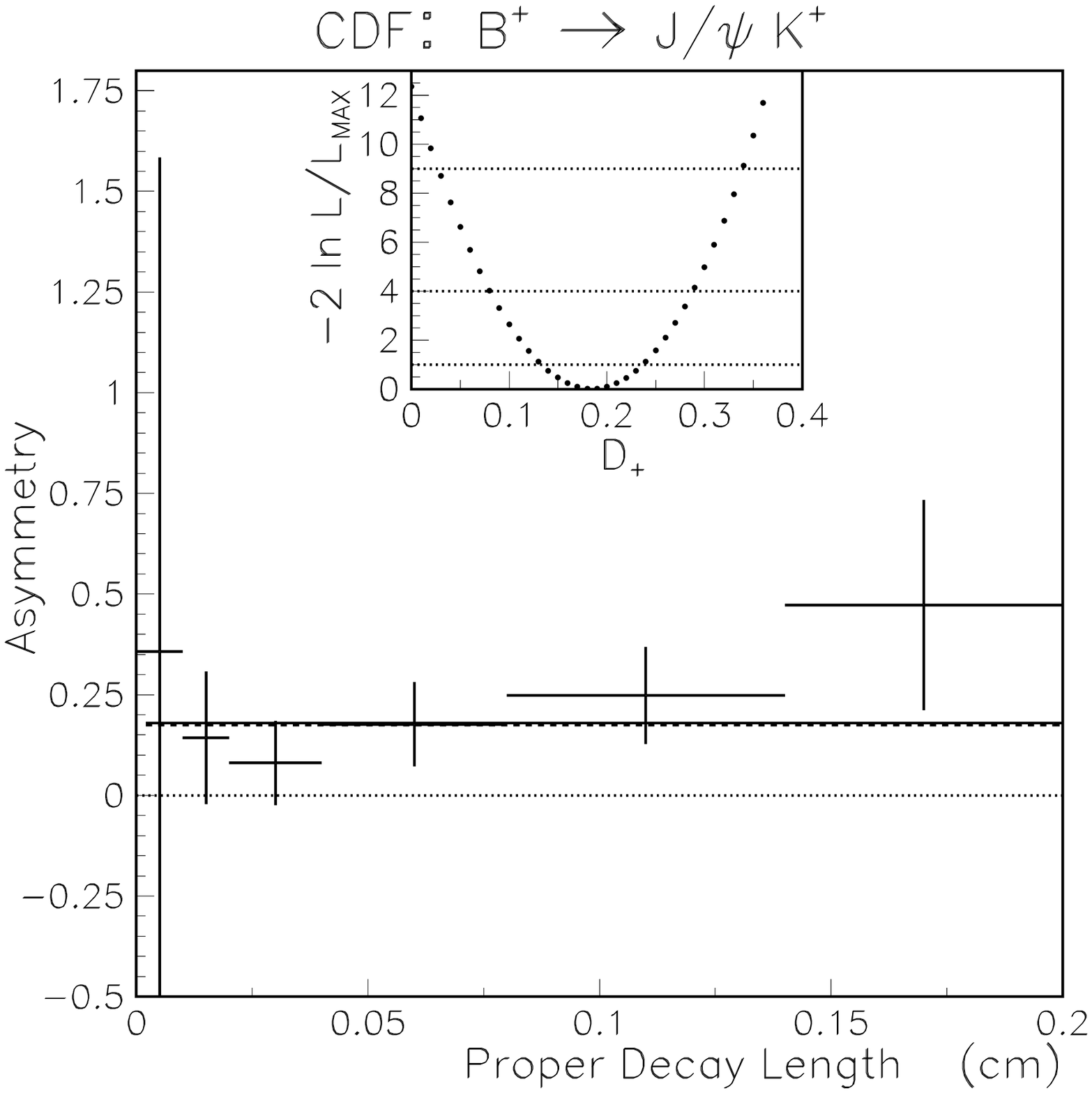}
\epsfysize 7.7cm 
\epsfbox{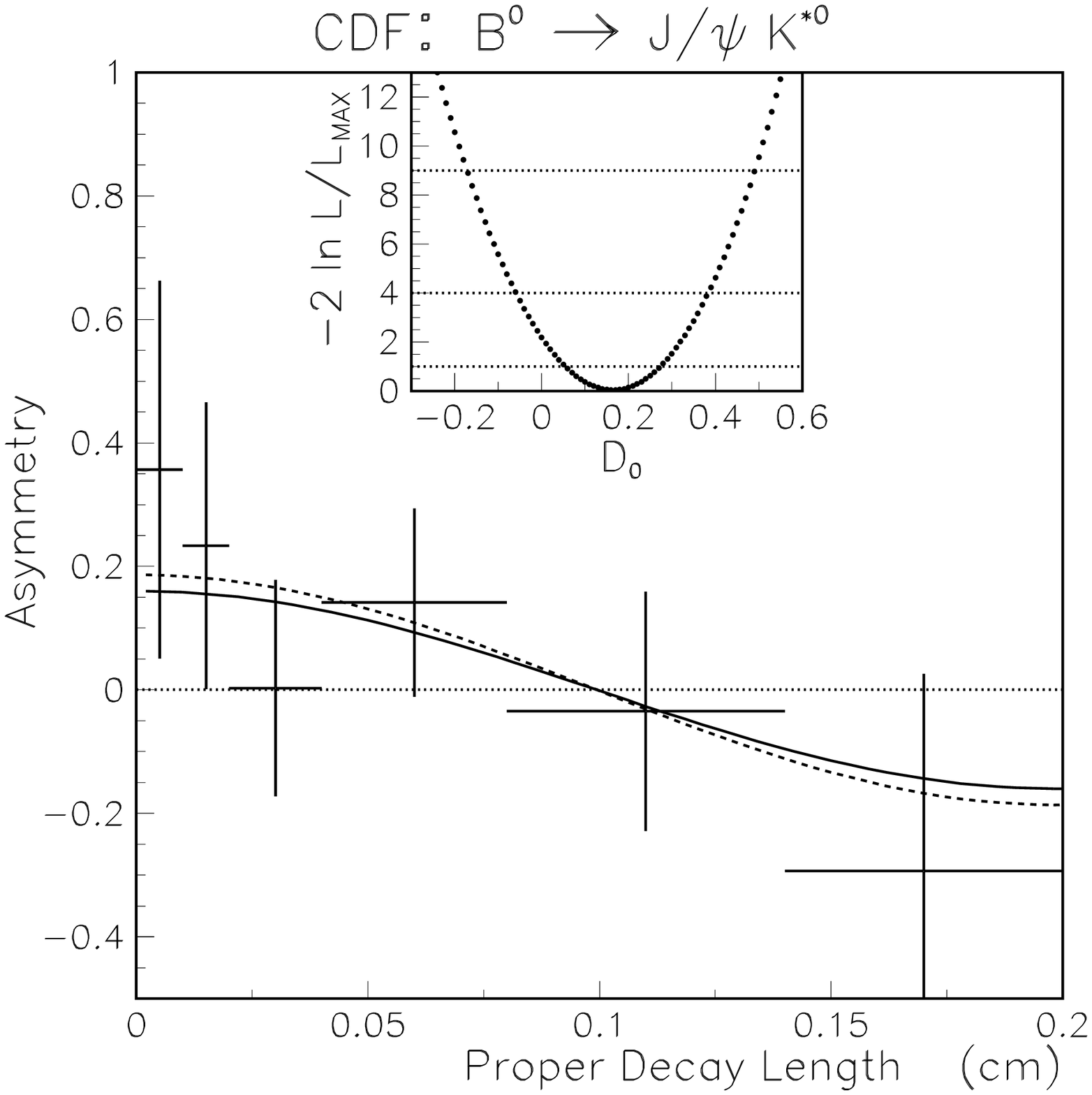}
}
\caption{The mass sideband subtracted flavor-charge asymmetry 
as a function of the reconstructed $ct$ (points):
left is $B^+ \rightarrow J/\psi K^{+}$, 
and $B^0 \rightarrow J/\psi K^{*0}$ is on the right.
Superimposed on the data are the likelihood fit results (solid lines).
The insets are scans through the log-likelihood functions
as the dilutions are varied about the fit maxima. 
Also shown in the main plots are the results of
simple $\chi^2$ fits to the points 
(dashed lines, partially obscured by the solid lines). 
}
\label{fig:PsiKfit}
\end{figure}

To display the flavor-charge asymmetries, we 
compute the mass sideband subtracted asymmetry 
(analogous to Eq.~(\ref{eq:A(RS-RS)}))
for the data in $ct$ bins. The results are shown
in Fig.~\ref{fig:PsiKfit} with
the likelihood fits superimposed (solid line).
The $J/\psi K^{+}$ plot shows a clear correlation,
consistent with being constant, and the $J/\psi K^{*0}$
data is in good agreement with the mixing hypothesis. 
Also shown in the figure insets are
$-2 \ln({\cal L/L}_{MAX})$ as a function of the dilution, 
where ${\cal L}$  is the value of the likelihood
for a given $\Dil_B$ after maximization with respect to all other
free parameters, and ${\cal L}_{MAX}$ is its value at the global maximum.
We see a well behaved, approximately parabolic, shape.
The 1, 2, and $3\sigma$ errors of the likelihood
are indicated in the inset by the three horizontal dotted lines.

As a simple check, the binned and mass sideband subtracted asymmetries
of Fig.~\ref{fig:PsiKfit} were fit to  a constant for $J/\psi K^+$, 
and $\Dil_0 \cos(\Delta m_d t)$ for $J/\psi K^{*0}$,
using a simple $\chi^2$ fit
without any additional corrections ({\it e.g.}, tagging asymmetries,
$t$ resolution, {\it etc.}). The results 
(dashed line in Fig.~\ref{fig:PsiKfit})
agree very well with the likelihood fits, 
indicating that the fits are 
driven by the basic asymmetries in the data and are not significantly
influenced by the refinements of the likelihood fit.
Of course, the fits are strongly dominated by the statistics.

The principal results of the likelihood fits are the dilutions
${\Dil}_B$ given in Table~\ref{tab:results},
along with the other fitted parameters.
A few remarks may be made on this table in passing.
The $t$ error scales $Y$ are virtually unity, indicating that
the lifetime modeling and error estimates describe the data well. 
The $X$ scales (the RMS of the signal $M_N$
Gaussian) are not 1.0, but are instead close to the known scale
of 1.3~\cite{CDFmass}. The tagging asymmetries $\alpha$ 
for the background terms are generally consistent
with the $(1.6 \pm 0.7)\%$ found as 
the average value from inclusive $J/\psi$'s
(Sec.~\ref{SubSubSec:TagBias}), 
with $\alpha_P$ for the $B^0$'s about $2.5\sigma$ high.
The reconstruction asymmetries $R$ are also not statistically 
significant beyond 1-2$\sigma$. The background dilutions 
are, not surprisingly, 
consistent with zero when selecting pairs of tracks 
with no net charge ($K^{*0}$) from the event, but significant, 
and anticorrelated, when selecting single charged particles ($K^\pm$).

As a subsidiary check, we replace the $\alpha_B(p_{T},n_{V})$
parameterization in the likelihood by
$\alpha_B(p_{T},n_{V}) + \alpha'$, where $\alpha'$ is a free
parameter.  The fit returns $\alpha' = -0.001 \pm 0.079$ for the $B^0$'s, and
$\alpha' = 0.036 \pm 0.052$ for the $B^+$ result.  If
$\alpha_B(p_{T},n_{V})$ is a good description of the data, then $\alpha'$
should be close to zero, as indeed they are.

The systematic uncertainties of the dilution measurements are 
determined by successively shifting the parameters fixed in the fit
up and down by  $1\sigma$ and repeating the fit.
The resulting shifts in the fitted dilution are taken as the
high and low systematic uncertainties due to that input parameter. 
Section~\ref{SubSec:inputs} discussed the 
uncertainties assigned to these input parameters. 
The results are shown in Table~\ref{tab:bk3} for  $J/\psi K^{+}$ and in 
Table~\ref{tab:bk2} for $J/\psi K^{*0}$, beginning with
the uncertainties associated with the $B$ decay properties.

Next are the systematic uncertainties arising from
the uncertainty in the parameters 
($a$'s and $b$'s of Eq.~(\ref{eq:alpha_param}))
describing the signal tagging asymmetry $\alpha_B$.
Because we used the  constraint on $\gamma_B/\alpha_B$
(App.~\ref{app:gamma_alpha}),
we cannot vary $\alpha_B$ and $\gamma_B$  independently.
We vary  $\alpha_B$ by varying the 
$a$ and $b$ parameters individually 
by $1\sigma$ for fixed ``central'' values of
$\gamma_B/\alpha_B =$ 0, 1, and 2.5, and remaximizing.
In this case the dilution shift is 
from the difference between the  nominal and shifted values
where both use the {\it same}  fixed 
value of $\gamma_B/\alpha_B$ in the fit.
The maximum excursion of the dilution
among the three $\gamma_B/\alpha_B$ combinations
is selected for each $a$ ($b$) parameter
as the uncertainty for that $a$ ($b$),
irrespective of the $\gamma_B/\alpha_B$ value used for the other  
$a$'s and $b$'s.
The tables list all variations, including those not used.
While this mixing of $\gamma_B/\alpha_B$'s is nominally inconsistent, 
it provides a conservative estimate.

The contributions from the uncertainty on $\gamma_B/\alpha_B$ 
follow in the tables using the nominal $\alpha_B$. 
Table~\ref{tab:bk2} also includes the effects
from the $K$-$\pi$ swap parameters.

The $J/\psi K^{+}$ systematic uncertainty is composed of roughly equal 
contributions from the $B$ lifetime, tagging charge asymmetry, 
and tagging efficiency asymmetry,
but overall has a small systematic uncertainty.
The largest effects for the  $J/\psi K^{*0}$ are due to the tagging asymmetry
and the width of the swapped $K$-$\pi$ mass distribution $X_{S}$,
which has a strong asymmetric effect. If the swapped $M_N$ distribution 
is broad there is little effect; however, as it gets narrower it is 
more difficult to distinguish the swapped from the
unswapped events and a larger uncertainty ensues.

The positive (negative) shifts of the dilution due to each parameter are 
added in quadrature to obtain the positive (negative) 
``combined uncertainty'' of the dilution in the Tables.
We thereby obtain the result,
\begin{eqnarray}
  {\Dil}_{+} & = & 0.185 \pm 0.052 \, ^{+0.003} _{-0.004} 
                                \label{eq:PsiKD+} \\
  {\Dil}_{0} & = & 0.165 \pm 0.112 \, ^{+0.018} _{-0.021}.
  \label{eq:PsiKD0}
\end{eqnarray}
These results are similar to the $\ell D^{(*)}$ results 
of Eqs.~(\ref{eq:final_lepD_Dx}) and~(\ref{eq:final_lepD_D0}).
In the next section a detailed comparison will be made
between the two sets of measurements. \\

\section{Checks and comparisons between $\ell D^{(*)}$ and
$J/\psi K$ data and Monte Carlo Simulation}   
\label{Sec:compr}

This section presents checks on the robustness of our results and
makes a closer comparison between tagging in our $\ell D^{(*)}$ and
$J/\psi K$ samples as a means of furthering our study of Same Side Tagging.
Although the physics processes believed to be responsible 
for the observed flavor-charge correlations
should not depend upon the $B$ decay mode,  potential
experimental biases could influence them differently, for instance
the fact that the $B$'s in the $\ell D^{(*)}$ data tend to have higher
$p_{T}$'s than those in the $J/\psi K$ data.
Given the limited statistical power of these measurements,
we complement the comparisons by also showing some results from
Monte Carlo simulations.  The simulation provides a further means to
study possible systematic differences between the two data samples,
and to gain some insight into underlying mechanisms.

It is not obvious to what extent one can
rely upon a given simulation to model particle distributions from
fragmentation and underlying parton interactions in a $p\overline{p}$ event.
While considerable effort has gone into developing and tuning simulations
for $e^{+}e^{-}$ collisions, the state of the art is somewhat
less well developed for the more complex high-energy hadron-hadron collisions.
We therefore
consider several basic comparisons between data and the simulation.
We find reasonable agreement and conclude that the simulation is a fair
model of the data, although our comparisons are necessarily rather coarse.
Having developed some confidence in the simulations, we proceed to
compare $\ell D^{(*)}$ and $J/\psi K$ dilutions.
It should be stressed that prior to this point of comparison
the analyses described in this paper have not depended upon accurately
simulating the tagging (though we have used simulations to model decay
kinematics, where the models are well established), 
and indeed, the analyses were designed with this independence in mind.

For the simulations discussed in this section
we have used the PYTHIA Monte Carlo generator, albeit tuned to
match the charged particle distributions in the $\ell D^{0}$ mode as
described in Appendix~\ref{sec:Tuned-PYTHIA-mc}.
Samples of $B^0$ and $B^+$ mesons were generated 
for the $\ell D^{(*)}$ and exclusive $J/\psi$ decay modes. 
The $B$ decay modes were forced via specific channels for
efficient generation, and the $\ell D^{(*)}$ events had a
sample composition approximating that found 
in the data. The events were passed through the CDF detector simulation,
reconstruction and selection code, and finally the 
SST algorithm. 

\subsection{General comparisons}

We begin our comparison of data and simulation by examining
distributions of some basic variables.
In all these comparisons we
use mass sideband-subtracted data samples, and we average over the decay
modes according to their contribution to the $\ell D^{(*)}$ or
$J/\psi K$ sample.  
Within each class of sample, the decay-mode specific distributions are 
very similar to one another.

\begin{figure}
\centerline{
\epsfysize 8.0cm
\epsfbox{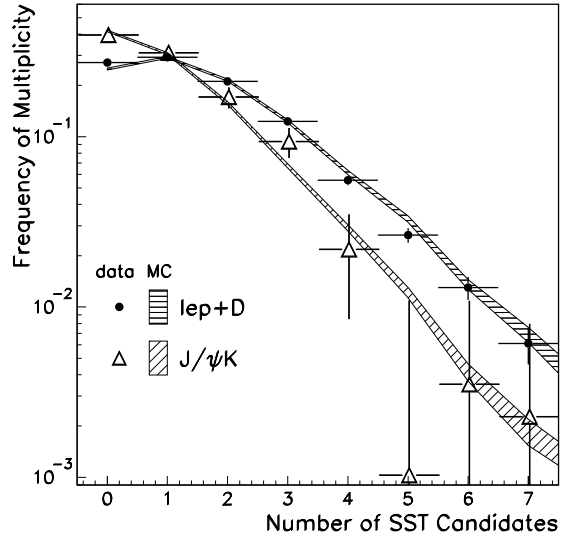}}
\caption{Number distributions of candidate tagging tracks 
in both $\ell D^{(*)}$ and $J/\psi K$ data (points) and simulation 
(shaded bands). The widths of the shaded bands are the statistical 
errors from the Monte Carlo sample size.
}
\label{fig:nSST}
\end{figure}

The multiplicity distribution of SST candidates per $B$
({\it i.e.}, tracks which satisfy all SST cuts except for 
the criterion of minimum $p_{T}^{rel}$)
is shown in Fig.~\ref{fig:nSST}
for both data and simulation.  There is general
agreement between the simulations and data for both $\ell D^{(*)}$ and
$J/\psi K$. The $\ell D^{(*)}$ channel tends to show on average
slightly more tagging candidates per $B$ than the $J/\psi K$.  
This difference is due to the higher energies which
characterize the $\ell D^{(*)}$ candidates and to the extra tracks present
from $B$ daughters which are not used in the partial $B$ reconstruction.
We note as a point of contrast that the corresponding distributions for 
the sidebands show some significant variations across the decay modes, so
the agreement seen in Fig.~\ref{fig:nSST} is not a trivial result.

\begin{figure}
\centerline{
\epsfysize 8.0cm
\epsfbox{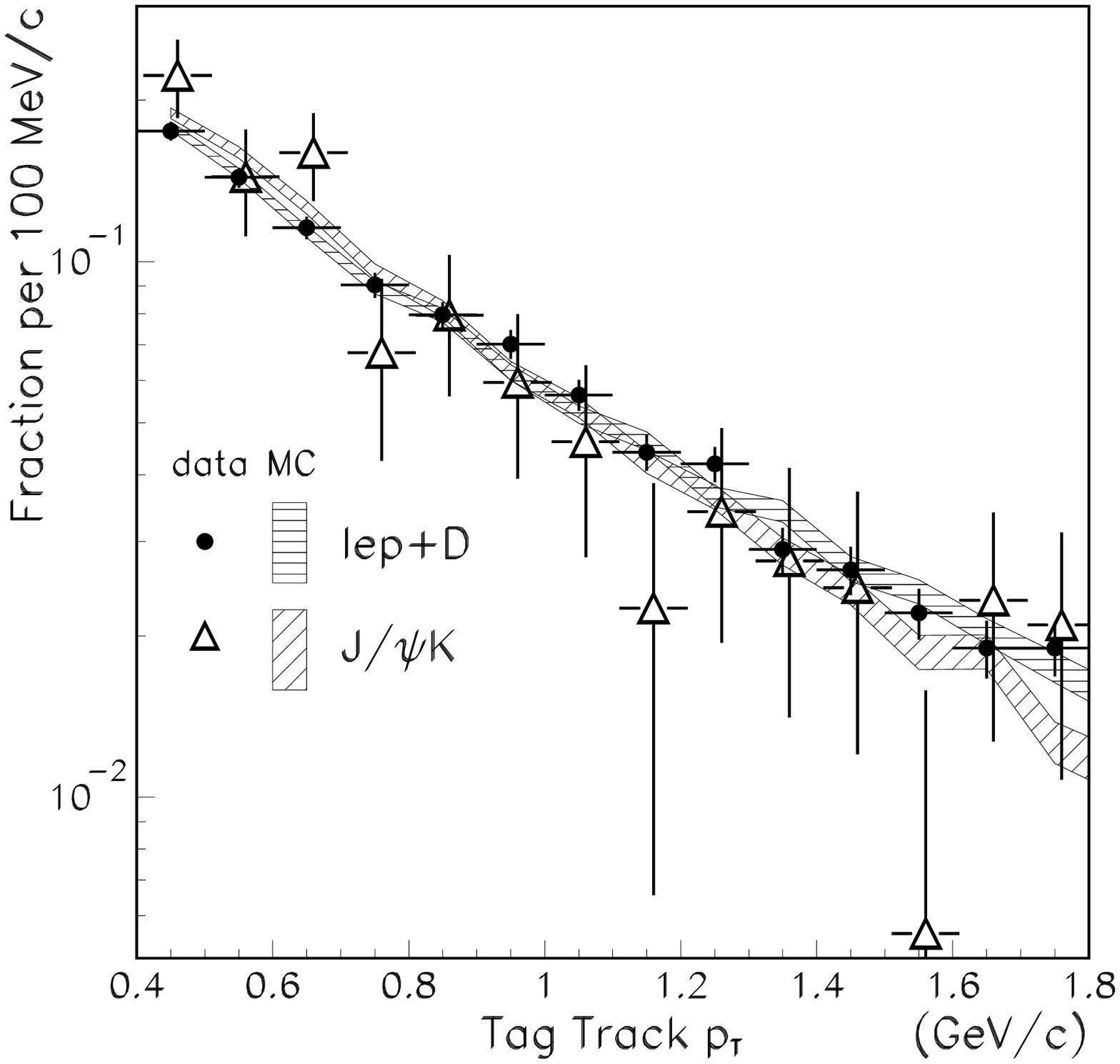}}
\caption{The $p_T$ distributions of the tag
tracks in both $\ell D^{(*)}$
and $J/\psi K$ data (points) and simulation (shaded bands,
width indicating the statistical error).}
\label{fig:ptSST}
\end{figure}

We impose the full SST criteria and show
the $p_{T}$ distributions of the SST tags  in
Fig.~\ref{fig:ptSST}.  Again, there is good agreement between the
data and simulation for the two types of data.  
The tag $p_{T}$ distributions of the two types of modes
are also very similar to each other.
The $p_{T}$ distribution for tagging {\it candidates} 
(not shown) have a somewhat harder track $p_{T}$ spectrum 
in the higher-energy $\ell D^{(*)}$ sample than in the $J/\psi K$ mode, 
but in both cases the data and simulation agree well there, too.

\begin{figure}
\centerline{
\epsfysize 8.0cm
\epsfbox{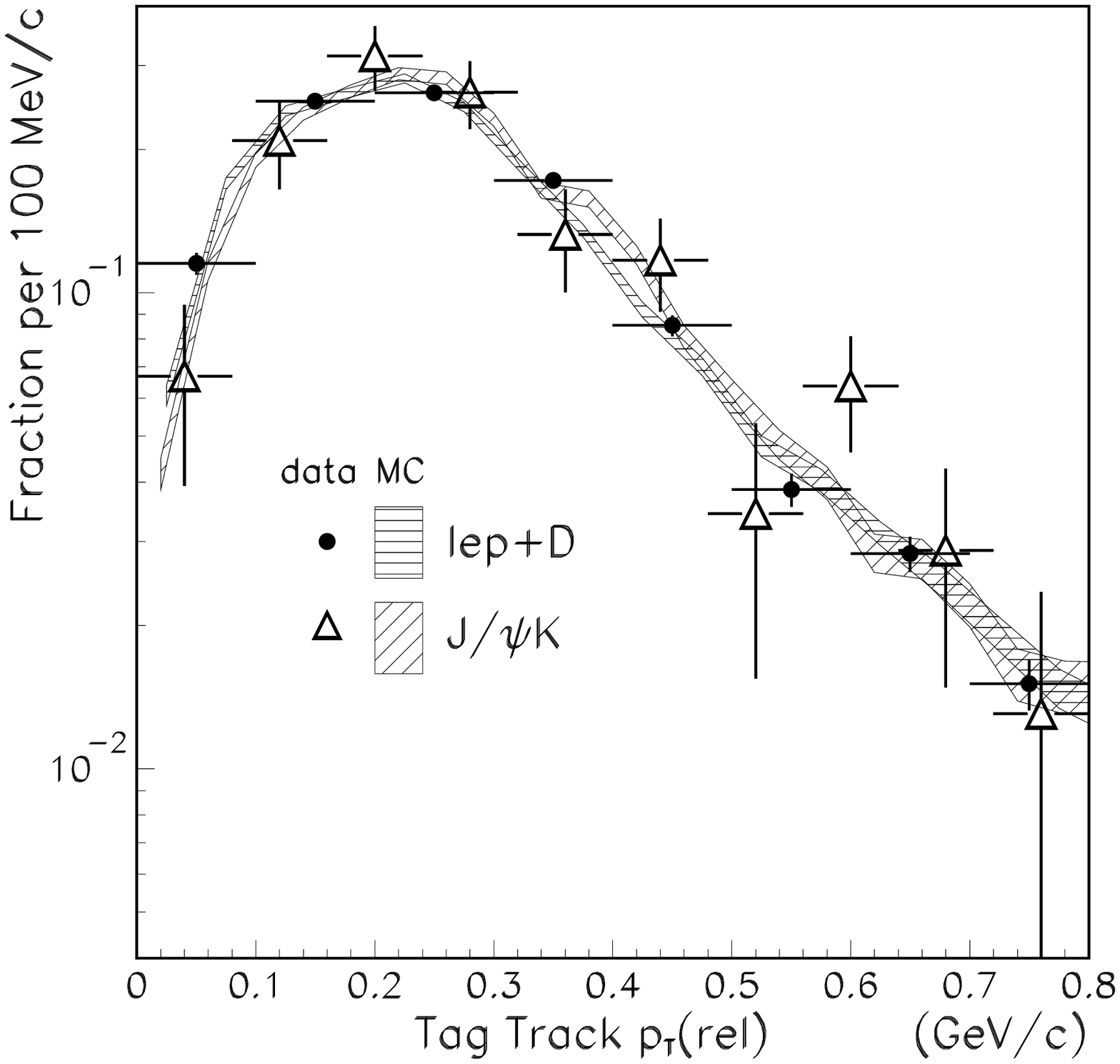}}
\caption{The $p_T^{rel}$ distributions of tag
tracks in both $\ell D^{(*)}$
and $J/\psi K$ data (points) and simulation (shaded bands,
width indicating the statistical error).}
\label{fig:ptrelSST}
\end{figure}

Finally, we compare the $p_{T}^{rel}$ distributions for the tag tracks
in Fig.~\ref{fig:ptrelSST}.  
The simulation again agrees well with the data, and in this case
the  $\ell D^{(*)}$ and $J/\psi K$ modes are also very similar.

Comparisons using these three variables only provide a limited test,
but they are closely related to our SST algorithm and indicate
that the simulation reproduces basic characteristics of the data.

\subsection{Influence of the tagging \boldmath $p_T$ threshold}
\label{SubSec:DvsPtsst}

Our SST algorithm demands that tag candidates
have a minimum $p_T$ of 400 MeV/c as a compromise
between the low-$p_T$ tracking asymmetry (Sec.~\ref{SubSubSec:TagBias}) 
and the declining tagging efficiency for an increasing threshold.
In this section we consider the influence of this cut,
in particular, the stability of our results when repeating the analyses
for a range of $p_{T}(SST)$ thresholds. This variable is also a useful
vehicle for exploring some of the features of SST.

Applying a different tagging cut to the same sample means 
that some $B$'s will be tagged by a different particle, 
others will keep the same tag, and others will no longer
be tagged at all. If a tag changes, the new tag is largely 
uncorrelated with the old, so that, with respect to the new tags, 
a statistically independent subset of the data is created.
Thus, repeating the analysis with different tagging cuts produces 
measurements which are partially correlated with each other.
The greater the change in the tagging cut, the weaker 
the relative correlation.
We do not attempt to unravel this complex pattern of correlations;
we merely show variations with the $p_{T}$ cut to show the dependence
of the tagging results on changes in this parameter.  We
quote the naive statistical errors from the fits of a $p_{T}(SST)$ scan
with the understanding that the various points and their errors 
are correlated in an unspecified fashion.

\begin{figure}
\centerline{
\epsfysize 8.5cm
\epsfbox{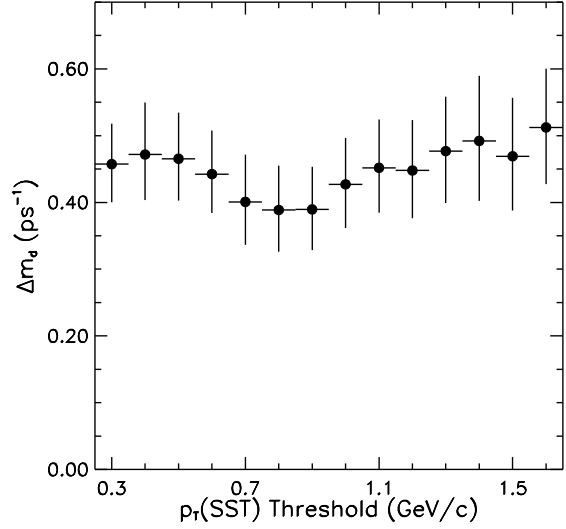}}
\caption{The extracted value of $\Delta m_d$ from the tagged 
$\ell D^{(*)}$ as a function of the tag $p_T$ threshold. 
Error bars are the naive statistical errors returned from the fit, 
and they are correlated with each other, 
as are the points themselves (see text).}
\label{fig:D0vsPtdmd}
\end{figure}

We first consider the stability of our main physics result, $\Delta m_d$.
Figure~\ref{fig:D0vsPtdmd} shows the variation of $\Delta m_d$
as the SST $p_T$ cut is varied from 0.3 to 1.6 GeV/c
in the $\ell D^{(*)}$ sample.
The results are fairly typical of those where the tagging $p_{T}$
threshold is being scanned.
The values are reasonably stable, {\it i.e.}, smoothly varying in
accordance with the subsample correlations mentioned above,
and quite consistent with a constant value.

\begin{figure}
\centerline{
\epsfysize 8.0cm
\epsfbox{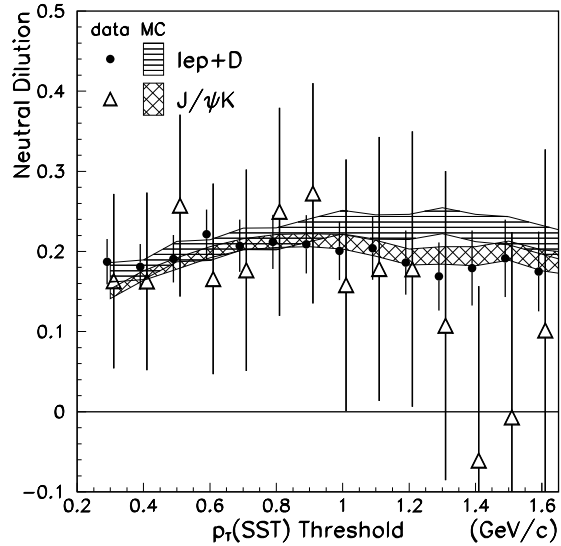}}
\caption{Tagging dilution as a function of tag $p_T$ cut for $B^0$.
Data are plotted with solid circles ($\ell D^{(*)}$) and
triangles ($J/\psi K$), and the corresponding simulations are
shown by the shaded bands (width indicates the statistical error).
The various points and their errors are correlated with each other (see text).}
\label{fig:D0vsPtsst}
\end{figure}

We also examine the effectiveness of the tagging algorithm as the
$p_{T}$ threshold is changed.  Figure~\ref{fig:D0vsPtsst} shows
the neutral dilutions for the $\ell D^{(*)}$ and $J/\psi K^{*0}$ data.
They are both relatively constant and are consistent with each other,
although the $J/\psi K^{*0}$ values offer little discrimination.
The simulation, also shown in Fig.~\ref{fig:D0vsPtsst},
agrees well with the data.

The $B^+$ dilutions are shown in Fig.~\ref{fig:DpmvsPtsst}
and display a striking rise as the tagging cut is increased. 
The $J/\psi K^{+}$ data, though with sizable statistical uncertainties,
show a rise similar to  $\ell D^{(*)}$
but perhaps offset to lower overall dilution.
There is an apparent shape discrepancy between the simulation and
the  $J/\psi K^{+}$ data. 
Appendix~\ref{sec:ken_statvar} describes a $\chi^2$-based comparison
between the simulation and data which indicates that statistical
fluctuations among these correlated measurements
can produce such disagreements (or larger) 
about 22\% of the time.

\begin{figure}
\centerline{
\epsfysize 8.5cm
\epsfbox{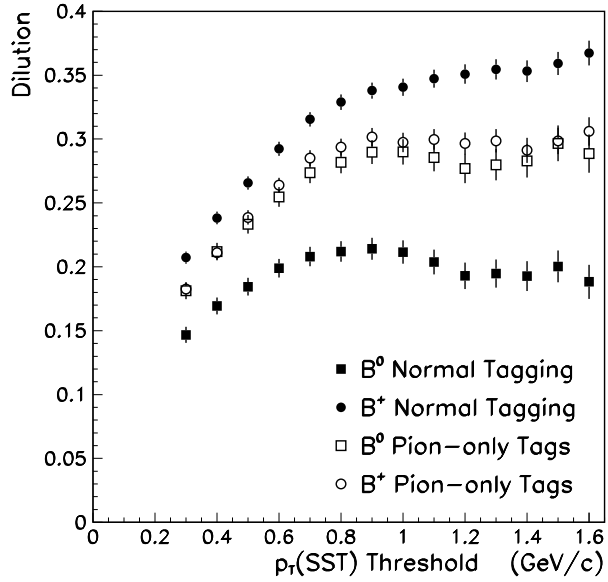}}
\caption{Dilution as a function of the tag $p_T$ cut 
from Monte Carlo simulation for the $J/\psi K$ modes of
$B^0$'s (solid square) and $B^+$'s (solid circle).
When the tagging is restricted to prompt pions only,
the neutral dilution becomes the open squares, and the charged the
open circles.}
\label{fig:MCD_vs_pt}
\end{figure}

The different magnitudes and behaviors of the charged and neutral dilutions,
as seen in Figs.~\ref{fig:D0vsPtsst} and \ref{fig:DpmvsPtsst}, may at
first be surprising in light of the isospin symmetry of the $B$-$\pi$ system.
However, these differences may result from the fact that tags
include not only pions, but kaons and protons as well~\cite{RosnerDun}.  
For instance, a $K^{-}$ would be
associated with a $B^{+}$, while a $B{^0}$ should be accompanied by a
$\overline{K}{^0}$, which cannot be a tag.  
The contrast between the charged and neutral dilutions is amplified
because when the associated kaon is a $K^*$, the final charged kaon is
always a $K^-$, {\it i.e.}, $B{^0}\overline{K}{^{*0}}$ followed by
$\overline{K}{^{*0}} \rightarrow K^- \pi^+$ versus
$B{^+}{K}{^{*-}}$ followed by
${K}{^{*-}} \rightarrow K^- \pi^0$.
A similar argument can be made that the tagging contribution 
from (anti)protons also degrades ${\cal D}_0$ and enhances ${\cal D}_+$.
We test this hypothesis in
the simulation by restricting SST to tag only on prompt pions.  
The predicted results are shown in Fig.~\ref{fig:MCD_vs_pt}, 
where it is seen that this restriction makes 
the charged and neutral dilutions nearly identical to one another. 
The ability of the simulation to reproduce the striking behavior 
of Figs.~\ref{fig:D0vsPtsst} and~\ref{fig:DpmvsPtsst} offers
indirect quantitative evidence that tagging on non-pions is
the effect in play here.\footnote{This observation indicates 
that the SST dilution for neutral $B$'s 
can be significantly improved with particle identification.}
A similar computation, with the same implications, 
was reported in Ref.~\cite{ALEPHSST} for a variant
of Same Side Tagging in $Z^0 \rightarrow b\bar{b}$;
although the measured charged and neutral dilutions showed 
a difference consistent with this effect, the uncertainties were so large 
that no definite conclusion was made there.

\begin{figure}
\centerline{
\epsfysize 8.8cm 
\epsfbox{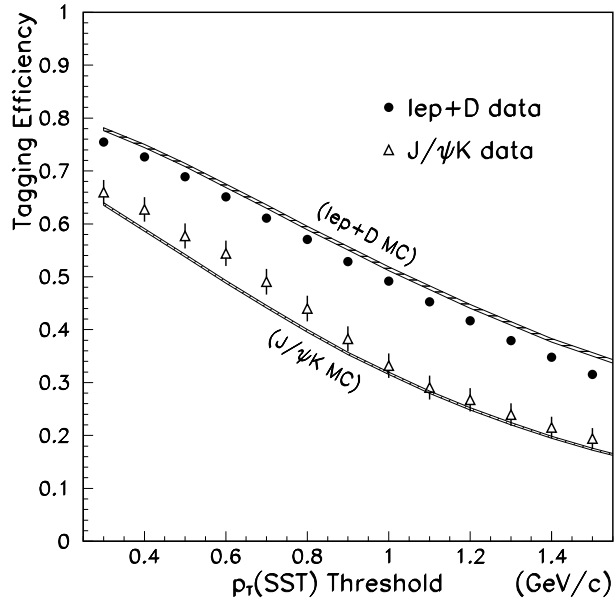}}
\caption{Tagging efficiency as a function of the tag $p_T$ threshold
for $\ell D^{(*)}$ (solid circles) and $J/\psi K$ (open triangles)
data. The results from the simulation are shown by the shaded
bands (width indicates the statistical uncertainty).
}
\label{fig:Eff_vs_pt}
\end{figure}

Finally, we show the tagging efficiencies in Fig.~\ref{fig:Eff_vs_pt}
as a function of $p_T(SST)$. We again average all the modes
of a given type. The data do not show a clear difference 
in the efficiencies among the separate modes, but 
the simulation indicates that the efficiency for charged $B$ mesons 
is shifted higher than the neutrals by $\sim 2$\% (absolute)
consistently over this $p_T(SST)$ range.
The calculated shapes agree fairly well
with the data, but in the case of $\ell D^{(*)}$ at least, 
there is a small systematic shift in the efficiency:
in the simulation there are too few cases where there is no
SST candidate associated with the $\ell D^{(*)}$.
This effect can also be seen in Fig.~\ref{fig:nSST} where
the simulation is slightly below the data in the zero bin.
However, the shape of the efficiency curve tracks the data well
in Fig.~\ref{fig:Eff_vs_pt}, reflecting the good description of
the track $p_T$ distribution. Also note that the efficiency 
falls off much more slowly than the tag $p_T$ distribution 
(Fig.~\ref{fig:ptSST}), because as the threshold is raised
and tags fail the $p_T$ cut, another track will often be available 
as a tag.
The $J/\psi K$ data shows some tendency to have a higher efficiency
than the calculations, but the difference is statistically marginal
(recall that the points are correlated and fluctuations
manifest themselves over a range of bins).

As noted in Sec.~\ref{Sec:mix}, the error on an asymmetry
measurement scales with the ``effective tagging efficiency''
\mbox{$\ed$} [Eq.~(\ref{eq:epsD2})]. The decline in efficiency
with increasing $p_T(SST)$, along with
the relatively constant neutral dilution, imply that the optimum
SST threshold should be relatively low for $B^0$'s and somewhat
higher for $B^+$'s. Our {\it a priori} choice of 400 MeV/c was a balance 
between the falling efficiency as the $p_T$ threshold is increased
and the inherent tracking asymmetry 
at low $p_T$ (Sec.~\ref{SubSubSec:TagBias}).
These simulations indicate that the maximum \mbox{$\ed_0$}
occurs for a $p_T$ threshold of about 400-500 MeV/c for 
our $\ell D^{(*)}$ samples,
and about 600-700 MeV/c for the $J/\psi K^{*0}$.
The rising charged dilutions compensate for the falling efficiency
such that \mbox{$\ed_+$} shows a broad plateau starting at about
600 MeV/c for the $\ell D^{(*)}$.
For the $J/\psi K^+$,  \mbox{$\ed_+$} reaches a maximum 
around 700 MeV/c, but then declines for large $p_T(SST)$.
Except for \mbox{$\ed_+$} from $\ell D^{(*)}$, the data are not
sufficiently precise to confirm these predictions for the optimum
threshold.
Although our SST threshold was not based on this 
{\it a posteriori} analysis, the $p_T$ cut is in fact close to
the optimum suggested by the Monte Carlo simulation.

\subsection{Dilution comparison between $\ell D^{(*)}$ and $J/\psi K$ data}

A cursory comparison of Eqs.~(\ref{eq:final_lepD_Dx}) 
and~(\ref{eq:final_lepD_D0}) 
to Eqs.~(\ref{eq:PsiKD+}) and (\ref{eq:PsiKD0}) already shows that
the dilutions measured in the $\ell D^{(*)}$ and $J/\psi K$
samples are very similar. We consider here how well they 
should agree.

\begin{figure}
\centerline{
\hspace{-0.0cm}
  \epsfysize 10.4cm 
  \epsffile{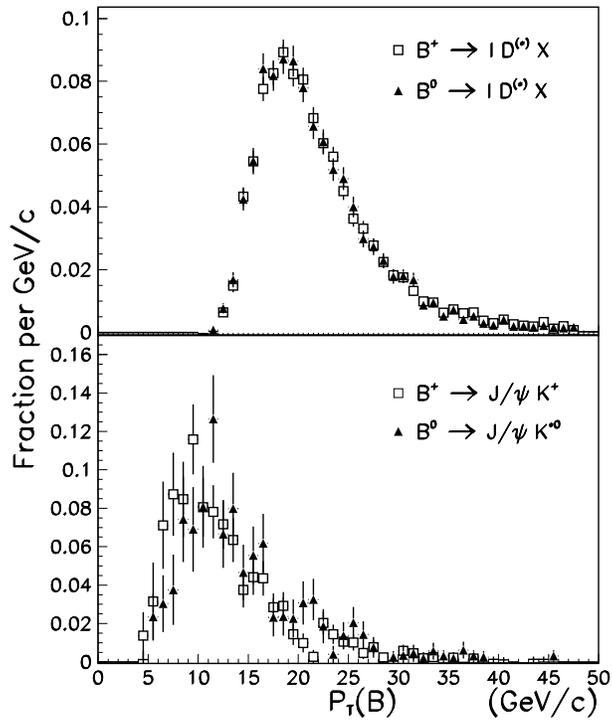}
}
\caption[]
{\small{
The sideband-subtracted $p_{T}(B)$ distributions 
for the (corrected) $\ell D^{(*)}$ (top)
and $J/\psi K$ (bottom) samples.  The charged and neutral $\ell D^{(*)}$
spectra are very similar, but the two $J/\psi K$ spectra are slightly
different due to the different cuts on the kaon momenta.
}}
\label{fig:ptb_psik}
\end{figure}

The main difference between the $\ell D^{(*)}$ and $J/\psi K$
data samples lies in their different $p_{T}(B)$ ranges.  The $\ell D^{(*)}$
sample requires a single-lepton trigger which has a higher lepton $p_{T}$
threshold than the two-lepton trigger used in the $J/\psi K$ samples.
The average $p_{T}$ of the $B$ mesons in the $\ell D^{(*)}$ sample 
(based on the corrections of Sec.~\ref{sec:bgcorr_resolutions}) is
about $21\;{\rm GeV}/c$, whereas it is about $12\;{\rm GeV}/c$ in the
$J/\psi K$ samples.  The spectra are shown in Fig.~\ref{fig:ptb_psik}.

\begin{figure}
\centerline{
\epsfysize 8.0cm 
\epsfbox{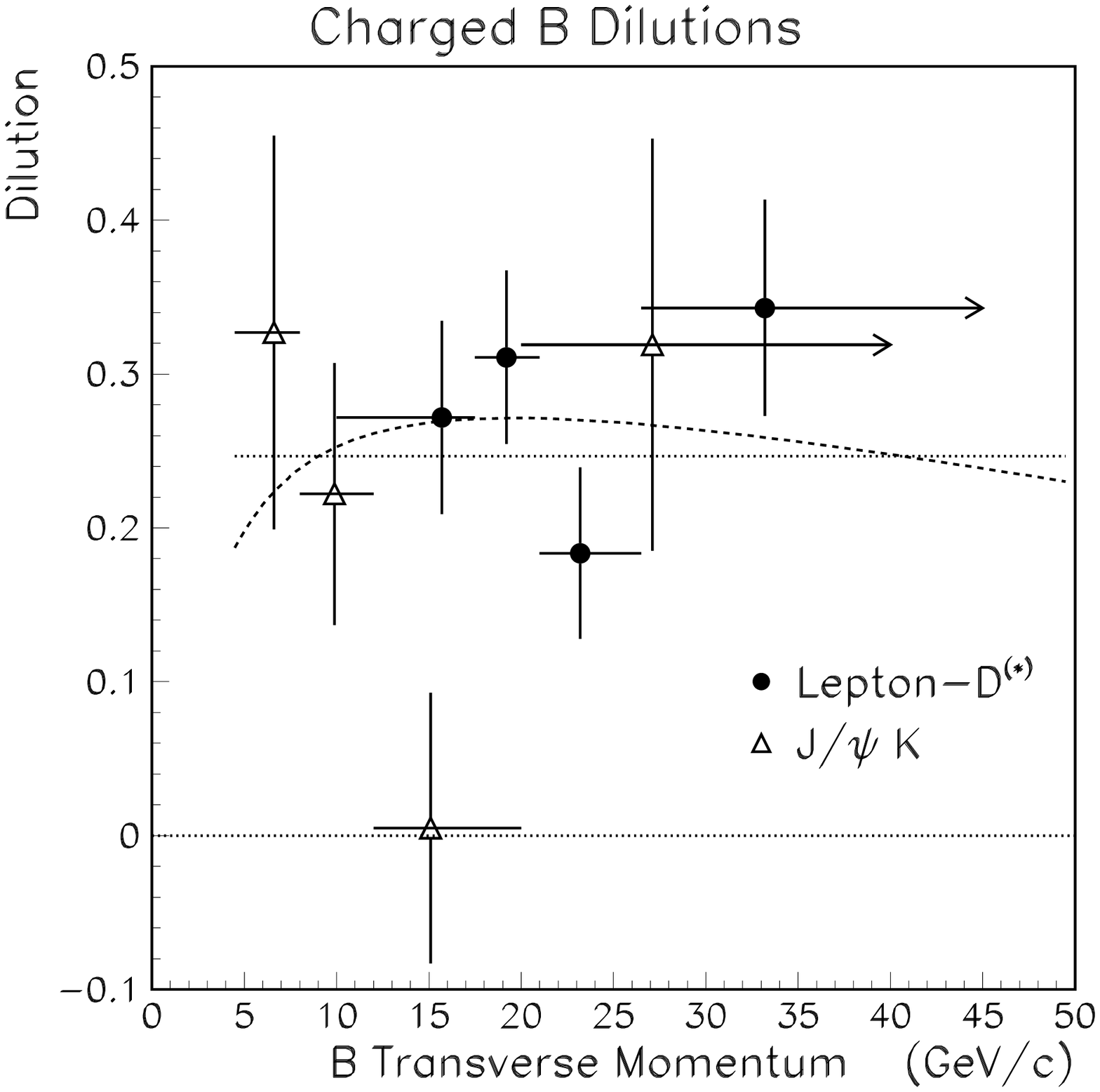}
\epsfysize 8.0cm 
\epsfbox{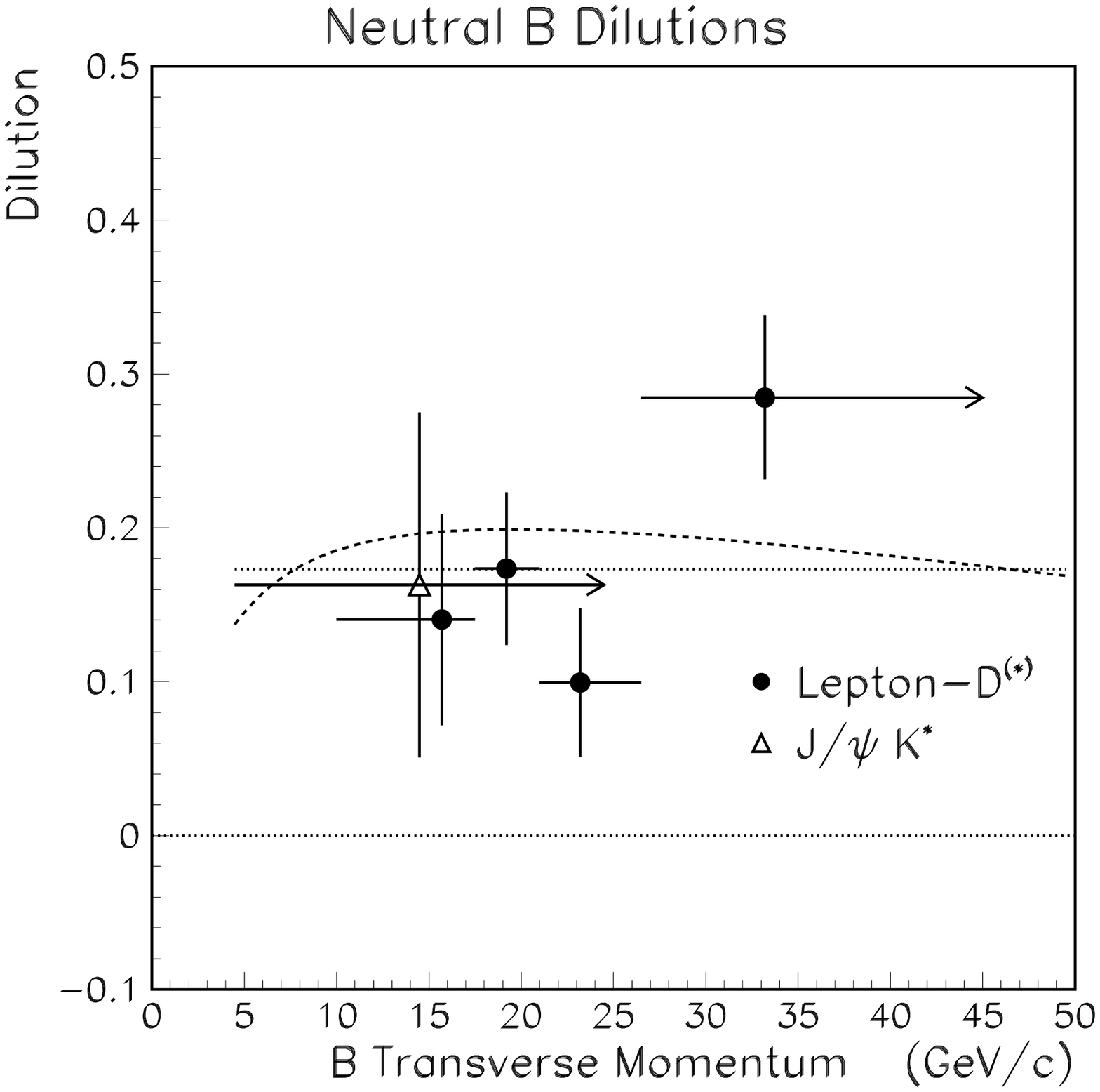}}
\caption{Tagging dilution as a function of $B$ meson $p_T$ for
charged (left) and neutral (right) mesons. The dilution measurements
are plotted at the $p_T(B)$-weighted centroid of each bin, and
the horizontal error bars span the width of the bin (arrows
indicate that a bin is unbounded). The closed
circles are the $\ell D^{(*)}$ data, and the open triangles are $J/\psi K$'s.
The dashed curves are Monte Carlo calculations of the $p_{T}$ dependence,
and the dotted lines mark the average dilution for the data points.}
\label{fig:DvsPt}
\end{figure}

We look for a $p_{T}(B)$ dependence by dividing the data samples
into $p_{T}(B)$ bins and repeating the analysis separately for each
bin.\footnote{In the case of the $\ell D^{(*)}$ analysis, we now fix
$\Delta m_{d}$ to the world average as in the $J/\psi K$ analysis.}
The results are shown in Fig.~\ref{fig:DvsPt}.  
No apparent $p_{T}(B)$ dependence is observed, 
though the statistical sensitivity of the data is  very limited.
The dilution from the $J/\psi K^{+}$ point around $15\;{\rm GeV}/c$ is
anomalously low, but the other measurements are consistent
with the $\ell D^{(*)}$ values, suggesting that the low point is simply
an unusually large fluctuation.  

Since the data samples are too small to be sensitive to a $p_{T}(B)$
dependence in the dilution, we turn to Monte Carlo simulation.  We again
use our tuned PYTHIA generator; however, in order to generate the very
large samples needed for an accurate study we dispense with the full
detector simulation and instead make simple fiducial cuts and apply a
$p_{T}$-dependent track efficiency parameterization.  We remove all
non-prompt particles from consideration as tags in lieu of the SST impact
parameter significance cut.  The dilutions calculated from this
simulation are also shown in Fig.~\ref{fig:DvsPt},
and they exhibit a common shape, rising with $p_{T}(B)$ up to about
$15\;{\rm GeV}/c$, above which they fall slowly.  The ratio of charged
to neutral dilutions, around 1.35, shows no significant
dependence on $p_{T}(B)$.\footnote{The dilutions are also observed to be
insensitive to the $B$ pseudorapidity, where acceptance and trigger effects
could be important.}

We can use the calculated $p_{T}(B)$-dependent dilution along with the
$p_{T}(B)$ spectra from data to compare the data and simulation without
having to subdivide the data into even smaller subsamples, as was
done for Fig.~\ref{fig:DvsPt}.
The $p_{T}(B)$-weighted average dilutions appropriate to each data sample 
are shown in Table~\ref{tab:d_mc}.  The simulation reproduces the
data measurements quite well.  We also calculate the ratios between
the data and simulation values; we find that the ratios are all consistent
with one another, and also with 1.0 (the measured $J/\psi K^{+}$ dilution
being $1.3\sigma$ low; see Fig.~\ref{fig:DvsPt}).

\subsection{Extrapolating dilutions}
\label{sec:extrap}

The fact that the simulation agrees well with the data, as exemplified 
by the comparisons between the calculated and measured dilutions
in Table~\ref{tab:d_mc},
suggests a method by which the SST dilution in any $B$ sample of a
similar $p_{T}$ range can be determined.  Such knowledge is essential
for measurements involving $B$ tagging where, unlike in the
$B^{0}$	-$\overline{B}{^0}$ mixing case, the dilution is not given by
the analysis itself.

The average dilution for a given sample of $B$'s is
calculated by weighting the Monte Carlo dilution shape by
the sideband-subtracted $p_{T}(B)$ spectrum for that sample, as above.
The dilution extrapolation is obtained by multiplying this average 
by the factor ${\cal D}_{data}/{\cal D}_{MC}$.
Since the simulation describes both the neutral and charged dilutions
well, one can incorporate both of these dilutions
in determining the ratio ${\cal D}_{data}/{\cal D}_{MC}$.
This factor is $0.906\pm 0.101$ when averaging all the ratios 
in Table~\ref{tab:d_mc} (including
the correlations between the $\ell D^{(*)}$ measurements).

The uncertainties on such a dilution extrapolation come from both the
measurement uncertainties of the $\ell D^{(*)}$ and $J/\psi K$
dilutions (shown above) and from the modeling uncertainties of the
simulation.  To estimate the latter, we vary the parameters that
control the simulation over a fairly wide range.  The variations used
for the tuned PYTHIA are described in App.~\ref{sec:Syst-PYTHIA}.
We note, however, that the
Monte Carlo-derived dilutions always enter the above calculation in ratios,
{\it i.e.}, the relative variation of
$\ell D^{(*)}$ to $J/\psi K$, and charged to neutral, dilutions
as a function of $p_T(B)$.
We have studied the variations of these ratios as we change
the inputs to PYTHIA.
The largest change in the ratio of $\ell D^{(*)}$ (high-$p_{T}$) to
$J/\psi K$ (low-$p_{T}$) dilutions is $8\%$ from changing the
fragmentation $p_{T}$ width to $360\;{\rm MeV}/c$.  The ratio of
charged to neutral dilutions shifts by at most $4\%$, also when the
fragmentation $p_{T}$ width is set to the low value.

We have seen (Fig.~\ref{fig:MCD_vs_pt}) that the dilutions are also 
affected differently by tagging on pions and non-pions. This difference 
introduces another source of uncertainty, especially when relying 
on ${\cal D}_+$ to constrain ${\cal D}_0$ as suggested above. 
One could forgo this additional constraint and accept a somewhat 
larger error on ${\cal D}_0$, but the extrapolation using ${\cal D}_+$ 
and ${\cal D}_0$ is not unduly sensitive to the fraction of non-pion tags. 
As discussed in App.~\ref{sec:Syst-PYTHIA}, we estimate the uncertainty
due to this effect by allowing the $K^+$ to $\pi^+$ ratio of tags to vary 
by  $\pm 30$\%, and the $p$ to $\pi^+$ ratio by $\pm 50$\%, and find that 
the ratio ${\cal D}_0/{\cal D}_+$ changes by $\pm0.084$ due to the kaons 
and $\pm0.045$ for protons. A simple extrapolation from a ${\cal D}_+$ 
measurement to ${\cal D}_0$ would translate into a neutral dilution 
uncertainty of $\sim 0.02$ (given a ${\cal D}_+$ of $0.25$). However, 
since we propose extrapolating from both charged and neutral dilution 
measurements to a neutral dilution, the species-sensitive scaling 
factor (${\cal D}_0/{\cal D}_+$) only applies to the charged measurement. 
The significance of the uncertainty on the fraction of non-pion tags 
is thereby reduced in this application.

As an example, if we applied the above prescription to calculate the dilution
appropriate for our $J/\psi K^{*0}$ sample using all four measurements
of Table~\ref{tab:d_mc}, then the dilution obtained would 
be ${\cal D}_0 = 0.171 \pm 0.019 \pm 0.013~\cite{Ken}$.  The first 
uncertainty is from the statistical uncertainty on the scale factor above.  
The other is the systematic uncertainty from the extrapolation.
The latter uncertainty is the quadrature sum of the uncertainties obtained 
from varying the PYTHIA parameters ($\pm0.010$) and from varying the 
fraction of non-pion tags ($\pm0.008$). The systematic uncertainty
from the non-pion
contribution is one of the larger uncertainties, and could be largely 
eliminated by not using the ${\cal D}_+$ measurements. 
This, however, would result in an overall larger uncertainty due 
to the increased statistical error.
On the other hand, the systematic uncertainty is relatively small
even using ${\cal D}_+$,
and the dilution determination is not very sensitive to the simulation.

This general method can be used to estimate the dilutions in a variety of
$B$ meson samples of interest for precision measurements of CKM
parameters that will be performed in the future.  In the upcoming Run~II,
CDF expects to have tens of thousands of exclusively reconstructed $B^{0}$
and $B^{+}$ decays through various channels, including
$B^{0}\rightarrow J/\psi K_{S}^{0}$, $\pi\pi$,
$K\pi$, and $B^{0,+}\rightarrow D^{(*)}\pi$.  
The above recipe can be applied to all of them in spite of their likely
differences in selection criteria.  Moreover, further dilution
measurements can easily be incorporated into ${\cal D}_{data}/{\cal D}_{MC}$,
thereby facilitating increasingly precise measurements.
The individual large exclusive samples will yield good dilution determinations
relatively quickly and easily, and refined determinations combining
different modes can follow.

\subsection{Comparison Summary}
\label{SubSec:CompSum}

We have made several checks on the tagging characteristics
of SST in our data.  The $\ell D^{(*)}$ measurement of $\Delta m_d$ 
is largely insensitive to the $p_T$ threshold of the tagging algorithm.
However, the charged dilution shows a dependence on the
$p_{T}$ threshold of the tagging algorithm, in contrast with the largely
threshold-independent behavior of the neutral dilution.
This pattern is apparent in both $\ell D^{(*)}$ and $J/\psi K$ analyses,
and is also reproduced by the Monte Carlo simulation.
The simulation indicates that this difference
is due to tagging on charged kaons and protons.

Comparisons with other variables
show considerable consistency in the characteristics
of the tagging across decay modes and with Monte Carlo simulation.
The general agreement with our tuned PYTHIA is good,
although the current data samples are insufficient to confirm some 
of the more subtle behavior suggested by the simulation.
In particular, the tagging does not appear to be particularly 
sensitive to the different kinematics of the exclusive and 
semi-exclusive samples in the variables examined, 
such as the $p_T$ of the $B$, suggesting that SST has more general
applicability than to the decay modes examined here.

\section{Summary}
\label{Sec:sumry}

We have developed a Same Side Tagging (SST) method
based on the flavor-charge correlations between $B$ mesons
and nearby charged particles (``$\pi^\pm$'') at production. 
We have used SST to tag the initial $B$ flavor 
in two classes of $B$ reconstruction, 
while the (nominal) decay flavor 
was obtained from the $B$ reconstruction.
Comparison of the initial and decay flavors allows one to quantitatively
study the strength ({\it i.e.}, dilution) 
of the $B$-$\pi^\pm$ correlation, and 
to observe the $B^0$-$\overline{B}{}^0$ flavor oscillations.

The first sample consisted
of $B \rightarrow \ell D^{(*)}X$ decays, which,
because it is only a partial reconstruction,
involved additional complications.
We have discussed extensively these complications,
including the separation of $B^+$ and $B^0$ decays 
as well as the corrections for tagging on $B$ decay products.
We observed $B$-$\pi^\pm$ correlations, used them to
reveal the time-dependent flavor oscillation of $B^0_d$'s,
and measured its  frequency to be 
\begin{eqnarray}
\Delta m_d = 0.471 ^{+0.078}_{-0.068} \pm 0.034 {\; \rm ps}^{-1}.
\end{eqnarray}
This result is comparable to other single tagging
measurements, and agrees well with a recent world average of
$ 0.484 \pm 0.026 \; {\rm ps}^{-1}$
from 6 measurements~\cite{PDG97}.

The dilution of the flavor-charge correlation 
for this $\ell D^{(*)}$ sample was found to be
$0.27 \pm 0.03 \pm 0.02$ for the charged, and
$0.18 \pm 0.03 \pm 0.02$ for the neutral mesons.
The effective tagging efficiencies $\epsilon D^2$ are
\begin{eqnarray}
\epsilon D_+^2 & = & 5.2 \pm 1.2  ^{+0.9}_{-0.6} \; \%\\
\epsilon D_0^2 & = & 2.4 \pm 0.7  ^{+0.6}_{-0.4} \; \%, 
\end{eqnarray}
which are the largest values demonstrated to date for tagging methods 
applied to high energy hadron collider data~\cite{CDFJETQ}.
Although other tagging methods may actually have higher dilutions
than SST, the combination of good dilution and very high
tagging efficiency for SST results in the largest $\epsilon D^2$.

This SST method was further tested in the exclusively reconstructed
$B^+ \rightarrow J/\psi K^+$ and  $B^0 \rightarrow J/\psi K^{*0}$ decays.
The flavor-charge correlations were observed, and the
flavor oscillation was again seen with the $B^0$'s; however,
the small sample size did not permit an accurate determination of
$\Delta m_d$. The dilutions measured in these samples agree well with
those obtained from the $\ell D^{(*)}$ data,
although with much less precision.

The behavior of SST was also studied by comparing 
the two classes of data samples to a version of the PYTHIA Monte Carlo
tuned to charged particle distributions
from our $\ell D^0$ data.
Comparing the behavior of several kinematic quantities,
the data and the simulation 
both portray a consistent picture, indicating 
that the simulation captures
the basic features of this SST. Of particular note, 
the differences in the charged and neutral dilutions are 
principally due---according to the simulation---to tagging on kaons. 
Also, despite the different kinematics of
our $\ell D^{(*)}$ and $J/\psi K$ selections, 
the tagging largely behaves the same way for both, as exemplified 
by the weak dependence of the dilutions on the $p_T$ of the $B$.
Furthermore, we have developed a general method to estimate the
dilution in a sample of $B$ mesons starting from the dilution 
measurements in $\ell D^{(*)}$ and $J/\psi K$ samples.  
We also expect this dilution to be fairly close to 
the dilution observed in $\ell D^{(*)}$, provided
that the average $B$ momentum is not vastly different.

This Same Side Tagging method has been demonstrated to be a powerful
means to tag the initial $B$ flavor, even in the complex environment
of a hadron collider.  
In the upcoming Run II of the Tevatron we expect to collect
$\sim 2 fb^{-1}$ of data with the upgraded CDF detector.
This should result in tens of thousands of exclusively reconstructed
$B^0$ and $B^+$ decays in various channels that can be used for
precision measurements of CKM parameters.  
The Same Side Tagging technique will be useful for those measurements
where initial flavor determination is  critical.

\section*{Acknowledgments}

     We thank the Fermilab staff and the technical staffs of the
participating institutions for their vital contributions.  
We also thank T.~Sj\"{o}strand for his suggestions in tuning PYTHIA.
This work was
supported by the U.S. Department of Energy and National Science Foundation;
the Italian Istituto Nazionale di Fisica Nucleare; the Ministry of Education,
Science and Culture of Japan; the Natural Sciences and Engineering Research
Council of Canada; the National Science Council of the Republic of China; 
the A. P. Sloan Foundation; and the Swiss National Science Foundation.

\appendix

\section{Monte Carlo simulations}
\label{app:mc-sample}

In this paper, two types of Monte Carlo simulations are used.
Calculations depending only on the production and decay
of $B$ mesons employ
a Monte Carlo generator that simulates only a single $B$. 
Situations which depend upon the fragmentation particles
resulting from the hadronization of the $b$ quark,
as well as the ``underlying event'' particles, use the full event 
generator PYTHIA.

The entire $\ell D^{(*)}$ analysis uses the single $B$ generator
simulation, with the one exception of the determination 
of the $\xi_{MC}(ct)$ shape (Sec.~\ref{sec:determine-xi}), 
which uses the default PYTHIA simulation. The $J/\psi K$ analysis
also uses the single $B$ generator, and
the comparisons made in Sec.~\ref{Sec:compr} rely 
on a specially tuned variant of PYTHIA.

\subsection{Simulation of a single $B$ meson}
\label{sec:bgenerator-mc}

Monte Carlo simulation of only a single $B$ meson is 
based on the following elements. 
Single $b$ quarks are generated using the inclusive
$b$-quark production calculation
of  Nason, Dawson and Ellis~\cite{ref:NDE}, 
and the MRSD0~\cite{ref:MRSD0} parton distribution functions.
The $b$ quark is then transformed
into a $B$ meson, with no additional hadronization products,
using the Peterson fragmentation model 
($\epsilon = 0.006$)~\cite{ref:Peterson}.
The $B$ meson is decayed using the QQ program 
(Version 9.1)~\cite{ref:QQ}
developed by the CLEO Collaboration.
The sample composition parameters governing the $B$ decay are
listed in Table~\ref{tab:sc-param-mc}.

\subsection{Monte Carlo simulation of the whole event}
\label{sec:PYTHIA-mc}

\subsubsection{``Default'' PYTHIA}
\label{sec:default-PYTHIA-mc}

The PYTHIA Monte Carlo (PYTHIA 5.7/JETSET 7.4)~\cite{ref:PYTHIA} 
is used in instances 
where more than just a single decaying $B$ meson is required.
PYTHIA simulates a complete $p\bar{p}$ interaction: the $b\bar{b}$ pair, 
the hadronization products, and the remaining beam fragments 
(``underlying event'').  
PYTHIA uses an improved string fragmentation model
tuned to experimental data, mostly from high energy $e^+e^-$ collisions.

Our PYTHIA generation uses most of the typical default parameters.
The CTEQ2L~\cite{ref:CTEQ} parton distribution functions are used, and
the $b$ quarks are fragmented using the Peterson fragmentation model 
($\epsilon = 0.006$)~\cite{ref:Peterson}.
$B^{**}$ states are also generated by the fractions
listed in Table~\ref{tab:sc-B**}.
However, we suppress the actual $B$ decay performed 
by PYTHIA and instead invoke the QQ program with the same
parameters in Table~\ref{tab:sc-param-mc}.
In this way we maintain a consistent decay model across
the two different generators.

\subsubsection{``Tuned'' PYTHIA}
\label{sec:Tuned-PYTHIA-mc}

The PYTHIA generator is controlled by a series 
of parameters whose default values have been adjusted to achieve
good agreement with,  primarily,  high energy $e^+e^-$ data.
Discrepancies between the ``default'' PYTHIA (as defined above)
and CDF $p\bar{p}$ data 
are apparent, especially when considering particle production
that does not originate from the $b$ hadronization, 
{\it i.e.}, the ``underlying event.''
We have made a separate study~\cite{Dejan}
of the fidelity of the  ``default'' PYTHIA 
generator (after detector simulation)
by comparing it to the $\ell D^0$ data (Sec.~\ref{SubSec:lepd_recon}).
This comparison studied track multiplicities (with SST quality cuts)
in $\Delta R$ and $\Delta \phi$ intervals 
around the $B$ direction, and
in several $p_T$ bins. The data are found to have a higher
multiplicity of underlying event tracks
(as measured away from the $B$, {\it e.g.}, $\Delta R > 0.4$)
than PYTHIA predicts.

We may obtain a good description of the charged particle
multiplicities and $p_T$ distributions
by adjusting several PYTHIA parameters.
The properties of multiple interactions and
beam remnants are controlled primarily through the 
multiple interaction cross section [PARP(31)], the model for
their generation [MSTP(82)], the ratio of $gg$ and
$q\bar q$ multiple interactions [PARP(85,86)], and the
width of the Gaussian $p_T$ spread  of
particles produced in the breakup of color strings [PARJ(21)].
Once these parameters are adjusted to obtain agreement
with the data away from the $b$-jets, 
we assume the underlying event is well modeled.
We then adjust the Peterson constant PARJ(55)
so that the generated multiplicity of tracks inside 
the $|\Delta R| < 1$ cone around the $b$ matches the observed one.
Table~\ref{tab:tune-PYTHIA} lists the default and tuned values 
of the relevant PYTHIA parameters.  More details may be found
in Ref.~\cite{Dejan}.

\begin{figure}
\centerline{
\epsfysize 8.7cm
\epsfbox{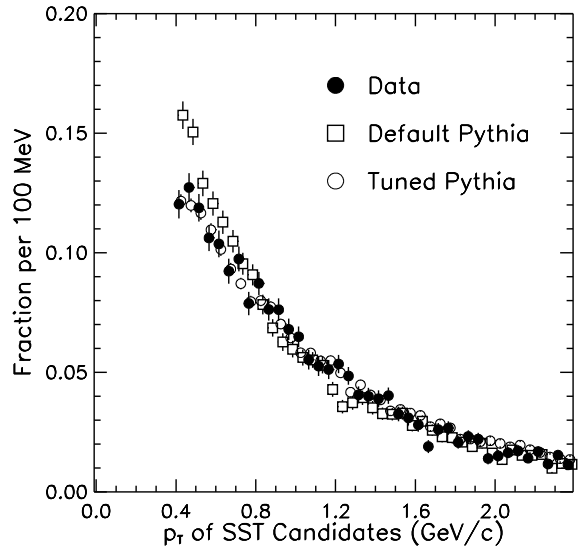}}
\caption{The $p_T$ distribution of tracks satisfying SST selection
cuts except for the $p_T^{rel}$ requirement:
average for $\ell D^{(*)}$ data (solid circles),
default PYTHIA (open squares), and
tuned PYTHIA (open circles).
}
\label{fig:tune_vs_untune}
\end{figure}

As an example of the effects of the tuning, we show in 
Fig.~\ref{fig:tune_vs_untune}
the $p_T$ distribution of SST candidates ({\it i.e.},
tracks that satisfy the SST selection
cuts except for the $p_T^{rel}$ requirement [Sec.~\ref{SubSec:sstag}]) 
in $\ell D^{(*)}$ data and the two simulations. 
The tuning procedure uses this distribution from the $\ell D^0$
subsample---except
that the tracks were 
not restricted to $\Delta R < 0.7$ around the $B$ as they are 
in Fig.~\ref{fig:tune_vs_untune}---so the agreement of the tuned
version with the data is to be expected. The shape 
of the default PYTHIA  $p_T$ spectrum
shows a clear disagreement with the data,
with a large excess of tracks at low $p_T$. 
While there is much better agreement between data 
and the tuned PYTHIA in the shapes 
of the $p_T$ and the frequency distribution 
of SST candidates (see Fig.~\ref{fig:nSST}), 
the tuned Monte Carlo underestimates 
the number of $\ell D^{(*)}$ events that fail to tag by a few percent
(see Fig.~\ref{fig:Eff_vs_pt}).
The charged and neutral dilutions as a function 
of the SST $p_T$ threshold 
(similar to Figs.~\ref{fig:D0vsPtdmd} and \ref{fig:D0vsPtsst}) 
for the two versions of the simulation differ by not more than
$\sim 2\sigma$ of the Monte Carlo statistical uncertainty;
despite the better description of the data by the tuned PYTHIA,
the tagging results are not much different between the two simulations.

We then have a variant of PYTHIA tuned to our $\ell D^0$ data.
This was done for only one $\ell D^{(*)}$ mode, and
comparisons with the others, 
or the $J/\psi K$ modes, are independent of the tuning. 
We find the tuned version generally provides better
agreement than the default version
with all the data samples considered in this paper,
in spite of the fact that the tuning used only
global multiplicity and $p_T$ distributions and did not
consider particle correlations.

\subsubsection{Systematic uncertainties for the dilutions derived from PYTHIA}
\label{sec:Syst-PYTHIA}

For the study presented in Sec.~\ref{sec:extrap}, we rely
on the tuned PYTHIA to calculate the dependence of the dilution on the 
$p_T$ of the $B$~meson being tagged.  In order
to determine a systematic uncertainty on the dilution extrapolation
due to the simulation, we regenerated Monte Carlo samples 
varying selected PYTHIA input parameters.

The four parameters we varied were 
the string fragmentation model parameter $\sigma_{p_T}^{frag}$ 
which describes the distribution of particle momenta transverse
to the string direction,
the underlying event cross section scale factor [PARP(31)], the
Peterson fragmentation parameter $\epsilon_b$, 
and the combined contribution of the $B^{**}$ modes
(see Tables~~\ref{tab:sc-B**} and~\ref{tab:tune-PYTHIA}).
These four were selected as parameters that most directly influence  
the track momentum and multiplicity of potential tags, 
and hence the dilution.

We varied $\sigma_{p_T}^{frag}$ from our tuned value down to 0.36 (the
default value) and up to 0.8, though the statistical uncertainty from
tuning this parameter on the data was only $0.02$~\cite{Dejan}.
Likewise, we varied the cross section scale factor 
from 1.0 (the default value) up to 2.5,  even though 
its tuned  uncertainty  was only 0.04.
The large ranges we used for these two parameters
were chosen as  conservative allowances
for the applicability of this model.\footnote{
The PYTHIA default, $\sigma_{p_T}^{frag} = 0.36$, results from tuning to
LEP data.  This ``string-breaking'' parameter should be, 
to first order, the same for $e^+e^-$ and $p\overline{p}$ colliders.  
The sizeable difference with the LEP value may 
signal a limitation of the tuning procedure, 
or be a hint that the model is inadequate.
}
The range was selected by varying the parameter to (approximately) 
span a symmetric range about the tuned value
that included the default PYTHIA value.

The parameter $\epsilon_B$ and the fraction of $B$ mesons originating
from $B^{**}$ have been measured elsewhere, and their effects on
the model are better understood.  We varied $\epsilon_B$ from
$0.004$ to $0.008$ and the $B^{**}$ fraction up and down by
$25$\%.  These ranges are indicative of the statistical
uncertainties derived from the tuning studies~\cite{Dejan}.

As an additional systematic uncertainty on the behavior of the dilutions,
we varied the fractions of kaon and proton tags in our samples.
Section~\ref{SubSec:DvsPtsst} indicates that the difference
between ${\cal D}_+$ and ${\cal D}_0$ in the simulation
is due to tagging on kaons and protons. 
We varied the kaon fraction by $\pm 30\%$ and
the proton fraction by $\pm 50\%$ to evaluate this uncertainty.

\subsection{Detector simulation}
\label{app:mc-det}

The outputs of the physics  simulations
are passed through the standard CDF fast detector simulation.
This simulation is based on parameterizations of detector
responses determined from data, often test beam measurements. 
The detector simulation output can be
reconstructed using standard CDF software.
These reconstructed Monte Carlo events may then be treated as
real data in the analysis.

\begin{figure}
\postscript{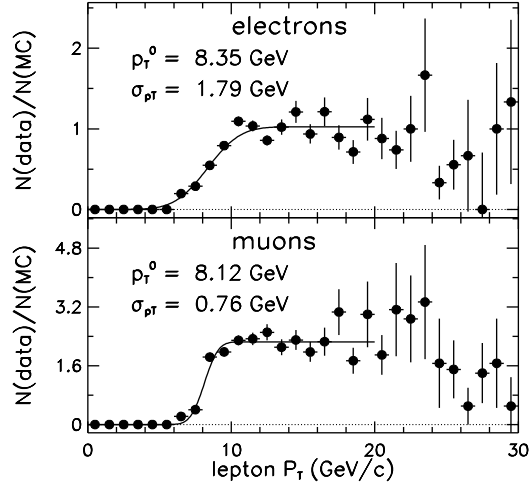}{0.45}
	\begin{center}
	\parbox{5in}{\caption{\small 
		The ratios of the $p_T(\ell)$ distributions 
		of the data to the Monte Carlo simulation, for
		electrons (top) and muons (bottom), for the \kpp\ signature.
		The distributions are fit with the error function,
		where  $p_T^0$ is the midpoint $p_T(\ell)$ and
		$\sigma_{p_T}$ the ``width'' of the turn-on.
		The overall normalizations are immaterial
		since the $\ell D^{(*)}$ analysis requires only relative
		efficiencies.
		\label{fig:turn-on}
	}}
	\end{center}
\end{figure}

The inclusive lepton trigger introduces a strong kinematic bias 
in the $\ell D^{(*)}$ analysis. This bias must be well modeled 
in the simulation to obtain the proper relative reconstruction efficiencies
and $ct$ corrections, otherwise an incorrect sample composition 
will result.
We take an empirical approach rather than simulate 
the trigger directly.
The trigger is modeled by a simple error function parameterization of
the ratio of 
the observed lepton $p_T$ distribution in the data 
to that generated by the simulation~\cite{Petar}.
Examples of such ratios and the error function
fits are shown in Fig.~\ref{fig:turn-on} for one signature.  Only the region
$0< p_T(\ell) < 20 \mbox{ GeV}/c$ is fit, since this is where the effect
of the trigger turn-on is the most pronounced.  
Fits are performed on all five decay signatures, and the 
sample-weighted average of the five sets of fit parameters 
is used to describe the electron and muon trigger efficiencies.
These parameterizations are then applied to the Monte Carlo events
to obtain simulated data sets with the correct trigger turn-on.

\begin{figure}
 \postscript{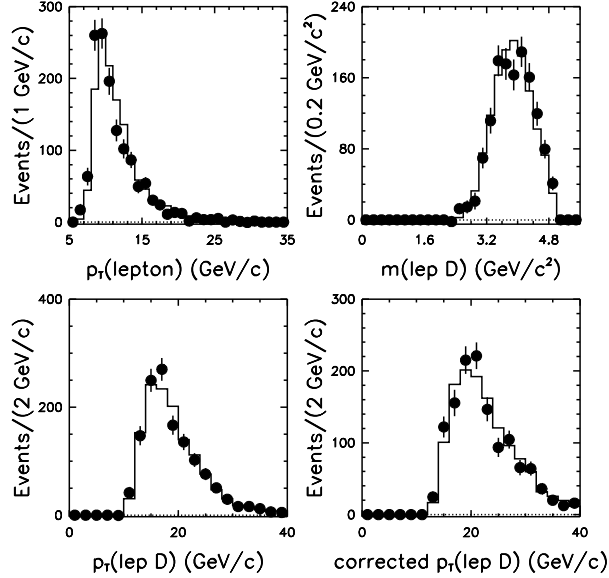}
  {0.5}
	\begin{center}
	\parbox{5in}{\caption{\small 
		A comparison between the data and the single-$B$
		Monte Carlo simulation for the decay signature \kpp.
		The distributions compared 
		are: $p_T$ of the lepton, $e$ and $\mu$ combined (top left),
		mass of the $\ell D^{-}$ system (top right),
		the $p_T$ of the $\ell D^{-}$ system (bottom
		left), and  $p_T(\ell D^{-})$ after correcting 
		for the missing neutrino (bottom
		right).  Only the agreement in the corrected
		$p_T(\ell D^{-})$ distribution relates directly 
		to the analysis.
		\label{fig:gb_mccmp-kps}
	}}
	\end{center}
\end{figure}

A comparison of some kinematic distributions from the data and 
the simulation is given in Fig.~\ref{fig:gb_mccmp-kps}
for a sample decay signature. 
As can be seen, this procedure
provides a fairly accurate representation of the data.

For the $J/\psi K$ we found that the $p_T(B)$ distribution 
that results after detector simulation and selection cuts 
compares fairly well with the data 
without additional trigger simulation.
The trigger turn-on
at very low $p_T(\mu)$ is largely governed by the energy
loss in the material before the muon chambers. This effect is 
already included in the detector simulation, and thus
no specific trigger simulation is done for
the $J/\psi K$ Monte Carlo samples.

\section{Charge Asymmetry Tag Corrections for the $J/\psi K$ samples}
\label{app:tagasym}

A charge bias in the SST algorithm could fake an asymmetry.  
The maximum likelihood
approach described in Sec.~\ref{SubSec:fitter} provides a natural
tool for the parameterization of a charge bias and 
its effect on data.  

A tagging algorithm is characterized by the probability, $P(s|p)$,
that a given
production flavor $p$ yields a tag of charge $s$.  
The production flavor $p$ follows the same
convention as the reconstructed flavor $r$:
$p=+1$ for $B^{+}$ and
$B^{0}$, and $-1$ for $B^{-}$ and $\overline{B}{^0}$.  The tag, $s$,
takes the value of $+1$ for tagging on a positive track,  $-1$ for
a negative track, and $0$ if there is no tag.  
This probability can be written in a form
similar to the expression of  $P(s|r)$ used in our 
likelihood function ({\it e.g.}, Eqn.~(\ref{eqn:L_B})), namely
\begin{equation}
  P_{\phi}(s|p) = \left(\frac{1+p\kappa_{\phi}(s)\Dil_{\phi}}{2}\right)
               {\cal E}_{\phi}(s),
\label{app:psp}
\end{equation}
which is characteristic  of an  asymmetry in $p\kappa_{\phi}(s)$ with
amplitude $D_{\phi}$
($\phi=$ $S$, $P$, or $L$ for the type of event).

The six $P_\phi(s|p)$ describing a tagging method (for a given $\phi$)
are reduced to four by the two constraints
\begin{equation}
  P_\phi(+|p) + P_\phi(-|p) + P_\phi(0|p) \equiv 1
\end{equation}
for either  $p$.  Four independent variables may be chosen to describe
the tagging in terms of these probabilities as 
\begin{eqnarray}
  {\Dil}_{\phi} & = &
    \frac{P_\phi(+|+)+P_\phi(-|-)-P_\phi(-|+)-P_\phi(+|-)}{P_\phi(+|+)+P_\phi(-|-)+P_\phi(-|+)+P_\phi(+|-)} \label{eqn:D_def_by_P} \label{eqn:P_Dil}\\
  \epsilon_{\phi} & = &
    \frac{P_\phi(+|+)+P_\phi(+|-)+P_\phi(-|+)+P_\phi(-|-)}{2}  \label{eqn:P_Eff}\\
  \alpha_{\phi} & = &
    \frac{P_\phi(+|+)+P_\phi(+|-)-P_\phi(-|+)-P_\phi(-|-)}{P_\phi(+|+)+P_\phi(+|-)+P_\phi(-|+)+P_\phi(-|-)}   \label{eqn:P_alpha} \\
  \delta_{\phi} & = &
    \frac{P_\phi(+|+)+P_\phi(-|+)-P_\phi(+|-)-P_\phi(-|-)}{P_\phi(+|+)+P_\phi(+|-)+P_\phi(-|+)+P_\phi(-|-)}.  \label{eqn:P_delta}
\end{eqnarray}
The first quantity ${\Dil}_{\phi}$ is the usual 
dilution,\footnote{With this convention, the
dilution is positive for tagging $B^0$'s and
negative for $B^+$'s where the sign correlation is reversed.
However, we explicitly invert the sign in front 
of the ``$r\kappa_{B}(s)\Dil_{B}$'' term in Eq.~(\ref{eqn:L_B})
for $J/\psi K^+$,
so that $\Dil_B$ is positive in Table~\ref{tab:results} for {\it both}
$B^0$ and $B^+$. All the background dilutions 
follow the nominal convention of Eq.~(\ref{eqn:D_def_by_P}).} 
and the second quantity $\epsilon_{\phi}$ is the charge-averaged tagging
efficiency.  The charge bias in the tagging algorithm is given as
$\alpha_{\phi}$ [Eq.~(\ref{eq:alpha})],
and $\delta_{\phi}$ is the flavor asymmetry in the
tagging efficiency [Eq.~(\ref{eq:delta})].
We find it more convenient to express the latter asymmetry 
as $\gamma_{\phi} \equiv \delta_{\phi}/\Dil_{\phi}$.

Solving for  $P_\phi(s|p)$  in terms of
$\Dil_{\phi}$, $\epsilon_{\phi}$, $\alpha_{\phi}$, and
$\gamma_{\phi}$, and then casting the
expressions in the form of Eq.~(\ref{app:psp}), one derives the
following expressions for $\kappa_{\phi}(s)$, the charge 
asymmetry corrected tag, and ${\cal E}_{\phi}(s)$, the corrected
efficiency:
\begin{eqnarray}
  \kappa_{\phi}(s) & = &
    \left\{
      \begin{array}{cl}
        s\left(\frac{\textstyle  1+s\gamma_{\phi}}
                    {\textstyle  1+s\alpha_{\phi}}\right) &
            {\rm for}\; s=\pm 1 \\
        -\frac{\textstyle \gamma_{\phi}\epsilon_{\phi}}
              {\textstyle 1-\epsilon_{\phi}} &
            {\rm for}\; s=0
      \end{array}
    \right.  \label{eqn:Corr_tag}\\
  {\cal E}_{\phi}(s) & = &
    \left\{
      \begin{array}{cl}
        \epsilon_{\phi}(1+s\alpha_{\phi}) & {\rm for}\; s=\pm 1 \\
        2(1-\epsilon_{\phi}) & {\rm for}\; s=0.
      \end{array}
    \right. \label{eqn:Corr_eff}
\end{eqnarray}
Notice that the untagged events may actually have a small finite dilution
since  $\kappa_{\phi}(0)$ need not be zero.
This non-zero dilution arises because the untagged events contain 
a greater number of events which {\em should} have been 
tagged with the sign against which the tagging efficiency is biased.

These equations provide us with a formulation to incorporate tagging asymmetries
in the likelihood function (Sec.~\ref{SubSec:fitter}).

\section{Constraints on the tagging efficiency asymmetry}
\label{app:gamma_alpha}

As discussed in the latter part of Sec.~\ref{SubSubSec:TagBias},
a tagging method may not tag on $b$ and $\bar{b}$ mesons
with equal efficiencies. The efficiency asymmetry $\delta_{\phi}$
is given by Eqs.~(\ref{eq:delta}) or~(\ref{eqn:P_delta}), and 
appears in the likelihood function 
via $\gamma_{\phi} \equiv \delta_{\phi}/\Dil_{\phi}$
in Eq.~(\ref{eqn:Corr_tag}).
We determine $\delta_{P,L}$ for the $J/\psi K$ backgrounds 
by letting them float in the likelihood fit. However, for 
the $B$ signal, we independently constrain $\gamma_{\phi}$
as explained here.

With an ideal detector
the tagging method would be described by some ``true'' dilution
and efficiency, and the $\alpha$ and $\delta$ asymmetries 
would be zero. 
A detector bias could alter this situation by adding
or losing tracks based on their charge. For example,
positive tracks may be added to the event by proton spallation
from the beam pipe. This effect is actually very small, but in any case,
it adds tracks equally around both $b$ and $\bar{b}$ mesons.
This generates a non-zero $\alpha$ (more positive tags than 
negative) in what was an ideal detector, 
but $\delta$ remains zero ($b$ and $\bar{b}$ mesons
have the same positive tag excess).
On the other hand, preferential loss of one charge 
makes $\delta \not=0$.
An efficiency asymmetry is created since $\overline{B}{^0}$
events are more likely to tag on negative tracks than 
${B}{^0}$'s---this correlation, after all, is why SST works.
The CTC has such a reduced efficiency for low $p_T$ negative tracks
(Sec.~\ref{SubSubSec:TagBias}).

We consider the situation where we have a net
loss of negative tracks, as is actually observed in our 
data.
Losing a track has one of three
outcomes.  First, if there was no other SST candidate in the event,
the tag would simply be lost, giving a net positive tagging asymmetry.  
If, on the other hand, the SST tagged on
another negative track, then the loss has no effect, since it
is only the sign of the tag which matters.  However, if the SST tagged
instead on a positive track, then the tagging asymmetry would be enhanced
over that from simply losing negative tags.

If $\Dil_{\phi}'$ and $\epsilon_{\phi}'$ are
the nominal tagging dilution and efficiency in the absence 
of negative track loss ($\alpha' = \delta' =0$), 
then the probabilities $P(s|p)$ [Eq.~(\ref{app:psp})]
{\it with} the negative track loss
can be rewritten in terms of the nominal quantities as
\begin{eqnarray}
P(+|+) & = & \epsilon_{\phi}' \left(\frac{1+\Dil_{\phi}'}{2}\right) +
             \epsilon_{\phi}' \left(\frac{1-\Dil_{\phi}'}{2}\right)\eta f_{1} \label{eqn:Ppp_const} \\
P(-|+) & = & \epsilon_{\phi}' \left(\frac{1-\Dil_{\phi}'}{2}\right)(1-\eta) \label{eqn:Pmp_const} \\
P(+|-) & = & \epsilon_{\phi}' \left(\frac{1-\Dil_{\phi}'}{2}\right) +
             \epsilon_{\phi}' \left(\frac{1+\Dil_{\phi}'}{2}\right)\eta f_{2} \label{eqn:Ppm_const}\\
P(-|-) & = & \epsilon_{\phi}' \left(\frac{1+\Dil_{\phi}'}{2}\right)(1-\eta),  \label{eqn:Pmm_const}
\end{eqnarray}
where $\eta$ is the fraction of negative tags which are lost but
{\it not} counting those which re-tag on another negative track, 
and $f_{1(2)}$ is the fraction of $p = +1$ ($p = -1$)
[or $B^{0}$ ($\overline{B}{^0}$)] 
which, having lost a negative tag, re-tag on a positive track.

\begin{figure}
\centerline{
\epsfysize 7.5cm
\epsfbox{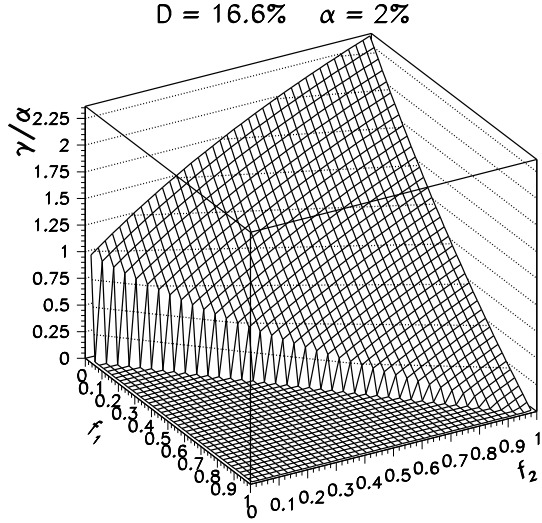}}
\caption{Variation of $\gamma_{\phi}/\alpha_{\phi}$ versus $f_{1}$ and 
$f_{2}$ for $\alpha_{\phi} = 2\%$ and nominal 
dilution $\Dil_{\phi}' = 16.6\%$.  
Note that (0,0) is the leftmost corner and (1,1) the rightmost, 
and that the function is not defined for the region
$\Delta f < 0$.}
\label{ga:ga}
\end{figure}

We can calculate the ratio $\gamma_{\phi}/\alpha_{\phi}$  
by substituting Eqs.~(\ref{eqn:Ppp_const})-(\ref{eqn:Pmm_const})
into Eqs.~(\ref{eqn:P_Dil}), (\ref{eqn:P_alpha}), and (\ref{eqn:P_delta}),
and obtain
\begin{eqnarray}
  \frac{\gamma_{\phi}}{\alpha_{\phi}} & = &
  \frac{\{2-\eta(1-\overline{f}+\Dil_{\phi}'\Delta f)\}
        \{\Dil_{\phi}'[1-\overline{f}]+\Delta f\}}{
        \{1+\overline{f}-\Dil_{\phi}'\Delta f\}
        \{2\Dil_{\phi}'-\eta(\Dil_{\phi}'[1+\overline{f}]-\Delta f)\}},
\end{eqnarray}
where  $\overline{f} = (f_{1} + f_{2})/2$ and 
$\Delta f = (f_{1} - f_{2})/2$.
The behavior of $\gamma_{B}/\alpha_{B}$ is shown in
Fig.~\ref{ga:ga} for values of $\alpha_{B}=2\%$ and $\Dil_{B}'=16.5\%$, 
values which are close to what is observed in data.
We also use
\begin{eqnarray}
  \eta & = & \frac{2\alpha_\phi}
                  {(1+\alpha_\phi) + (1-\alpha_\phi)(\overline{f}-\Dil'_\phi\Delta f)},
\end{eqnarray}
obtained from Eq.~(\ref{eqn:P_alpha}).
The nominal dilution used
is actually the observed dilution in $J/\psi K$ data
after negative track loss 
(Section~\ref{SubSec:inputs}), but the two dilutions are expected to be
similar in light of the small charge asymmetry of
$\alpha_{B}\sim 2\%$ (Sec.~\ref{SubSubSec:TagBias}).  

The largest $\gamma_{\phi}/\alpha_{\phi}$ is
achieved with $\overline{f}=\Delta f = 0.5$, which gives
\begin{eqnarray}
   \frac{\gamma_{\phi}}{\alpha_{\phi}} =
   \frac{(\Dil'_{\phi}+1)(4-(1+\Dil'_{\phi})\eta)}{(3-\Dil'_{\phi})(4D'_{\phi}+(1-3\Dil'_{\phi})\eta)}. 
\end{eqnarray}
For values of $\alpha_{B}$ and $D_{B}'$ close to
what is expected in signal events, this maximal value is about 2.5.
The maximum corresponds to the unrealistic situation where 
all the lost negative tags (which do not re-tag negative)
{\it always} re-tag on a positive track for  $B^0$'s and
{\it never}  re-tag on a positive track for  $\overline{B}{^0}$'s.
For the likelihood fit we choose the nominal value 
of this ratio to be 1.0, and
for the purpose of evaluating systematic uncertainties 
this ratio is varied between 0 and 2.5.
Since $\alpha_{B}$ is on the order of 2\%, $\gamma_{B}$ lies
between 0 and 5\%.

\section{Statistical significance of \boldmath ${\cal D}_+$ {\it vs.} $p_T(SST)$
shape differences}
\label{sec:ken_statvar}

\begin{figure}
\centerline{
\epsfysize 8.0cm
\epsfbox{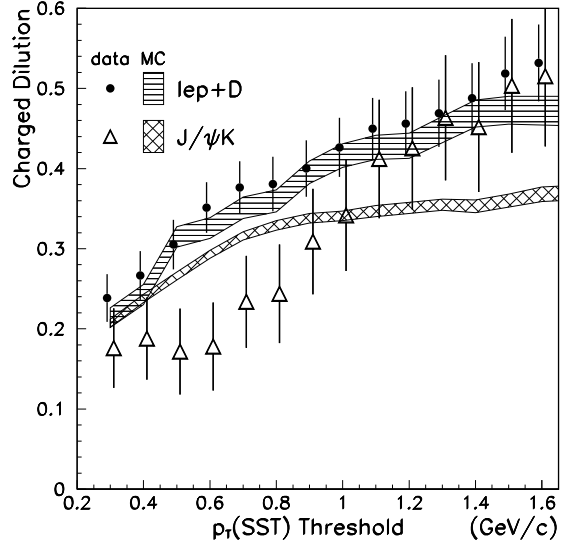}}
\caption{Tagging dilution as a function of tag $p_T$ cut for $B^+$
data in the $\ell D^{(*)}$ (solid circles) and
$J/\psi K$ (triangles) mode. The corresponding simulations are
shown by the shaded bands (width indicates the statistical error).
The various points and their errors are correlated with each other (see text).}
\label{fig:DpmvsPtsst}
\end{figure}

The variation of the  $J/\psi K^+$  charged dilution versus 
the $p_T$ cutoff shown in Fig.~\ref{fig:DpmvsPtsst} 
apparently does not agree very well with either 
the $\ell D^{(*)}$ data or the tuned PYTHIA simulation 
around $p_T(SST) \sim 0.6$~GeV/c. 
As noted in Sec.~\ref{SubSec:DvsPtsst}, 
neighboring points are highly correlated and 
it is difficult to judge the significance of trends 
{across several points} from the drawn error bars.
The correlation is complicated 
because events which lose their tags as the $p_T(SST)$ cutoff
is raised will sometimes re-tag on another, higher $p_T$, track
in the event.  This effect causes the fluctuations to be larger
than might be naively expected.
We consider here a test to gauge the statistical significance
of the shape differences.

The dilution differences between adjacent $p_T(SST)$ cut-offs
(say from 0.4 to 0.5 GeV/c) are much less correlated than 
the absolute dilutions.
The common components largely cancel in the differences.
We estimate the probability to obtain
dilution differences similar to the $J/\psi K^+$ data
and use that estimate as a measure of the statistical likelihood to obtain
shape disagreements like the data.

We calculate a $\chi^2$ comparing the dilution differences
between the data and the Monte Carlo simulation, {\it i.e.},
\begin{eqnarray}
\zeta^2 & = & \sum_{i}{{\left({\delta_i - \overline{\delta}_i \over 
	\sigma(\delta_i)}\right)}^2}
\label{eq:z-deltaD}
\end{eqnarray}
where $i$ is the index of the $p_T$-cut 
(13 values in $0.1$ GeV/c increments starting from $0.3$ GeV/c), 
$\delta_i = \Dil_{i+1}-\Dil_i$,
$\Dil_i$ is the measured dilution at $i$, $\overline{\delta}_i$ is the
corresponding difference from the tuned PYTHIA
(Sec.~\ref{sec:Tuned-PYTHIA-mc}), 
and $\sigma(\delta_i)$ is the statistical uncertainty on $\delta_i$.
We calculate these differences relative to the PYTHIA value 
$\overline{\delta}_i$ since $\Dil_+$ varies with $p_T(SST)$ 
(Fig.~\ref{fig:DpmvsPtsst}) and would otherwise introduce 
an unwanted systematic contribution to $\zeta^2$.

\begin{figure}
  \vspace{-0.5cm}
\centerline{
\hspace{-0.0cm}
  \epsfysize 7.4cm
  \epsffile{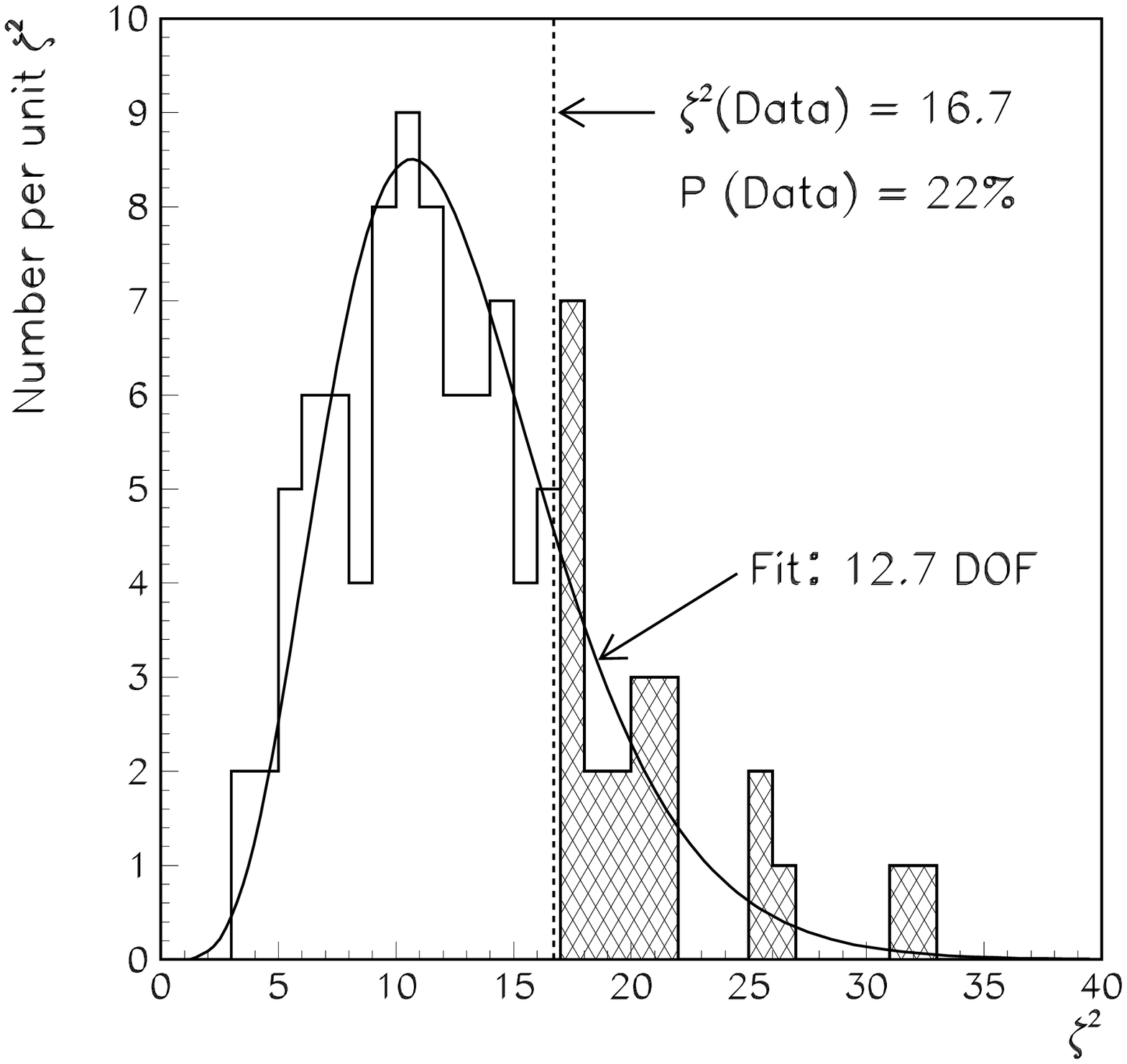}
}
\caption[]
{\small{
The distribution of $\zeta^2$ values obtained from  $J/\psi K^+$ 
Monte Carlo samples equivalent to the data. 
A fit to the standard $\chi^2$ distribution
is shown by the solid curve and yields  12.7 degrees of freedom.
The vertical line indicates the $\zeta^2$ value obtained from the $J/\psi K^+$ 
data sample, and Monte Carlo samples with larger values of $\zeta^2$ 
are shaded.
}}
\label{fig:kpm_vary_chisq}
\end{figure}

We subdivide the PYTHIA sample into 100 subsamples, 
each with statistics equivalent to the $J/\psi K^+$ sample, 
and compute $\zeta^2$ for each~\cite{Ken}.
These subsamples should have the same sort of statistical
fluctuations as the data.
The $\zeta^2$ distribution for the Monte Carlo subsamples is shown
in Fig.~\ref{fig:kpm_vary_chisq}.
The value obtained from the  $J/\psi K^+$ data,
$\zeta^2 = 16.7$, is marked by the vertical line. The data is higher
than typical, but well within the spread of the Monte Carlo samples.

The distribution of the $\zeta^2$'s should,
if the dilution differences are truly uncorrelated,
follow the standard $\chi^2$ distribution 
for $n$ degrees of freedom, which in this case
is the number of differences.
A fit of the $\chi^2$-distribution to the 100 subsamples,
with $n$ as a free parameter,
is also shown in Fig.~\ref{fig:kpm_vary_chisq}.
The fit yields \mbox{$n = 12.71 \pm 0.48$}, in good agreement
with there having been 13 differences in the $\zeta^2$ sum.

We compute from this fit the probability for a sample
the size of the  $J/\psi K^+$ data to yield a $\zeta^2$ at, 
or above, the 16.7 observed in the data to be $22\%$.
Thus, we conclude that the observed differences in the dilution shape 
with $p_T(SST)$ are not statistically unusual.

\narrowtext

\begin{table}
\begin{center}
\begin{tabular}{lcccccc} 
\multicolumn{2}{c}{$\,$} & \multicolumn{5}{c}{Decay Signatures} \\
\cline{3-7}
\multicolumn{2}{c}{Selection Cuts} & $\ell \overline{D}{^0}$ &  $\ell D^-$ 
                                   &\multicolumn{3}{c}{$\ell D^{*-}$} \\
\cline{5-7}
                   &   & $K^+\pi^-$ & $K^+\pi^-\pi^-$ 
                   & $K^+\pi^-$ & $K^+\pi^-\pi^+\pi^-$ & $K^+\pi^-\pi^0$ \\
\hline
$p_T(\ell) >$              & (GeV/c)     & 6.0 & 6.0 & 6.0 & 6.0 & 6.0 \\
$p_T(K)   > $              & (GeV/c)     & 0.7 & 0.6 & --- & --- & 1.0 \\
$p_T(\pi) > $              & (GeV/c)     & 0.5 & --- & --- & --- & 0.8 \\
$p_T(D)   > $              & (GeV/c)     & 2.0 & 3.0 & --- & --- & --- \\
$d_0/\sigma_0 >$           &             & 3.0 & 2.0 & 1.0 & 0.5 & 1.0 \\
${\lxy(D)}/\sigma_{\lxy} >$&             & 3.0 & 5.0 & 1.0 & 1.0 & 1.0 \\
$|\Delta m(D^*)| <$        & (MeV/c$^2$) & --- & --- & 3.0 & 2.0 & --- \\
$m(\ell D) <$              & (GeV/c$^2$) & 5.0 & 5.0 & --- & --- & --- \\
$-0.5 < ct_D <$            &  (mm)       & 1.0 & 2.0 & 1.0 & 1.0 & 1.5 \\ 
\end{tabular}
	\parbox{5in}{\caption{\small 
                Kinematic and geometric selection cuts 
		for the five decay signatures.
		The impact parameter significance cut $d_0/\sigma_0$ 
		is applied to $D$ daughter tracks. 
		$\lxy(D)/\sigma_{\lxy}$ is the $D$ decay length significance 
		 relative to the primary, while $ct_D$ is 
		the proper decay length of the $D$ with respect 
		to the $B$ vertex, and $\Delta m(D^*)$
                is the mass difference between the $D^*$ candidate and
                the $D$ candidate plus pion mass.
	\label{tab:Selection-Cuts}
	}}
\end{center}
\end{table}

\begin{table}
\begin{center}
\begin{tabular}{|c|c|c|c|} 
\hline
Name            &  $J^P$   &  Width    &  Decay Modes \\    
\hline
$D_0^*$         &  $0^+$   &  wide     &   $D\pi$ \\
$D_1^*$         &  $1^+$   &  wide     &   $D^*\pi$ \\
$D_1(2420)$     &  $1^+$   &  narrow   &   $D^*\pi$ \\
$D_2^*(2460)$   &  $2^+$   &  narrow   &   $D\pi$, $D^*\pi$ \\
\hline
\end{tabular}
	\parbox{5in}{\caption{\small 
		The expected $D^{**}$ states and properties. 
		The classification into wide and narrow states
		follows the predictions of
		Heavy Quark Effective Theory.
		\label{tab:d**-states}
		}}
\end{center}
\end{table}

{\tightenlines
\begin{table}
\begin{center}
\begin{tabular}{lccc}
\multicolumn{1}{c}{$\,$} & \multicolumn{3}{c}{Decay Signatures} \\
\cline{2-4}
\multicolumn{1}{c}{$B^0$ Decay Chains } 
                 & \multicolumn{1}{c}{$\ell D^{*-}$} 
                      & \multicolumn{1}{c}{$\ell D^-$}
                             & \multicolumn{1}{c}{$\ell \overline{D^0}$}  \\ 
\hline
$\rightarrow {D}^{**-}\ell^+\nu$  & & &  \\
     \hspace*{0.4cm}${D}^{**-}\rightarrow \overline{D}^{*0}\pi^-_{**}$  
                 &
                      &
                            &   \\ 
     \hspace*{1.7cm}$\overline{D}^{*0}\rightarrow \overline{D}^{0}\pi^0_*$  
                  &  ---
                      & ---
                            & $f^{**}P_V(2/3){\cal B}^*(D^{-}\pi^0_*)$ \\
     \hspace*{1.7cm}$\overline{D}^{*0}\rightarrow \overline{D}^{0}\gamma$  
                 & ---
                      & ---
                            & $f^{**} P_V(2/3) \, 
                                        {\cal B}^*(\overline{D}^{0}\gamma)$ \\ 
     \hspace*{0.4cm}${D}^{**-}\rightarrow D^{*-}\pi^0_{**}$  
                                               & & &   \\
     \hspace*{1.7cm}${D}^{*-}\rightarrow \overline{D}^{0}\pi^-_*$  
                 & $f^{**}P_V(1/3){\cal B}^*(\overline{D}^{0}\pi^-_*)
                                               \epsilon(\pi_*)$ 
                      & ---
                            & $f^{**} P_V(1/3) \, {\cal B}^*
        (\overline{D}^{0}\pi^-_*)(1-\epsilon(\pi_*))$ \\ 
     \hspace*{1.7cm}${D}^{*-}\rightarrow D^{-}\pi^0_*$  
                 & ---
                      & $f^{**} P_V(1/3) \, {\cal B}^*(D^{-}\pi^0_*)$ 
                            & --- \\
     \hspace*{1.7cm}${D}^{*-}\rightarrow {D}^{-}\gamma$  
                 & ---
                      & $f^{**} P_V(1/3) \, {\cal B}^*(D^{-}\gamma)$ 
                            & --- \\
     \hspace*{0.4cm}${D}^{**-}\rightarrow \overline{D}^{0}\pi^-_{**}$  
                 & ---
                      &  ---
                            & $f^{**} (1-P_V)(2/3) $   \\  
     \hspace*{0.4cm}${D}^{**-}\rightarrow {D}^{-}\pi^0_{**}$  
                 & ---
                      & $f^{**} (1-P_V)(1/3)$
                            & ---  \\  
\hline
$\rightarrow {D}^{*-}\ell^+\nu$  & & &  \\
     \hspace*{0.4cm}${D}^{*-}\rightarrow \overline{D}^{0}\pi^-_{*}$  
                 &  $f^* {\cal B}^*(\overline{D}^{0}\pi^-_{*})\epsilon(\pi_*)$ 
                      & ---
                           & $f^* {\cal B}^*( \overline{D}^{0}\pi^-_{*}) 
                                                    (1-\epsilon(\pi_*))$  \\ 
     \hspace*{0.4cm}${D}^{*-}\rightarrow D^{-}\pi^0_{*}$  
                 & ---
                      & $f^* {\cal B}^*( D^{-}\pi^0_{*})$ 
                           & ---  \\
     \hspace*{0.4cm}${D}^{*-}\rightarrow D^{-}\gamma$  
                 & ---
                      & $f^* {\cal B}^*( D^{-}\gamma)$ 
                           & ---  \\
\hline
$\rightarrow {D}^{-}\ell^+\nu$ 
                 &  ---
                      & $f$  
                           &  --- \\
\hline
\hline
\multicolumn{1}{c}{$B^+$ Decay Chains} 
                 & \multicolumn{1}{c}{ $\,$ } 
                      & \multicolumn{1}{c}{ }
                             & \multicolumn{1}{c}{ }  \\ 
\cline{1-1}
$\rightarrow \overline{D}^{**0}\ell^+\nu$  & & &  \\
     \hspace*{0.4cm}$\overline{D}^{**0}\rightarrow {D}^{*-}\pi^+_{**}$  
                 & 
                      & 
                            &   \\ 
     \hspace*{1.7cm}${D}^{*-}\rightarrow \overline{D}^{0}\pi^-_*$  
                 & $f^{**}P_V(2/3){\cal B}^*(\overline{D}^{0}\pi^-_*)
                                                         \epsilon(\pi_*)$ 
                      & ---
                            & $f^{**}P_V(2/3){\cal B}^*(\overline{D}^{0}
                                           \pi^-_*) (1-\epsilon(\pi_*))$  \\ 
     \hspace*{1.7cm}${D}^{*-}\rightarrow {D}^{-}\pi^0_*$  
                 &  ---
                      & $f^{**}P_V(2/3){\cal B}^*({D}^{-}\pi^0_*)$ 
                            & --- \\ %
     \hspace*{1.7cm}${D}^{*-}\rightarrow \overline{D}^{0}\gamma$  
                 & ---
                      & $f^{**}P_V(2/3){\cal B}^*({D}^{-}\gamma)$
                            & --- \\%
     \hspace*{0.4cm}$\overline{D}^{**0}\rightarrow \overline{D}^{*0}
                                                                \pi^0_{**}$  
                 & & &   \\
     \hspace*{1.7cm}$\overline{D}^{*0}\rightarrow \overline{D}^{0}\pi^0_*$  
                 & ---
                      & ---
                            & $f^{**}P_V(1/3){\cal B}^*(\overline{D}^{0}
                                                               \pi^0_*)$ \\
     \hspace*{1.7cm}$\overline{D}^{*0}\rightarrow \overline{D}^{0}\gamma$  
                 & ---
                      & ---
                            & $f^{**}P_V(1/3){\cal B}^*
                                               (\overline{D}^{0}\gamma)$ \\
     \hspace*{0.4cm}$\overline{D}^{**0}\rightarrow {D}^{-}\pi^+_{**}$  
                 & ---
                      & $f^{**}(1-P_V)(1/3)$  
                            &  ---   \\  
     \hspace*{0.4cm}$\overline{D}^{**0}\rightarrow \overline{D}^{0}
                                                             \pi^0_{**}$  
                 & ---
                      & ---
                            &  $f^{**}(1-P_V)(1/3)$   \\  
\hline
$\rightarrow \overline{D}^{*0}\ell^+\nu$  & & &  \\
     \hspace*{0.4cm}$\overline{D}^{*0}\rightarrow \overline{D}^{0}\pi^0_{*}$  
                 & ---
                      & ---
                           & $f^{*}{\cal B}^*(\overline{D}^{0}\pi^0_{*})$   \\
     \hspace*{0.4cm}$\overline{D}^{*0}\rightarrow \overline{D}^{0}\gamma$  
                 & ---
                      & --- 
                           &  $f^{*}{\cal B}^*(\overline{D}^{0}\gamma)$   \\
\hline
$\rightarrow \overline{D}^{0}\ell^+\nu$ 
                 &  ---
                      & ---  
                           &  $f$   \\
\end{tabular}
\end{center}
\caption{Table of the various $B$ decay chains and
their contributions to the sample composition
of the three general categories of decay signatures 
($\ell D^{*-}$, $\ell D^-$, and $\ell \overline{D}^0$).
Dashes indicate that a particular decay chain makes no contribution.
The total contribution to a given sample is simply the sum over
the entries in a vertical column. The branching ratio 
of $D^* \rightarrow XY$ is denoted by ``${\cal B}^*(XY)$.''
All $\pi^0_{(*)}$'s, $\pi_{**}$'s, and $\gamma$'s 
are lost from the reconstruction,
and $\pi^-_*$'s are identified with efficiency $\epsilon(\pi_*)$.
See the text for a discussion of the other parameters.
}
\label{tab:all-decay}
\end{table} 
}

\begin{table}
\begin{center}
\begin{tabular}{lcc} 
Signature    	& $\sigma^{ct}(0)$ ($\mu$m) &  $b$ \\
\hline 
\kpp\   	& $39$ & $0.108$\\
\kps\   	& $52$ & $0.075$\\
\ktps\  	& $49$ & $0.073$\\
\kpzs\  	& $62$ & $0.070$\\
\kp\    	& $45$ & $0.092$\\
\end{tabular}
\parbox{5in}{\caption{\small 
	The parameters of the linear model of the $ct$ resolution
	[Eq.~(\ref{eq:sigma(ct)})] for the five direct decay chains.
	\label{tab:sigma0+bct}
	}}
\end{center}
\end{table}

\begin{table}
\begin{center}
\begin{tabular}{lc} 
\multicolumn{1}{c}{Decay signature} & $R^{**\>(meas)}$ 	\\
\hline
\kpp\  	&  $0.056 \pm 0.022$	\\
\kps\   &  $0.003 \pm 0.029$	\\
\ktps\  & $-0.016 \pm 0.026$	\\
\kpzs\  &  $0.034 \pm 0.021$	\\
\kp\ 	&  $0.029 \pm 0.018$	\\
\end{tabular}
	\parbox{5in}{\caption{\small
		The fractions of tags (with no $d_0/\sigma_{0}$ cut)
		identified as $\pids^\pm$ candidates,
		$R^{**}$, measured in the five decay signatures.
		\label{tab:measured-R**}
		}}
\end{center}
\end{table}

\begin{table}
\begin{center}
\begin{tabular}{cccccc} 
Input&\multicolumn{2}{c}{Fit Input} &$\,$& \multicolumn{2}{c}{Fit Output} \\
\cline{2-3} \cline{5-6}
Parameter   & Value    & Error       & & Value    &  Error   \\
\hline
$f^{**}$    & $0.360 $ & $\pm 0.120$ & & $0.309 $ & $ \pm {0.100}   $ \\
$R_f$       & $2.50  $ & $\pm0.60  $ & & $2.51  $ & $ \pm {0.60}    $ \\
c\tbo\  ($\mu$m)
  & $468$ & $\pm 18$ & & $ 468$ & $\pm{18}$ \\
\tbx/\tbo\ 	    
	    & $1.020 $ & $\pm0.050 $ & & $1.021 $ & $\pm{0.049}  $ \\
\end{tabular}
	\parbox{5in}{\caption{\small
		The fit input and output values of the measured 
                sample composition parameters.
	\label{tab:in-out}
	}}
\end{center}
\end{table}

\begin{table}
\begin{center}
\begin{tabular}{lrrr} 
\hline
 name         &    \dmd\   &     \Dx\    &    \Do\    \\ \hline
 \dmd\        &  $ 1.000$  &  $ -0.372$  &  $-0.172$  \\
 \Dx\         &   $-0.372$ &  $  1.000$  &  $ 0.372$  \\
 \Do\         &   $-0.172$ &  $  0.372$  &  $ 1.000$  \\
 $R_f$        &   $-0.020$ &  $  0.126$  &  $ 0.007$   \\   
 $f^{**}$     &   $-0.385$ &  $  0.406$  &  $ 0.504$   \\ 
 $P_V$        &   $-0.326$ &  $ -0.284$  &  $-0.310$   \\ 
 $\epsilon(\pi_*)$  
              &   $-0.031$ &  $ -0.082$  &  $ 0.100$   \\      
 $\xi_{norm}$ &   $ 0.304$ &  $ -0.355$  &  $-0.445$   \\   
 \tbo\        &   $-0.005$ &  $  0.002$  &  $-0.001$   \\  
 \tbx/\tbo\   &   $-0.051$ &  $ -0.157$  &  $ 0.009$   \\ 
\hline
\end{tabular}
	\parbox{5in}{\caption{\small 
	The correlation coefficients for
	\dmd,	\Dx, and \Do, with respect to all ten fit parameters.
	\label{tab:correl-coef}
	}}
\end{center}
\end{table}

\begin{table}
\begin{center}
\begin{tabular}{lccc} 
Source 	 & $\sigma(\dmd)$ & $\sigma(\Dx)$ & $\sigma(\Do)$ \\
\hline
\hline
Sample Composition & 
 $^{+0.0295} _{-0.0310}$ & $^{+0.0216} _{-0.0131}$ & $^{+0.0225} _{-0.0131}$\\
$b$-quark spectrum  & $\pm 0.0060$ & $\pm 0.0052$ & $\pm 0.0017$  \\
$e$ isolation cuts  & $\pm 0.0045$ & $\pm 0.0036$ & $\pm 0.0047$  \\
Decay model         & $\pm 0.0115$ & $\pm 0.0005$ & $\pm 0.0045$ \\
\lxy\ resolution    & $\pm 0.0033$ & $\pm 0.0003$ & $\pm 0.0000$  \\
$\xi_{MC}(ct)$ shape 
                    & $\pm 0.0035$ & $\pm 0.0002$ & $\pm 0.0015$  \\
$B \to D_s^{(*)}D^{(*)}X$ 
                    & $\pm 0.0010$ & $\pm 0.0006$ & $\pm 0.0004$ \\
$B_s \to \nu\ell D_s^{**}$
	            & $\pm 0.0010$ & $\pm 0.0019$ & $\pm 0.0008$ \\
$g \to c\bar{c} \to \ell D^{(*)}$
		    & $\pm 0.0006$ & $\pm 0.0012$ & $\pm 0.0025$ \\
\hline
Total  & 
$^{+0.0329} _{-0.0343}$ & $^{+0.0226} _{-0.0147}$ & $^{+0.0237} _{-0.0150}$ \\ 
\hline
\end{tabular} 
\parbox{5in}{\caption{\small
	Table of the systematic uncertainties.
	\label{tab:systematics}
	}}
\end{center}
\end{table}

\begin{table}
\begin{center}
\begin{tabular}{lccc} 
\hline
\multicolumn{1}{c}{Decay} & \multicolumn{3}{c}{Fractional Contribution} \\
\multicolumn{1}{c}{Signature} & $B \to D_s^{(*)}D^{(*)}$ & $B_s \to \nu\ell 
           D_s^{**}$ & $g \to \ell D^{(*)}$ \\
\hline
 \kpp\    & $0.017$  & $ 0.011$  & $0.005$ \\
 \kps\    & $0.005$  & $ 0.008$  & $0.002$ \\
 \ktps\   & $0.005$  & $ 0.008$  & $0.005$ \\
 \kpzs\   & $0.006$  & $ 0.009$  & $0.019$ \\
 \kp\     & $0.009$  & $ 0.011$  & $0.005$ \\
\hline
\end{tabular} 
\parbox{5in}{\caption{\small 
	Fractional contribution of $B \to D_s^{(*)}D^{(*)}X$,
	$B \to D_s^{(*)}D^{(*)}$, and $g\to c\bar{c} \to \ell D^{(*)}$
	 to the $\ell D^{(*)}$ samples.
	\label{tab:combined-frac}
	}}
\end{center}
\end{table}
 
{\tightenlines
\begin{table}
\begin{center}
\begin{tabular}{lccc}    
\multicolumn{2}{c}{Output Parameters}
    & $B^+ \rightarrow J/\psi K^+$    &  $B^0 \rightarrow J/\psi K^{*0}$  \\ 
                                                               \hline\hline
Frac. Signal& $f_B$          & $  0.067 \pm  0.003 $ & $  0.156 \pm  0.009 $\\
Frac. L-Lived Back.& $f_L$   & $  0.160 \pm  0.008 $ & $  0.222 \pm  0.023 $\\
Mass Error Scale   & $X$     & $   1.34 \pm   0.05 $ & $   1.54 \pm   0.10 $\\
$c\tau$ Error Scale& $Y$     & $   0.99 \pm   0.01 $ & $   1.06 \pm   0.03 $\\
\multicolumn{1}{c}{\underline{Prompt:}} & \multicolumn{3}{c}{  } \\
Mass Slope ($\times 10^{-3}$) & $\zeta_P $
                             & $   11.0 \pm    0.9 $ & $    4.4 \pm    2.4 $\\ 
Back. Tag Eff. & $\epsilon_P$& $  0.703 \pm  0.005 $ & $  0.830 \pm  0.011 $\\
Tag Asym.      &  $\alpha_P$ & $  0.033 \pm  0.013 $ & $  0.092 \pm  0.031 $\\
Eff. Asym.     & $\delta_P$  & $ -0.002 \pm  0.007 $ & $  0.012 \pm  0.013 $\\
Recon. Asym.   &  $R_P$      & $  0.003 \pm  0.011 $ & $  0.036 \pm  0.029 $\\
Dilution       & $\Dil_P$    & $ -0.069 \pm  0.013 $ & $ -0.003 \pm  0.031 $\\
\multicolumn{1}{c}{\underline{Long-Lived:}} & \multicolumn{3}{c}{ }\\ 
1st Lifetime ($\mu m$)   
               & $\tau_{L1} $   & $   595. \pm   53. $ & $  371. \pm   65.$ \\ 
2nd Lifetime ($\mu m$)   
               & $\tau_{L2} $   & $   135. \pm    9. $ & $   99. \pm   21.$ \\ 
Frac. Neg. Back.      & $f_N$   & $  0.137 \pm 0.014 $ & $ 0.096 \pm 0.029$ \\ 
Frac. 2nd Lifetime  &$f_{\tau2}$& $  0.781 \pm 0.030 $ & $ 0.626 \pm 0.104$ \\ 
Mass Slope ($\times 10^{-3}$)
                     & $\zeta_L$& $  -12.3 \pm   2.7 $ & $ -17.6 \pm   5.3$ \\ 
Back. Tag Eff. & $\epsilon_L$   & $  0.771 \pm 0.014 $ & $ 0.778 \pm 0.031$ \\ 
Tag Asym.      & $\alpha_L$     & $  0.015 \pm 0.037 $ & $-0.044 \pm 0.079$ \\ 
Eff. Asym.     & $\delta_L$     & $ -0.026 \pm 0.018 $ & $-0.029 \pm 0.038$ \\
Recon. Asym.   & $R_L$          & $  0.030 \pm 0.034 $ & $ 0.095 \pm 0.070$ \\ 
Dilution       & $\Dil_L$       & $ -0.089 \pm 0.038 $ & $-0.050 \pm 0.079$ \\ 
\multicolumn{1}{c}{\underline{$B$ Signal:}} & \multicolumn{3}{c}{ }\\ 
Tag Eff.       & $\epsilon_B$   & $  0.624 \pm 0.020 $ & $ 0.635 \pm 0.030$ \\ 
Recon. Asym.   & $R_B$          & $  0.077 \pm 0.041 $ & $-0.086 \pm 0.068$ \\ 
{\bf Dilution} &\boldmath $\Dil_B$&\boldmath$0.185 \pm  0.052 $ & 
                                          \boldmath$0.165 \pm  0.112 $  \\ 
\end{tabular}
\vspace*{-0cm}
\end{center}
\caption{\small Likelihood fit results for $B^+ \rightarrow J/\psi K^+$
and $B^0 \rightarrow J/\psi K^{*0}$. The parameter
of central interest is the signal dilution in the bottom line.
}
\label{tab:results}
\end{table}
}

{\tightenlines
\begin{table}
\begin{center}
\begin{tabular}{lcccrr}  
\multicolumn{2}{c}{\raisebox{-1.0ex}{Parameter}} & Central & 
    \raisebox{-1.0ex}{Variation} &\multicolumn{2}{c}{Shift in $\Dil_+$} \\ 
     && Value   &           & Negative  & Positive \\ \hline
$\tau_B$ & $(\mu m)$ & $486$ & $\pm 18$ & $   -0.0012$&$    0.0011$\\
\multicolumn{6}{c}{\underline{$\gamma_B/\alpha_B = 1$}}\\
$a_1$& $\times 10^{-4}\; (GeV/c)^4$ & $3.9$ & $\pm 1.8$ &
$   -0.0001$&$    0.0001$\\
$b_1$& $\times 10^{-3}\; (GeV/c)^4$ & $1.3$ & $\pm 0.4$ & 
$    0.0003$&$   -0.0003$\\
$a_2$& $\times 10^{-2}$             & $1.4$ & $\pm 0.4$ &
$    0.0001$&$   -0.0002$\\
$b_2$& $\times 10^{-2}$             & $2.6$ & $\pm 0.8$ &
$   -0.0005$&$    0.0005$\\
\multicolumn{6}{c}{\underline{$\gamma_B/\alpha_B = 0$}}\\ 
$a_1$& $\times 10^{-4}\; (GeV/c)^4$ & $3.9$ & $\pm 1.8$ &
$    0.0002$&$   -0.0003$\\
$b_1$& $\times 10^{-3}\; (GeV/c)^4$ & $1.3$ & $\pm 0.4$ &
$    0.0007$&$   -0.0007$\\
$a_2$& $\times 10^{-2}$             & $1.4$ & $\pm 0.4$ &
$   -0.0004$&$    0.0003$\\
$b_2$& $\times 10^{-2}$             & $2.6$ & $\pm 0.8$ &
$   -0.0010$&$    0.0010$\\
\multicolumn{6}{c}{\underline{$\gamma_B/\alpha_B = 2.5$}}\\
$a_1$& $\times 10^{-4}\; (GeV/c)^4$ & $3.9$ & $\pm 1.8$ &
$   -0.0010$&$    0.0009$\\
$b_1$& $\times 10^{-3}\; (GeV/c)^4$ & $1.3$ & $\pm 0.4$ &
$   -0.0005$&$    0.0004$\\
$a_2$& $\times 10^{-2}$             & $1.4$ & $\pm 0.4$ &
$    0.0015$&$   -0.0019$\\
$b_2$& $\times 10^{-2}$             & $2.6$ & $\pm 0.8$ &
$    0.0005$&$   -0.0008$\\
\multicolumn{6}{c}{\underline{$\alpha_B =$ Central Value}}\\
$\gamma_B/\alpha_B$ & & $1$ & $^{1.5} _{1.0}$ &
$    0.0015$&$   -0.0034$\\
\hline
\multicolumn{4}{l}{\bf Combined Uncertainty} &
  \multicolumn{2}{c}{$\begin{array}{c} \boldmath +0.003 \\ \boldmath -0.004
                    \end{array}$} \\
\end{tabular}
\end{center}
\caption{{The fixed inputs for the
$B^+ \rightarrow J/\psi K^+$ fit, their central values,
$1\sigma$ variations, and the resulting shifts of the central value
of $\Dil_+$. The shifts are combined in quadrature (see text) 
to obtain the combined systematic uncertainty.
}}
\label{tab:bk3}
\end{table}
}

{\tightenlines
\begin{table}
\begin{center}
\begin{tabular}{lcccrr}  
\multicolumn{2}{c}{\raisebox{-1.0ex}{Parameter}}& Central & 
      \raisebox{-1.0ex}{Variation} &\multicolumn{2}{c}{Shift in $\Dil_0$} \\
      && Value   &           & Negative  & Positive \\ \hline
$\tau_B$ & $(\mu m)$ & $468$ & $\pm 18$ &
$   -0.0002$&$    0.0002$\\
$\Delta m$ & $(ps^{-1})$ & $0.474$ & $\pm 0.031$ &
$   -0.0005$&$    0.0003$\\
\multicolumn{6}{c}{\underline{$\gamma_B/\alpha_B = 1$}}\\ 
$a_1$& $\times 10^{-4}\; (GeV/c)^4$ & $3.9$ & $\pm 1.8$ &
$   -0.0008$&$    0.0007$\\
$b_1$& $\times 10^{-3}\; (GeV/c)^4$ & $1.3$ & $\pm 0.4$ &
$    0.0024$&$   -0.0025$\\
$a_2$& $\times 10^{-2}$             & $1.4$ & $\pm 0.4$ &
$    0.0013$&$   -0.0014$\\
$b_2$& $\times 10^{-2}$             & $2.6$ & $\pm 0.8$ &
$   -0.0039$&$    0.0038$\\
\multicolumn{6}{c}{\underline{$\gamma_B/\alpha_B = 0$}}\\
$a_1$& $\times 10^{-4}\; (GeV/c)^4$ & $3.9$ & $\pm 1.8$ &
$   -0.0008$&$    0.0008$\\
$b_1$& $\times 10^{-3}\; (GeV/c)^4$ & $1.3$ & $\pm 0.4$ &
$   -0.0004$&$    0.0004$\\
$a_2$& $\times 10^{-2}$             & $1.4$ & $\pm 0.4$ &
$   -0.0002$&$    0.0001$\\
$b_2$& $\times 10^{-2}$             & $2.6$ & $\pm 0.8$ &
$   -0.0002$&$    0.0001$\\
\multicolumn{6}{c}{\underline{$\gamma_B/\alpha_B = 2.5$}}\\ 
$a_1$& $\times 10^{-4}\; (GeV/c)^4$ & $3.9$ & $\pm 1.8$ &
$   -0.0011$&$    0.0010$\\
$b_1$& $\times 10^{-3}\; (GeV/c)^4$ & $1.3$ & $\pm 0.4$ &
$    0.0063$&$   -0.0065$\\
$a_2$& $\times 10^{-2}$             & $1.4$ & $\pm 0.4$ &
$    0.0040$&$   -0.0044$\\
$b_2$& $\times 10^{-2}$             & $2.6$ & $\pm 0.8$ &
$   -0.0092$&$    0.0088$\\
\multicolumn{6}{c}{\underline{$\alpha_B =$ Central Value}}\\ 
$\gamma_B/\alpha_B$ & & $1$ & $^{+1.5} _{-1.0}$ &
$    0.0031$&$   -0.0057$\\
$f_{S}$ & & $0.1$ & $\pm 0.1$ &
$    0.0086$&$   -0.0160$\\
$X_{S}$ & & $5.0$ & $\pm 2.0$ &
$    0.0111$&$   -0.0003$\\
$\mu_{S}$ & & $-0.5$ & $\pm 0.5$ &
$   -0.0008$&$    0.0007$\\
\hline
\multicolumn{4}{l}{\bf Combined Uncertainty} &
  \multicolumn{2}{c}{$\begin{array}{c}\boldmath +0.018\\ \boldmath -0.021
                    \end{array}$} \\
\end{tabular}
\end{center}
\caption{{The systematic uncertainties 
from the fixed $B^0 \rightarrow J/\psi K^{*0}$ fit parameters. 
The table is similar to that for $J/\psi K^+$
(Table~\ref{tab:bk3}) except for the addition 
of the oscillation frequency $\Delta m_d$,  and
$f_{S}$, $\mu_{S}$ and $X_{S}$, which model the
$K$-$\pi$ swapping in the $K^{*0}$ reconstruction.
The shifts in $\Dil_0$ are combined in quadrature (see text) 
to obtain the combined systematic uncertainty.
}}
\label{tab:bk2}
\end{table}
}

\begin{table}[htb]
\begin{center}
\begin{tabular}{lccc}    \hline
Sample & MC Calculation & Data Meas. & Data/MC Ratio\\ \hline
$B^0\rightarrow \ell D^{(*)}$& 0.196 & $0.181\pm 0.035$ & $0.923 \pm 0.179$ \\ 
$B^0\rightarrow J/\psi K^{*0}$ & 0.189 & $0.165 \pm 0.112$ & $0.873\pm 0.593$\\
$B^+\rightarrow \ell D^{(*)}$& 0.266 & $0.267\pm 0.037$ & $1.004 \pm 0.139$ \\
$B^+\rightarrow J/\psi K^{+ }$ & 0.254 & $0.185 \pm 0.052$ & $0.728\pm 0.205$\\
\end{tabular}
\end{center}
\caption{
\label{tab:d_mc}
Calculated and measured dilutions for the $\ell D^{(*)}$ 
and $J/\psi K$ samples.
The calculated values are from the simplified simulation.
The ratios are of the measured values divided 
by the Monte Carlo calculations.
}
\end{table}

\begin{table}
\begin{center}
\begin{tabular}{cc} 
Parameter       &  Value \\
\hline
$R_f$              &  $2.722$ \\
$f^{**}_{res}$     &  $0.231$  \\
$f^{**}_{non}$     &  $0.125$  \\
$f^{**}$           &  $0.356$  \\
$P_V$              &  $0.687$  \\
$\tbx/\tbo$        &  $1.014$  \\
\end{tabular}
\end{center}
		\caption{
                The values of the sample composition parameters
                used in the QQ (V9.1) $B$ decay program.
		The resonant ($B \rightarrow \nu \ell D^{**}$) and
		non-resonant ($B \rightarrow \nu \ell D^{(*)}\pi_{**}X$) 
		fractions, $f^{**}_{res}$ and $f^{**}_{non}$ respectively,
   		sum by definition to $f^{**}$.
                \label{tab:sc-param-mc}
        }
\end{table}

\begin{table}
\begin{center}
\begin{tabular}{cc} 
Parameter       &  Value \\
$\sigma(B^*  )/\sigma(B+B^*+B^{**})$      & 0.7625 \\
$\sigma(B^{**}:\,^1P_1)/\sigma(B+B^{**}:\,S=0)$  & 0.320   \\
$\sigma(B^{**}:\,^3P_0)/\sigma(B^*+B^{**}:\,S=1)$  & 0.033  \\
$\sigma(B^{**}:\,^3P_1)/\sigma(B^*+B^{**}:\,S=1)$  & 0.099  \\
$\sigma(B^{**}:\,^3P_2)/\sigma(B^*+B^{**}:\,S=1)$  & 0.165  \\
\end{tabular}
\end{center}
		\caption{
                The $\sigma(B)$ ratios represent the
                relative production rates used in PYTHIA for the different
                $B$ mesons. The relative ratios are labeled 
		by spectroscopic notation or their spin $S$,
                {\it e.g.}, ``$\sigma(B+B^{**}:\,S=0)$'' represents
                the sum of cross sections for $B$ and $B^{**}$ states 
		with zero spin.
                \label{tab:sc-B**}
        }
\end{table}

\begin{table}
\begin{center}
\begin{tabular}{rccl}
 Parameter & Default & Tuned & Description\\ \hline
 MSTP(82)  &1        &3      & model of multiple interactions\\
 PARP(85)  &0.33     &1.0    & fraction of color-connected\\
           &         &       & $gg$ multiple interactions\\
 PARP(86)  &0.66     &1.0    & total fraction of $gg$\\
           &         &       & multiple interactions\\
 MSTP(33)  &No       &Yes    & multiply cross-sections by PARP(31)\\
 PARP(31)  &1.00     &1.66   & increase cross-sections by 66\%\\
 PARJ(21)  &0.36     &0.6    & $\sigma^{frag}_{p_T}$\\
 MSTJ(11)  &4        &3      & use Peterson frag. for $b, c$\\
 PARJ(55)  &-0.006   &-0.0063 & $-\epsilon_b$ \\
\end{tabular}
\end{center}
\caption{The PYTHIA parameters modified from their
defaults in order to agree with $B  \rightarrow  \nu \ell D^0$,
$D^0 \rightarrow K^+\pi^-$ data.
}
\label{tab:tune-PYTHIA}
\end{table}


\begin{references}
\bibitem{CKM} N.~Cabibbo, Phys. Rev. Lett. {\bf 10}, 531 (1963); 
 M.~Kobayashi and K.~Maskawa, 
 Prog. Theor. Phys. {\bf 49}, 652 (1973).
\bibitem{UA1mix} UA1 Collaboration, C.~Albajar {\it et al}., 
 Phys. Lett. B {\bf 186}, 247 (1987).
\bibitem{ARGUSmix} ARGUS Collaboration, H.~Albrecht {\it et al}., 
 Phys. Lett. B {\bf 192}, 245 (1987).

\bibitem{ref:mixing} 
For a recent review of experimental results on $B^0\overline{B}{^0}$
 oscillations
see O.~Schneider, ``Heavy Quark Spectroscopy, Oscillations, and Lifetime,''
presented at the 18th International Symposium on Lepton-Photon Interactions,
Hamburg, Germany, to appear in the proceedings. 

\bibitem{NewEx}
  HERA-B Collaboration,   DESY-PRC 95/01, 1995; 
  CLEO III Collaboration, Cornell CLNS 94/1277, 1994;
  BELLE Collaboration,    KEK Report 3-1995, 1995; 
  BaBar Collaboration,    SLAC-R-95-457, 1995; 
  D0 Collaboration,       FERMILAB-Pub-96/357-E, 1996; 
  CDFII Collaboration,    FERMILAB-Pub-96/390-E, 1996.
\bibitem{LepDPRL} CDF Collaboration, F.~Abe {\it et al}., 
 Phys. Rev. Lett. {\bf 80}, 2057 (1998). 
\bibitem{Rosner} M.~Gronau, A.~Nippe, and J.~Rosner, 
 Phys. Rev. D {\bf 47}, 1988 (1993);
M.~Gronau and J.~L.~Rosner, Phys. Rev. D {\bf 49}, 254
 (1994).
\bibitem{OPALCorr}  OPAL Collaboration, R.~Akers et al., 
  Z. Phys. C {\bf 66}, 19 (1995). 


\bibitem{ALEPHSST}
 ALEPH Collaboration, R.~Barate et al.,
   Phys. Lett. B {\bf 425}, 215 (1998).


\bibitem{OPALks}  OPAL Collaboration, K.~Ackerstaff  {\it et. al.},
   CERN preprint CERN-EP/98-001.


\bibitem{CDFpsiKs}   CDF Collaboration, F.~Abe {\it et al}., 
  FERMILAB-Pub-98/189-E, 
  submitted to Phys. Rev. Lett..
\bibitem{CDFDET}  CDF Collaboration, F.~Abe {\it et al}., 
 Nucl. Instrum. Methods  A {\bf 271}, 387 (1988).

\bibitem{SVX} 
  D.~Amidei {\em et al.}, Nucl. Instrum. Methods A {\bf 350}, 73 (1994);
  P.~Azzi {\em et al.},   {\it ibid.} {\bf 360}, 137 (1995).
 
\bibitem{CDFTopEvidncePRD}   CDF Collaboration, F.~Abe {\it et al}., 
    Phys. Rev. D {\bf 50}, 2966 (1994). 

\bibitem{CDFLifetimePRD}   CDF Collaboration, F.~Abe {\it et al}., 
   Phys. Rev. D {\bf 57}, 5382 (1998).  
 
\bibitem{ALEPH1stJetQ}
 ALEPH Collaboration, R.~Buskulic {\it et al}.,
 Phys. Lett. B {\bf 284}, 177 (1992). 


\bibitem{lotsOfOST} 
 ALEPH Collaboration, R.~Buskulic {\it et al}., 
 Z. Phys. C {\bf 75},  397 (1997);
 OPAL Collaboration, K.~Ackerstaff {\it et al}., 
  {\it ibid.}  {\bf 76},  401  (1997); 
  {\it ibid.}  {\bf 76}, 417  (1997); 
 DELPHI Collaboration, P.~Abreu {\it et al}., 
  {\it ibid.}  {\bf 76},  579 (1997);
 L3 Collaboration, M.~Acciarri, {\it et al}.,  
 CERN-EP/98-028, submitted to  E. Phys. J. C.

\bibitem{CDFJETQ} CDF Collaboration, F.~Abe {\it et al}., 
 to be submitted to Phys. Rev. D.

\bibitem{OPALSSTJetQ}  OPAL Collaboration, R.~Akers et al., 
  Phys. Lett. B {\bf 327}, 411 (1994).

\bibitem{FieldFeynman} R.D. Field and R.P. Feynman,
  Nucl. Phys. B {\bf 136}, 1 (1978).

\bibitem{RosnerDun} I. Dunietz and J.L. Rosner,
  Phys. Rev D {\bf 51},  2471 (1995).

\bibitem{PDG} Particle Data Group, R.M. Barnett {\it et al}.,
              Phys. Rev. D {\bf 54}, 1 (1996).

\bibitem{ref:exp-D**}
For a recent survey of $D^{**}$ production measurements, see
C.~J.~Kreuter, ``Measurements of $B \to D^{**}\ell\nu$,''
presented at the 2nd International Conference on B Physics and CP Violation,
Honolulu, Hawaii, March 24-27, 1997, to appear in the proceedings.


\bibitem{ref:HQET} D. Ebert, V.O. Galkin, and R.N. Faustov,
 Phys. Rev. D {\bf 57}, 5663 (1998).




\bibitem{f**-old-ref} CLEO Collaboration,
R. Fulton {\it et al}, Phys. Rev. D {\bf 43}, 651 (1991).

\bibitem{Pv_theory} N. Isgur, D. Scora, B. Grinstein, and M. Wise,
  Phys. Rev D {\bf 39},  799 (1989).

\bibitem{Petar} P. Maksimovic, Ph.D. dissertation, 
  Massachusetts Institute of Technology, 1998 (unpublished).



\bibitem{Fumi} CDF Collaboration, F. Abe {\it et al}., 
Phys. Rev. Lett. {\bf 76}, 4462 (1996).


\bibitem{PDG97} R.M. Barnett {\it et al.}, Phys. Rev. D {\bf 54}, 1 (1996),
    and 1997 off-year partial update for the 1998 edition available on 
    the PDG WWW pages (URL: http://pdg.lbl.gov/). 


\bibitem{CDFmass} CDF Collaboration, F. Abe {\it et al}., 
 Phys. Rev. D {\bf 53}, 3496 (1996).        

\bibitem{CDFlife} CDF Collaboration, F. Abe {\it et al}., 
 Phys. Rev. Lett. {\bf 72}, 3456 (1994).


\bibitem{CDFbr} CDF Collaboration, F. Abe {\it et al}., 
 Phys. Rev. Lett. {\bf 76}, 2015 (1996);
 Phys. Rev. D {\bf 54}, 6596 (1996).

\bibitem{CDFcross} CDF Collaboration, F. Abe {\it et al}., 
 Phys. Rev. Lett. {\bf 75}, 1451 (1995).


\bibitem{Ken} K. Kelley, Ph.D. dissertation, 
  Massachusetts Institute of Technology, 1998 (unpublished).



\bibitem{ref:NDE}
P.~Nason, S.~Dawson, and R.~K.~Ellis, Nucl. Phys. B {\bf 327}, 49 (1988).

\bibitem{ref:MRSD0} A.D. Martin, W.J. Stirling, and P.G. Roberts, 
  Phys.~Rev.~ D {\bf 47}, 867 (1993).

\bibitem{ref:Peterson}
C.~Peterson {\em et al}., Phys.~Rev.~ D {\bf 27}, 105 (1985).

\bibitem{ref:QQ}
P.~Avery, K.~Read, and G.~Trahern, Cornell Internal Note CSN-212,
1985 (unpublished). 


\bibitem{ref:PYTHIA} H.-U.~Bengtsson and T.~Sj\"{o}strand, 
	Computer Physics Commun. {\bf 46}, 43 (1987);
``PYTHIA 5.7 and JETSET 7.4: Physics and Manual,''
by T. Sj\"{o}strand (Lund U.) LU-TP-95-20, Aug 1995.


\bibitem{ref:CTEQ} H.L. Lai  {\it et al}., 
  Phys.~Rev.~ D {\bf 51}, 4763 (1995).

\bibitem{Dejan} D. Vucinic,  Ph.D. dissertation, 
  Massachusetts Institute of Technology, 1998 (unpublished).

\end{references}
\end{document}